\documentclass[apj]{emulateapj}
\usepackage{apjfonts}
\usepackage{graphicx}

\slugcomment{Last modified: \today}
\shorttitle{The Galactic Center: Not an AGN}
\shortauthors{An et~al.}

\begin{document}
\title{The Galactic Center: Not an Active Galactic Nucleus}

\author{Deokkeun An\altaffilmark{1},
Solange V.\ Ram\'irez\altaffilmark{2},
Kris Sellgren\altaffilmark{3}}

\altaffiltext{1}{Department of Science Education, Ewha Womans University,
Seoul 120-750, Republic of Korea; deokkeun@ewha.ac.kr.}
\altaffiltext{2}{NASA Exoplanet Science Institute,
California Institute of Technology, Mail Stop 100-22, Pasadena, CA 91125.}
\altaffiltext{3}{Department of Astronomy, Ohio State University,
140 West 18th Avenue, Columbus, OH 43210.}

\begin{abstract} We present $10\ \mu$m--$35\ \mu$m {\it Spitzer} spectra of the
interstellar medium in the Central Molecular Zone (CMZ), the central $210$~pc
$\times$ $60$~pc of the Galactic center (GC). We present maps of the CMZ in
ionic and H$_2$ emission, covering a more extensive area than earlier
spectroscopic surveys in this region.  The radial velocities and intensities of
ionic lines and H$_2$ suggest that most of the H$_2$ 0--0 S(0) emission comes
from gas along the line-of-sight, as found by previous work.  We compare
diagnostic line ratios  measured in the {\it Spitzer} Infrared Nearby Galaxies
Survey (SINGS) to our data. Previous work shows that forbidden line ratios can
distinguish star-forming galaxies from LINERs and AGNs. Our GC line ratios agree
with star-forming galaxies and not with LINERs or AGNs.  \end{abstract}


\section{Introduction} 

The Galactic center (GC) is the closest galactic nucleus. At a distance of
$7.9\pm0.8$~kpc \citep{reid:09}, $1$~pc in the GC corresponds to only
$26\arcsec$.  The GC provides an opportunity to unveil interactions between
various physical processes in a nuclear environment of a galaxy with excellent
spatial resolution unapproachable for other galaxies.

The central $170\arcmin \times 40\arcmin$ ($\sim400$~pc $\times 90$~pc) region
of the Galaxy is often called the Central Molecular Zone \citep[CMZ;
][]{morris:96}.  The CMZ is a massive molecular cloud complex in the Galaxy,
which contains about 10\% of the Galaxy's molecular gas and produces 5\%--10\%
of its infrared (IR) and Lyman continuum luminosity
\citep{smith:78,nishimura:80,bally:87, bally:88,morris:96}. Earlier radio
continuum surveys revealed that the CMZ is an active star forming region and
contains the most active star forming cloud in the entire Galaxy, the
Sagittarius~B2 (Sgr~B2) complex. {\it Herschel} observations revealed that the
dust emission from the CMZ mainly arises in a ring-like structure
\citep{molinari:11}.  In this paper, we use the terms CMZ and GC
interchangeably, referring to the same $\sim200$~pc region in the center of the
Galaxy.

Star formation in the CMZ is inevitably affected by the extreme physical
conditions of the natal clouds, which have an order of magnitude higher gas
density than in the disk, with high gas temperature, pressure, turbulence,
strong tidal shear, and milli-Gauss magnetic field strengths \citep{morris:96}.
Because of these unusual conditions not found in nearby normal star-forming
regions in the disk, the nature of star formation in the CMZ is a subject of
active research.

A key to unlocking the secrets of star formation activities in the CMZ is a
detailed spectroscopic study of its interstellar medium (ISM) over a wide range
of wavelengths.  Mid-IR emission lines are particularly useful in analyzing the
physical properties of the GC gas because it lies behind heavy dust obscuration
($A_V \sim 30$~mag).  Forbidden emission lines in the mid-IR are generally
insensitive to electron gas temperatures, and can be used to determine physical
properties such as the electron density and ionization parameters, and to
identify sources of ionization.  Furthermore, molecular hydrogen emission from
pure rotational transitions can be observed in the mid-IR, which may hold vital
clues to gas heating mechanisms in the CMZ.

A mid-IR spectroscopic survey of the GC was previously conducted at $2\
\mu$m--$196\ \mu$m based on ISO observations \citep{rf:01a,rf:04,rf:05}.  They
observed $15$ different lines of sight to molecular clouds in the CMZ \citep[see
Figure~1 in][]{rf:05}.  \citet{rf:01a,rf:04} used observations of molecular
hydrogen emission to discuss heating mechanisms of warm molecular gas in the
CMZ, and concluded that low-density photon dominated regions (PDRs) and
low-velocity shocks ($v < 10$~km~s$^{-1}$) are required to explain the
temperatures derived from the warm molecular gas.  \citet{rf:05} analyzed fine
structure lines in the GC far from thermal continuum sources and massive
clusters.  They concluded that the ionizing radiation field is rather constant
throughout the CMZ, suggesting ionization by relatively hot and distant stars,
and found that excitation ratios, temperatures, and ionization parameters of
ionized gas in the CMZ are similar to those found in some low-excitation
starburst galaxies.

A more recent survey in the mid-IR was carried out by \citet{simpson:07} using
the Infrared Spectrograph \citep[IRS;][]{houck:04} onboard the {\it Spitzer
Space Telescope} \citep{werner:04}, which has a higher sensitivity than ISO at
$10\ \mu$m--$38\ \mu$m.  The \citeauthor{simpson:07} survey consists of
high-resolution spectra of $38$ positions along a narrow $24\arcmin$ long strip
at a Galactic longitude $l \approx +0.1\arcdeg$.  \citeauthor{simpson:07}
measured several forbidden emission lines and molecular hydrogen lines to
constrain the physical conditions of clouds near the Quintuplet Cluster, the
Arches Cluster, the Radio Arc Bubble \citep{rf:01b}, and Arched Filaments.
They concluded from their observations that the main source of excitation in the
GC is photo-ionization from the massive star clusters and that multi-component
PDR models can explain the observed line emission from molecular hydrogen
\citep[see also][]{contini:09}.  \citeauthor{simpson:07} also concluded
that shocks in the Radio Arc Bubble are responsible for strong [\ion{O}{4}]
emission.

In our recent spectroscopic survey of massive young stellar objects (YSOs) in
the CMZ \citep{an:09,an:11}, we used {\it Spitzer}/IRS to collect an extensive
set of mid-IR spectra for $107$ YSO candidates. The goal of this survey was to
discover and characterize the spectroscopic properties of massive YSOs in the
GC.  To achieve this original goal, we spent half of our observing time on
background spectra near each YSO candidate because of the strong and spatially
variable background emission in the GC.  Our GC background spectra, which are
the by-product of our YSO observing program, now constitute the largest and most
comprehensive mid-IR spectroscopic data set available to study the properties of
the ionized and molecular gas in the star-forming nucleus of the Galaxy.

In this paper, we present panoramic mapping results from ionic and molecular
hydrogen emission lines throughout the entire CMZ.  Data acquisition and
spectral analysis are presented in \S~\ref{sec:method}.  Line intensities and
radial velocities in the CMZ are presented in \S~\ref{sec:results}.  Mid-IR line
ratio diagnostics are used to compare the physical properties of the Galactic
nucleus to those of other nearby galaxies in \S~\ref{sec:discussion}.  Our
results are summarized in \S~\ref{sec:summary}.

\section{Method}\label{sec:method} 

\subsection{Observations and Extraction of Spectra}\label{sec:obs} 

\begin{figure*}
\centering
\includegraphics[scale=0.74]{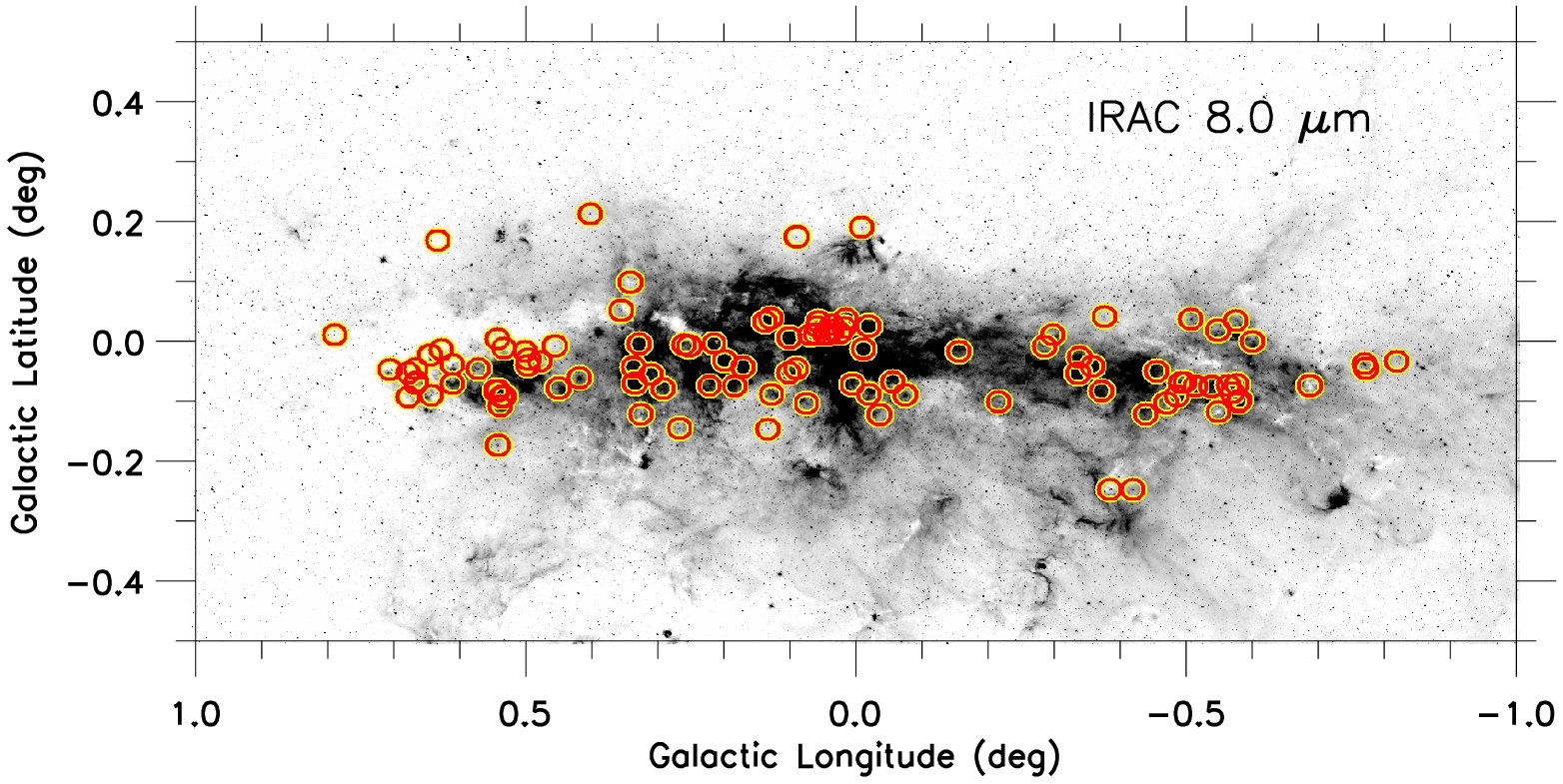}
\includegraphics[scale=0.74]{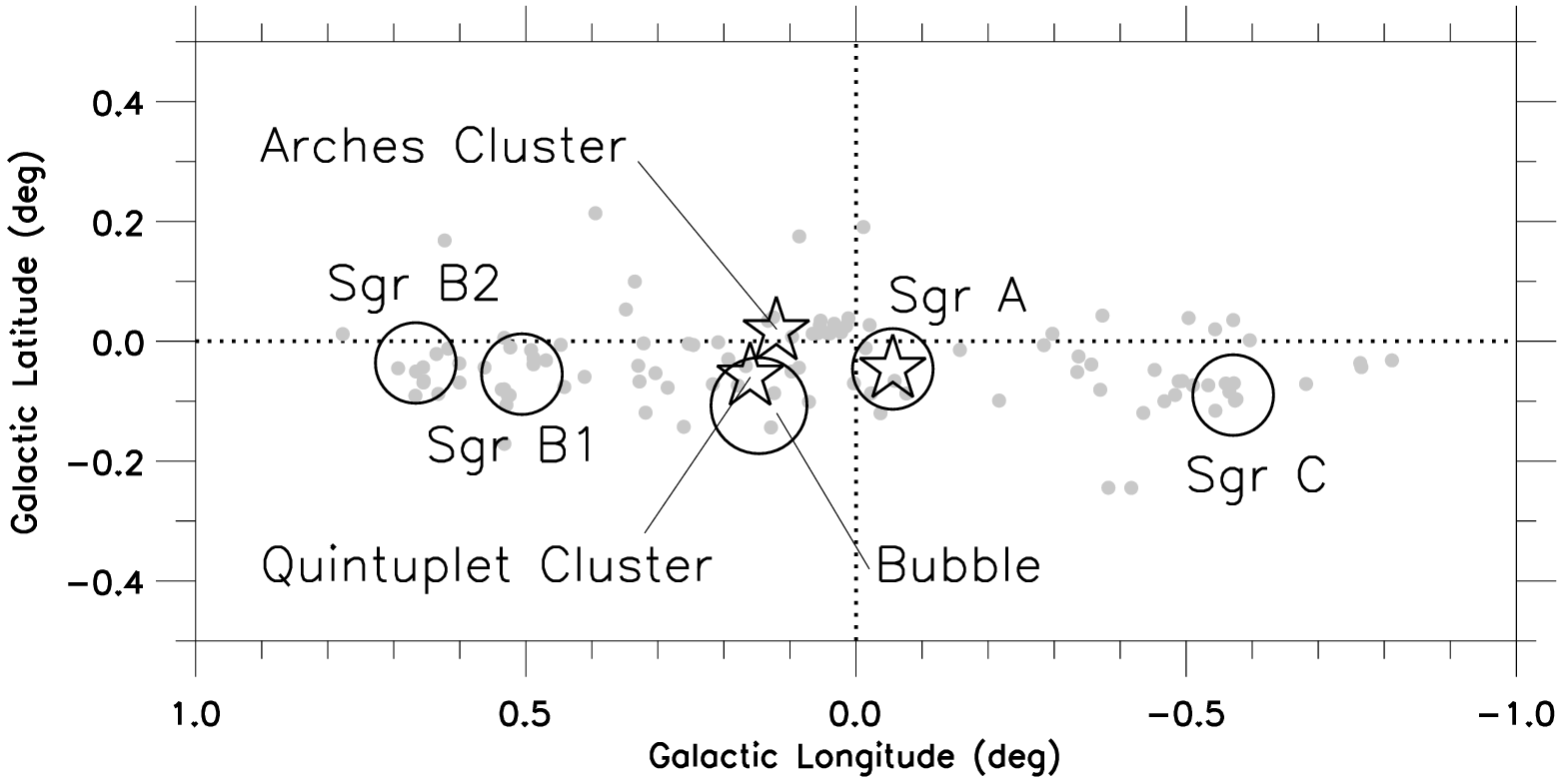}
\caption{ {\it Top:} Spatial distribution of 107 {\it Spitzer}/IRS targets on
the IRAC $8.0\ \mu$m image \citep{stolovy:06,ramirez:08} of the Galactic center
(GC).  The above image shows the entire Central Molecular Zone (CMZ), which
covers approximately $170\arcmin \times 40\arcmin$ ($\sim400$~pc $\times
90$~pc) centered on the GC.  Our IRS point-source targets \citep{an:09,an:11}
are marked with $1\arcmin$ ($2.3$~pc) radius circles. In this paper, we analyze
the GC interstellar medium (ISM) as measured by four background spectra that
are $\sim1\arcmin$ away from each target. {\it Bottom:} Schematic diagram of
locations of the CMZ molecular complexes (Sgr~A, Sgr~B1, Sgr~B2, and Sgr~C;
large open circles), the Radio Bubble (large open circle), and star clusters
(Quintuplet, Arches, and Central clusters; open star symbols).  The grey
circles represent the location of the {\it Spitzer}/IRS targets.
\label{fig:map}}
\end{figure*}

The IRS spectra presented in this paper were obtained in May and October 2008 as
part of {\it Spitzer} Cycle~4, during our observing program to identify massive
YSOs in the GC (Program ID: 40230, PI: S.\ Ram\'irez).  We targeted $107$ point
sources with extremely red colors on near- and mid-IR color-color diagrams as
YSO candidates \citep[see][for details]{an:11}.  The positions of our spectra
are shown in the top panel of Figure~\ref{fig:map}, on top of the {\it
Spitzer}/IRAC $8\ \mu$m image \citep{stolovy:06,ramirez:08}. Our survey
encompasses a large area in the flattened CMZ cloud complex ($|b| \la
0.24\arcdeg$), covering regions near strong radio continuum sources (Sgr~A,
Sgr~B1, Sgr~B2, and Sgr~C), massive star clusters (the Quintuplet, Arches, and
Central clusters), and the Radio Bubble (see the bottom panel of
Figure~\ref{fig:map}).

In our {\it Spitzer} program, massive YSO candidates were observed using both
high- and low-resolution IRS modules. In this work, however, we utilize only the
high-resolution observations taken with the short-high ({\tt SH}; $9.9\
\mu$m$-19.6\ \mu$m, $\lambda / \Delta \lambda \sim 600$; $4.7\arcsec \times
11.3\arcsec$ or $0.18$~pc $\times0.43$~pc slit entrance) and the long-high ({\tt
LH}; $18.7\ \mu$m$-37.2\ \mu$m, $\lambda / \Delta \lambda \sim 600$;
$11.1\arcsec \times 22.3\arcsec$ or $0.43$~pc $\times0.85$~pc slit entrance)
modules.  These two modules share the same slit centers on the sky.

Four background spectra were taken at $\sim1\arcmin$ ($2.3$~pc) away from each
of the YSO candidates; the positions of these points can be glimpsed by looking
at the open circles in the top panel of Figure~\ref{fig:map}, each of which has
a $1\arcmin$ radius. The exact locations of these spectra were carefully chosen
by visual inspection to avoid bright point sources on the IRAC $8\ \mu$m image
\citep{stolovy:06,ramirez:08}, and were intended to provide a level of
background mid-IR emission similar to that of the target position.  As shown in
Figure~\ref{fig:map}, our background spectra are somewhat uniformly distributed
over a $\sim90\arcmin\times25\arcmin$ ($\sim210$~pc $\times60$~pc) region in the
CMZ.

We reduced IRS spectra from the basic calibrated data (BCD) products version
{\tt S18.7.0}. We corrected for rogue pixel values using the software package
{\tt IRSCLEAN}\footnote{The SSC software packages can be found at\\ {\tt
http://irsa.ipac.caltech.edu/data/SPITZER/}.}
provided by {\it Spitzer} Science Center (SSC), and applied the {\tt DARKSETTLE}
software package to the {\tt LH} frames to correct for non-uniform dark current.
We extracted {\tt SH} and {\tt LH} spectra using the {\tt SPICE} tool in an
``extended'' extraction mode (with slit loss corrections for a source infinite
in extent), and further corrected for fringe patterns using the {\tt IRSFRINGE}
package. More information on our IRS observations and basic data reduction is
found in \citet{an:11}.

The contribution of zodiacal light is relatively high in the GC, typically
amounting to $\sim10\%$--$30\%$ of the continuum emission from the ISM in the
IRS spectral range. We used the zodiacal light estimator in the {\tt
SSC}-provided software {\tt SPOT}, which is based on COBE/DIRBE measurements, to
determine the contribution from the zodiacal light in each season.  The zodiacal
spectrum has a peak emission of $\sim40\ {\rm MJy\ sr^{-1}}$ at $20\ \mu$m, with
a $\sim10\%$ seasonal variation. We constructed a smoothed zodiacal spectrum
using quadratic interpolation, and then subtracted it from the extracted
spectra. Since our IRS targets are found within a degree of the GC, spatial
variations in the zodiacal emission are negligible. The correction for zodiacal
emission has no direct impact on emission line flux measurements, but has an
influence on the extinction correction since we measure foreground extinction
from the continuum emission near the $10\ \mu$m silicate feature (see
\S~\ref{sec:extinction}).

\begin{figure*}
\epsscale{0.95}
\plotone{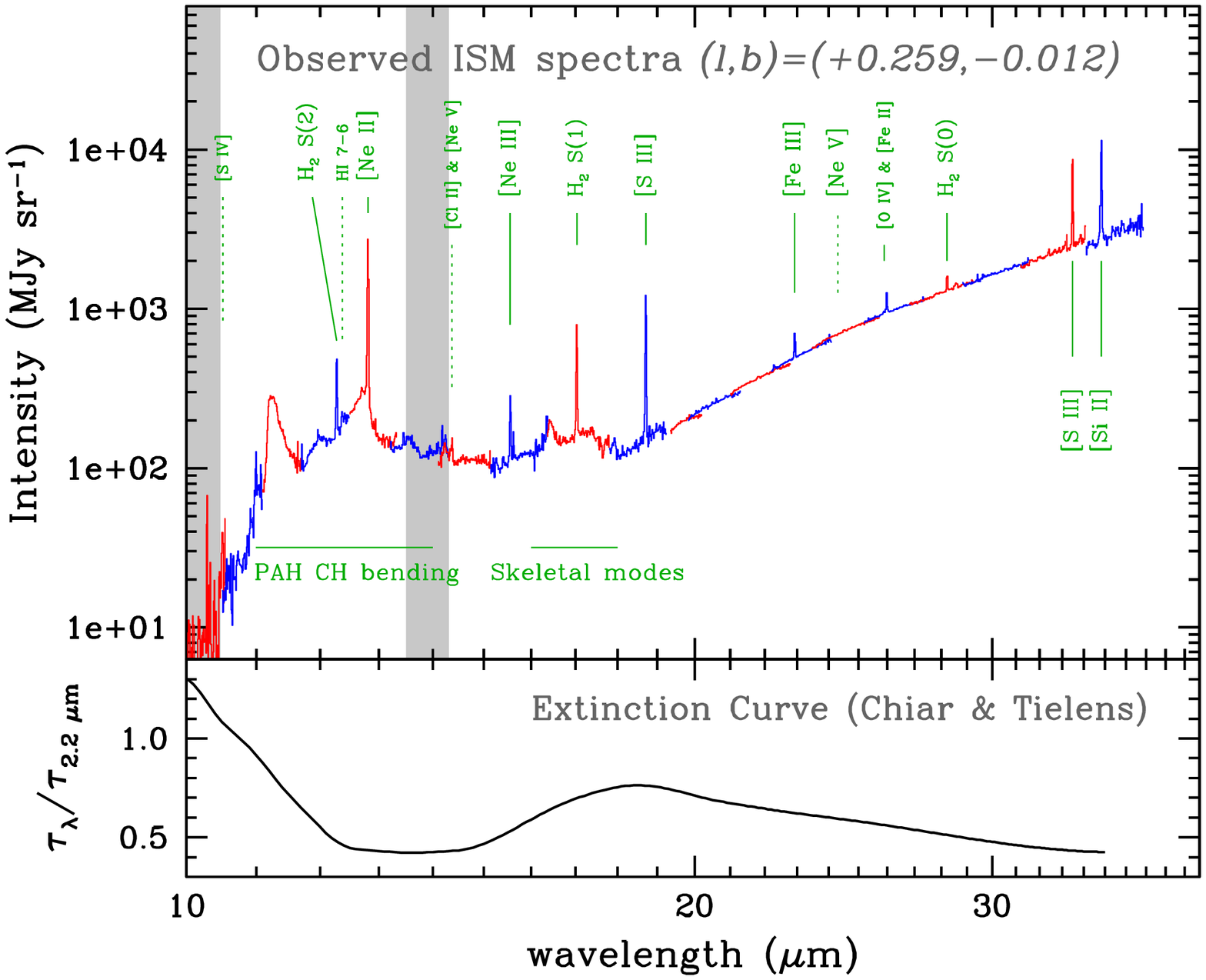}
\caption{{\it Top:} Observed {\it Spitzer}/IRS spectrum in the ISM of the GC,
measured at $(l,b) = (+0.2588\arcdeg, -0.0119\arcdeg)$, after subtracting the
zodiacal emission. Different spectral orders are shown in alternating colors.
Ionic forbidden emission lines and molecular hydrogen lines are marked with
vertical lines, with dotted lines indicating non-detected emission lines.
Vertical grey strips centered at $10.2\ \mu$m and $13.9\ \mu$m mark continuum
regions used to estimate the $9.7\ \mu$m silicate optical depth ($\tau_{9.7}$)
for each line of sight (see text).  {\it Bottom:} Extinction curve for the GC,
normalized to the extinction value in the $K$ passband \citep{chiar:06}.
\label{fig:spectra}}
\end{figure*}

We used the above procedure to obtain ISM spectra from {\tt SH} and {\tt LH} for
$428$ individual lines of sight in the GC (i.e., four background spectra for
each of $107$ point-source targets).  Figure~\ref{fig:spectra} shows one of the
observed spectra in the GC, after the zodiacal light correction.  Different IRS
spectral orders are shown in alternating colors. Forbidden emission lines and
molecular hydrogen lines are marked, and the wavelength ranges of strong
polycyclic aromatic hydrocarbon (PAH) emission features are indicated.

We identified spectral orders that best display individual lines, and used
Gaussian profile fits to compute line fluxes, as described in the following
section.  Figure~\ref{fig:spectra} shows that order tilts are present in some of
the IRS spectra.  In \citet{an:11}, we used low-resolution spectra to correct
for order tilts before we combined individual high-resolution spectra from
different orders.  In the following analysis, however, we combined spectra
without order-tilt corrections, because our low-resolution background slits are
not always co-spatial with the high-resolution slit positions.
Figure~\ref{fig:spectra} shows that the line fluxes should not be strongly
affected by the order tilts.

\subsection{Measurement of Emission Line Strengths and Radial
Velocities}\label{sec:fitting}

We measured emission line fluxes from several ionic fine-structure lines, as
well as molecular hydrogen lines from pure rotational transitions, H$_2$ 0--0
S(0), H$_2$ 0--0 S(1), and H$_2$ 0--0 S(2). A list of emission lines included in
this paper is shown in Table~\ref{tab:tab1}, in order of increasing wavelength.
In addition to these lines, we attempted to measure a flux from other ionic
features such as [\ion{Ar}{5}] $13.1\ \mu$m, [\ion{P}{3}] $17.89\ \mu$m,
[\ion{Fe}{2}] $17.94\ \mu$m, [\ion{Fe}{2}] $24.52\ \mu$m, and [\ion{Fe}{2}]
$35.35\ \mu$m, but none of these features were strong enough to be detected. The
third column in Table~\ref{tab:tab1} shows IRS modules and spectral orders, from
which individual line fluxes were measured (see below). The fourth column shows
normalized extinction coefficients from the GC extinction curve in
\citet{chiar:06}; see also the bottom panel of Figure~\ref{fig:spectra}.  The
ionization potential for each ion is listed in the last column in
Table~\ref{tab:tab1}.

\begin{deluxetable}{lcccc}
\tablewidth{0pt}
\tabletypesize{\scriptsize}
\tablecaption{Emission Lines Detected in {\it Spitzer}/IRS Spectra\label{tab:tab1}}
\tablehead{
  \colhead{Line} &
  \colhead{Wavelength} &
  \colhead{IRS Module/Order\tablenotemark{a}} &
  \colhead{$A_{\lambda}$/$A_K$\tablenotemark{b}} &
  \colhead{IP\tablenotemark{c}} \nl
  \colhead{} &
  \colhead{($\mu$m)} &
  \colhead{} &
  \colhead{} &
  \colhead{(eV)}
}
\startdata
{\rm [\ion{S}{4}]}   & $10.51$ & SH/19 & $1.080$ & $34.79$ \nl
{\rm H$_2$ S(2)}     & $12.28$ & SH/17 & $0.478$ & \nodata \nl
{\rm \ion{H}{1} 7-6} & $12.37$ & SH/17 & $0.462$ & \nodata \nl
{\rm [\ion{Ne}{2}]}  & $12.81$ & SH/16 & $0.435$ & $21.56$ \nl
{\rm [\ion{Ne}{5}]}  & $14.32$ & SH/14 & $0.430$ & $97.12$ \nl
{\rm [\ion{Cl}{2}]}  & $14.37$ & SH/14 & $0.431$ & $12.97$ \nl
{\rm [\ion{Ne}{3}]}  & $15.56$ & SH/13 & $0.533$ & $40.96$ \nl
{\rm H$_2$ S(1)}     & $17.04$ & SH/12 & $0.698$ & \nodata \nl
{\rm [\ion{S}{3}]}   & $18.71$ & SH/11 & $0.762$ & $23.34$ \nl
{\rm [\ion{Fe}{3}]}  & $22.93$ & LH/17 & $0.622$ & $16.19$ \nl
{\rm [\ion{Ne}{5}]}  & $24.32$ & LH/16 & $0.595$ & $97.12$ \nl
{\rm [\ion{O}{4}]}   & $25.89$ & LH/15 & $0.562$ & $54.94$ \nl
{\rm [\ion{Fe}{2}]}  & $25.99$ & LH/15 & $0.560$ & $7.90$  \nl
{\rm H$_2$ S(0)}     & $28.22$ & LH/14 & $0.513$ & \nodata \nl
{\rm [\ion{S}{3}]}   & $33.48$ & LH/12 & $0.433$ & $23.34$ \nl
{\rm [\ion{Si}{2}]}  & $34.82$ & LH/11 & $0.426$ & $8.15$  \nl 
\enddata
\tablenotetext{a}{IRS modules and spectral orders used in the extraction of line fluxes.}
\tablenotetext{b}{Extinction curve in \citet{chiar:06}.}
\tablenotetext{c}{Ionization potential for each ion.}
\end{deluxetable}

\begin{figure*}
\epsscale{1.0}
\plotone{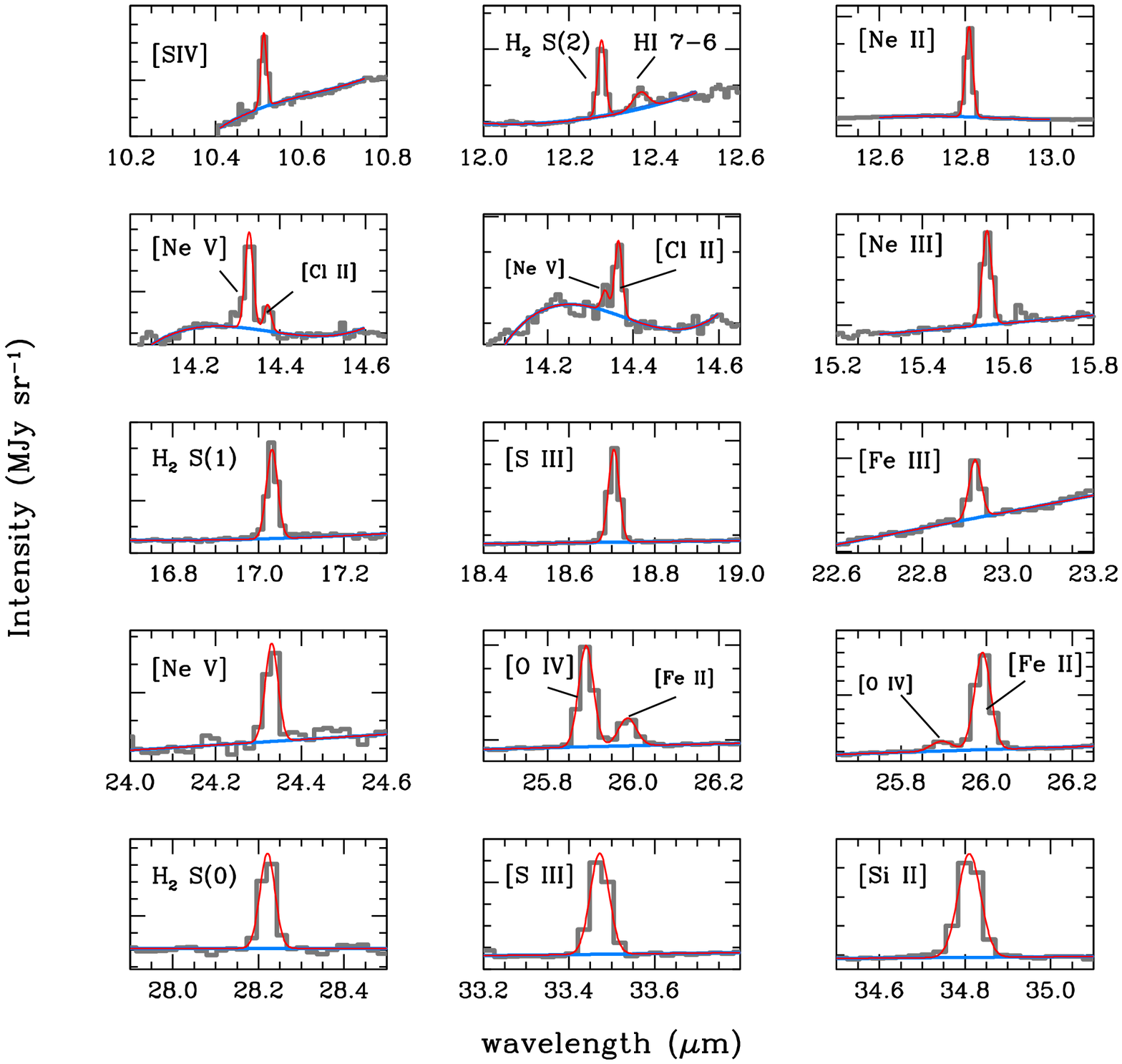}
\caption{Examples of line profile fits for the GC ISM spectra. Grey histograms
show observed {\it Spitzer}/IRS spectra, displayed over the same wavelength
interval, but with an arbitrary intensity scale on the vertical axis.  Ionic
emission lines as well as pure rotational lines from molecular hydrogen, such as
H$_2$ $0-0$ S(0), H$_2$ $0-0$ S(1), and H$_2$ $0-0$ S(2), are shown.  The
best-fitting Gaussian profile is shown as a red line on top of a local continuum
(blue line).
\label{fig:line}}
\end{figure*}

Figure~\ref{fig:line} shows examples of the line profile fits for the emission
lines listed in Table~\ref{tab:tab1}. Grey histograms are observed IRS spectra
from various lines of sight.  The red line represents the best-fitting Gaussian
profile, obtained using the non-linear least squares fitting routine MPFIT
\citep{markwardt:09}. The underlying blue line shows a local continuum
constructed from a 1$^{st}$--3$^{rd}$ order polynomial fit to the continuum
points on each side of the emission line. A standard deviation of data points
from the continuum line was used as an effective $1\sigma$ error per data point
in the profile fitting. High order polynomials were used to determine a local
continuum near PAH emission features. Simultaneous fits of two Gaussian profiles
were made for two sets of blended emission lines, [\ion{Ne}{5}] $14.32\ \mu$m
and [\ion{Cl}{2}] $14.37\ \mu$m, and [\ion{O}{4}] $25.89\ \mu$m and
[\ion{Fe}{2}] $25.99\ \mu$m.

The total line flux was computed by integrating the underlying flux of the best
fitting Gaussian profile in wavenumber space. Errors in these line fluxes were
estimated by adding in quadrature the $1\sigma$ flux uncertainties derived from
uncertainties in the height and width of the  Gaussian fit, or by propagating
$1\sigma$ errors in the local continuum, whichever is larger. The latter was
computed as \begin{equation}\label{eq:sigma} \sigma_{\rm flux} \approx F_{\rm
rms}\ \Delta \nu_{\rm max} \approx F_{\rm rms}\ \left(
\frac{c}{\lambda^2}\right) \Delta \lambda_{\rm max}, \end{equation} where
$F_{\rm rms}$ is the rms dispersion in the local continuum. The parameter
$\Delta \lambda_{\rm max}$ is the maximum value of the full-width at
half-maximum (FWHM) of the line profile set in the Gaussian line fitting, which
we assumed to be $\Delta v = 600$~km~s$^{-1}$ for all lines.  In most cases,
errors propagated from Gaussian fits were larger than those estimated using
Equation~\ref{eq:sigma}.

The line center was allowed to vary in our line fitting procedure, from which we
measured line-of-sight velocities ($v_r$) from individual emission lines.
Although high-resolution spectra from {\it Spitzer}/IRS have a relatively low
spectral resolution ($R \equiv \Delta \lambda / \lambda = 600$), radial
velocities can be obtained with a precision of a few tens of kilometers per
second for strong emission lines \citep[e.g.,][]{simpson:07}. We computed radial
velocities in the Local Standard of Rest (LSR), after correcting for the
spacecraft motion: the correction terms (keyword {\tt VLSRCORR} provided by the
SSC) for May and October 2008 observing runs were $+38.9$~km/s and $-18.5$~km/s,
respectively.

The accuracy of our radial velocity measurements was evaluated as follows.
First of all, we compared $v_r$ measurements from the same ionic species,
[\ion{S}{3}] $18.71\ \mu$m and [\ion{S}{3}] $33.48\ \mu$m, which are found on
two different IRS modules. We found that the mean difference in $v_r$ between
these lines is $\Delta v_r = 31$~km~s$^{-1}$, indicating the error from a
module-to-module and/or order-to-order change in the IRS wavelength calibration.
In addition, the scatter in the $v_r$ difference from these two lines is
$\sigma_{v} = 17$~km~s$^{-1}$, which is a measure of the precision in the $v_r$
determination.

The size of systematic errors in $v_r$ can be further examined by comparing
$v_r$ determined among strong ionic emission lines. We used [\ion{Ne}{2}]
$12.81\ \mu$m as a basis of comparison in $v_r$. We found a median difference in
$v_r$ for all of the IRS spectra at $l \leq +0.2\arcdeg$ as follows: $\Delta v_r
= +5.3$~km~s$^{-1}$ for [\ion{Ne}{3}] $15.56\ \mu$m, $\Delta v_r =
-10.8$~km~s$^{-1}$ for [\ion{S}{3}] $18.71\ \mu$m, $\Delta v_r =
+21.7$~km~s$^{-1}$ for [\ion{S}{3}] $33.48\ \mu$m, and $\Delta v_r =
+71.1$~km~s$^{-1}$ for [\ion{Si}{2}] $34.82\ \mu$m, respectively, in the sense
that a positive difference indicates a larger $v_r$ from a given line than from
[\ion{Ne}{2}].  We used these $\Delta v_r$ values to first shift all ionic lines
to the [\ion{Ne}{2}] velocity.  The [\ion{Ne}{2}] line has on average a smaller
$v_r$ by $\Delta v_r = 22.60$~km~s$^{-1}$ than the mean $v_{\rm LSR} =
+51.0$~km~s$^{-1}$ measured from a hydrogen recombination line in Sgr~B1
\citep{mehringer:92}.  We thus applied this second correction to shift all ionic
lines to match \citet{mehringer:92} in Sgr B1.

Molecular hydrogen lines from pure rotational transitions are also strong enough
to be detected in the GC and provide radial velocities. We compared $v_r$ from
individual IRS spectra to the peak $v_r$ of CO $J=4\rightarrow3$ emission
\citep{martin:04} at $-0.2\arcdeg < l < 1.0\arcdeg$, and found $\Delta v_r =
+62.5$~km~s$^{-1}$, $+51.2$~km~s$^{-1}$, and $+29.4$~km~s$^{-1}$, for H$_2$ S(2)
$12.28\ \mu$m, H$_2$ S(1) $17.04\ \mu$m, and H$_2$ S(0) $28.22\ \mu$m,
respectively. The sense of the difference is that radial velocities from these
lines are on average lower than those from the CO line.  We then shift all H$_2$
lines to the CO velocity.

The above systematic offsets in the measured $v_r$ indicate that the wavelength
calibration error in the {\it Spitzer}/IRS spectra is on the order of at least a
few tens of kilometers per second. To summarize, we put all $v_r$ measurements
from the IRS observations in the Local Standard of Rest (LSR) by applying
zero-point offsets to $v_r$.

\subsection{Foreground Extinction Estimates}\label{sec:extinction}

We corrected line flux measurements for foreground extinction between the GC and
the Sun. This was done on a spectrum-to-spectrum basis, because of the patchy
dust  extinction in the GC. We utilized two different approaches to correcting
for foreground extinction.

\begin{figure*}
\centering
\includegraphics[scale=0.64]{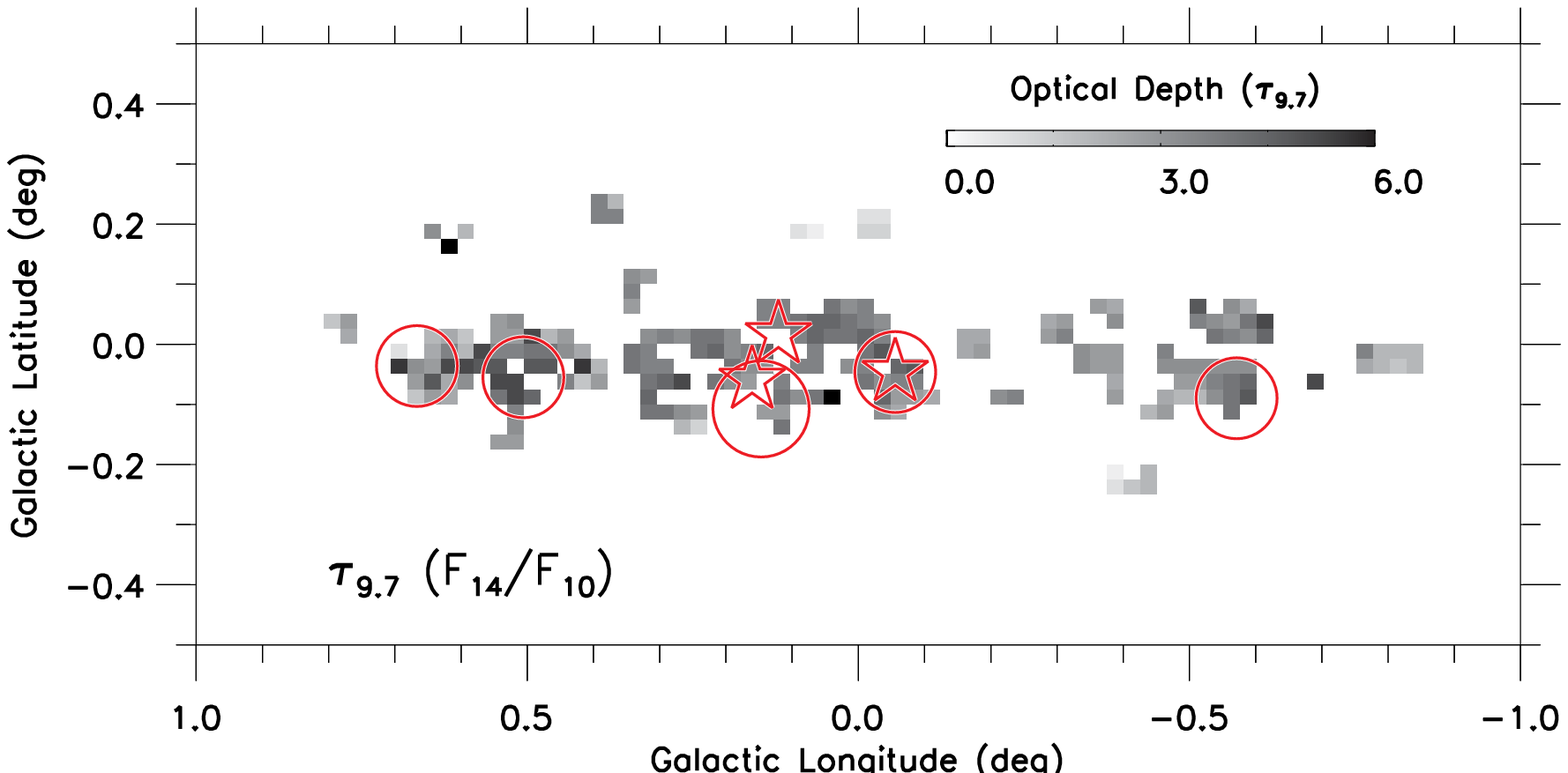}
\includegraphics[scale=0.64]{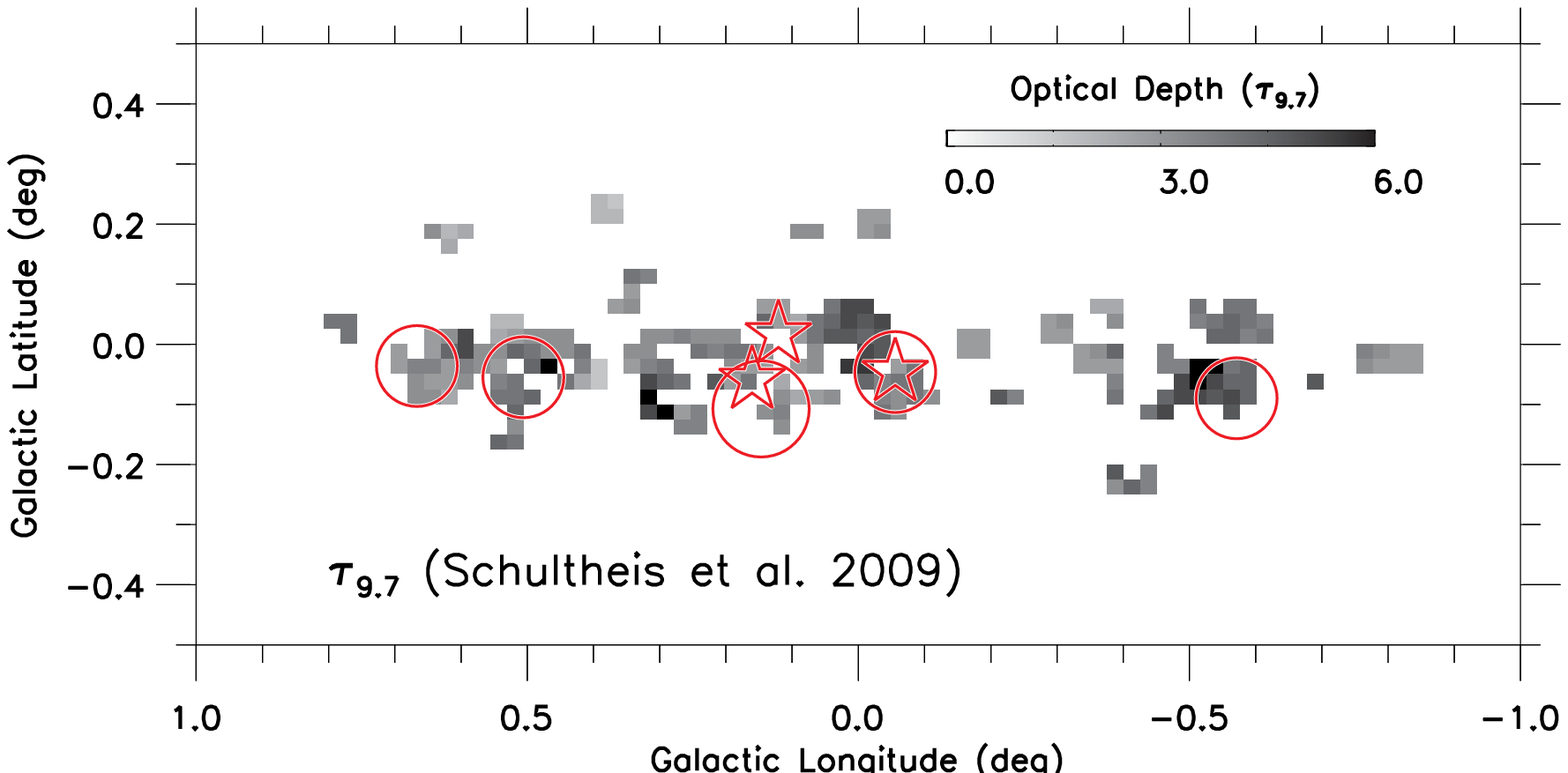}
\caption{Extinction maps shown as the silicate optical depth at $9.7\ \mu$m
($\tau_{9.7}$) in the line of sight to the GC. Each pixel covers
$1.5\arcmin\times1.5\arcmin$ ($\sim 3.5$~pc $\times 3.5$~pc), and key features
of the GC are overlaid (see Figure~\ref{fig:map}). {\it Top:} The $9.7\ \mu$m
silicate optical depth inferred from the ratio of $10\ \mu$m and $14\ \mu$m
continuum fluxes based on the \citet{simpson:07} method. {\it Bottom:} The $9.7\
\mu$m silicate optical depth from the \citet{schultheis:09} extinction map,
which is based on 2MASS and {\it Spitzer}/IRAC colors of GC giants.  We convert
$A_V$ from \citet{schultheis:09} into $\tau_{9.7}$ by adopting $A_V / \tau_{9.7}
= 9$ \citep{roche:85}.
\label{fig:tau}}
\end{figure*}

The first approach is based on the method developed in \citet{simpson:07}, who
determined  the flux ratio at $10\ \mu$m and $14\ \mu$m ($F_{14}/F_{10}$) and
used it to infer the optical depth of the $9.7\ \mu$m silicate absorption
feature ($\tau_{9.7}$) near the Radio Arc Bubble. Following the
\citeauthor{simpson:07} prescription, we estimated mean fluxes at $10.00\ \mu$m
$\leq \lambda \leq 10.48\ \mu$m and $13.50\ \mu$m $\leq \lambda \leq 14.30\
\mu$m from {\tt SH} spectra after a $3\sigma$ rejection.  As shown by vertical
grey strips in Figure~\ref{fig:spectra}, these two continuum wavelength ranges
are almost free of PAHs features and ionic emission lines.  Then we estimated
$\tau_{9.7}$ for each line of sight, by using a simple linear relationship
between the $9.7\  \mu$m silicate optical depth and the continuum flux ratio,
$\tau_{9.7} = (\ln{(F_{14}/F_{10})} - 0.809) / 0.560$, which was inferred from
tabulated values in \citet{simpson:07}.  The top panel in Figure~\ref{fig:tau}
shows the spatial distribution of  $\tau_{9.7}$ derived in this way for all
lines of sight in our program.

Alternatively, we utilized the $A_V$ extinction map toward the GC
\citep{schultheis:09} based on the observed stellar locus of red giant stars on
2MASS and {\it Spitzer}/IRAC color-color diagrams. The bottom panel in
Figure~\ref{fig:tau} shows the distribution of  $\tau_{9.7}$ from
\citet{schultheis:09}, after converting their $A_V$ values into $\tau_{9.7}$
using $A_V / \tau_{9.7} = 9$ \citep{roche:85}.

\begin{figure}
\epsscale{1.05}
\plotone{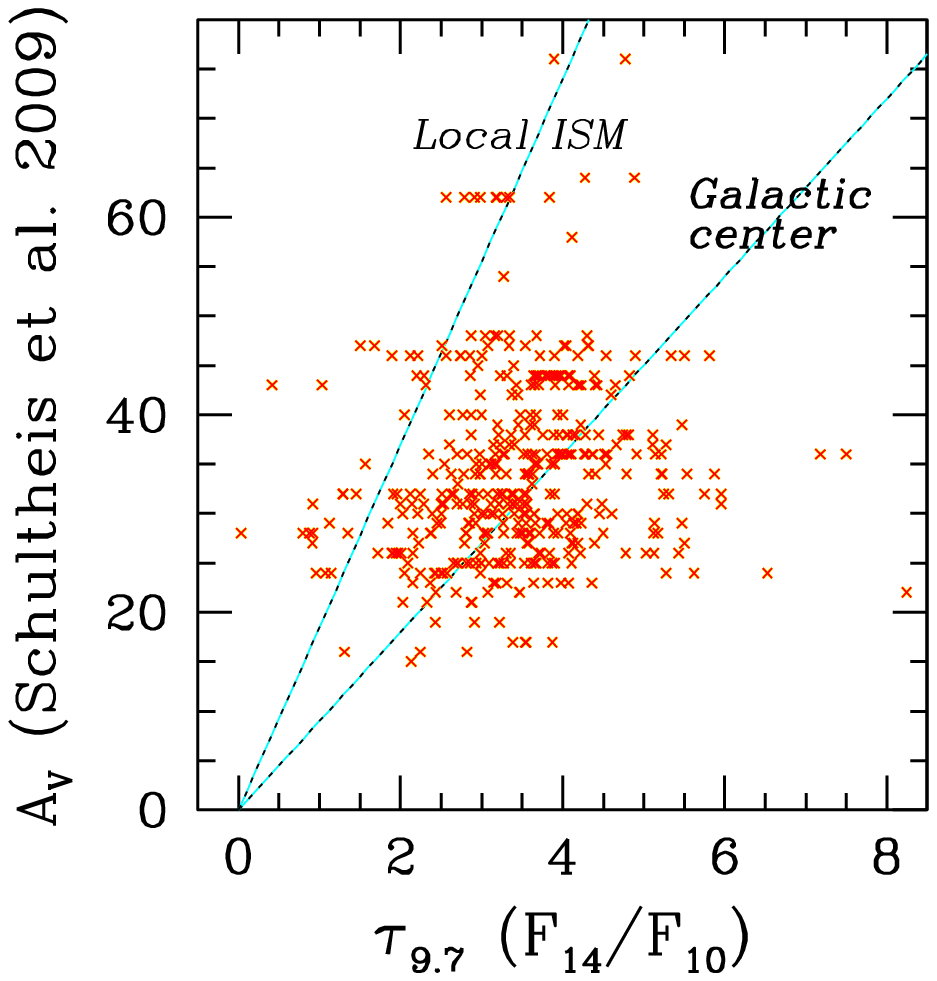}
\caption{Comparison between the silicate optical depth at $9.7\ \mu$m
($\tau_{9.7}$) derived using the \citet{simpson:07} method (F$_{14}$/F$_{10}$)
and the visual extinction ($A_V$) from the \citet{schultheis:09} extinction map
for all of the IRS background spectra. Dashed lines are previously measured
relationships in the line of sight to the GC \citep[$A_V / \tau_{9.7} =
9$;][]{roche:85} and for the local ISM \citep[$A_V / \tau_{9.7} =
18.5$;][]{roche:84}, respectively.
\label{fig:extinction}}
\end{figure}

Figure~\ref{fig:extinction} compares $\tau_{9.7}$ derived using the
\citet{simpson:07} method to the visual extinction ($A_V$) from
\citet{schultheis:09}.  Figure~\ref{fig:extinction} shows a large scatter
between these two independent extinction estimates.  However, their mean trend
agrees well with the GC relation \citep[][$A_V / \tau_{9.7} = 9$]{roche:85}.
For comparison, we also show the relationship for the local ISM \citep[][$A_V /
\tau_{9.7} = 18.5$]{roche:84} in Figure~\ref{fig:extinction}, which does not
match the observed trend in the GC at all. 

In the following analysis, we derived $\tau_{9.7}$ using \citet{simpson:07}
method and present results after correcting line fluxes for extinction, unless
otherwise stated.  However, we repeated our analysis with foreground extinction
from the \citet{schultheis:09} map, and compared results with each other. As
shown below, our main results are solid, and are only weakly dependent on the
specific choice of foreground extinction correction.

In addition to the above extinction corrections, we further corrected observed
line fluxes from [\ion{Ne}{3}] $15.56\ \mu$m for the absorption induced by
CO$_2$ ice grains on a spectrum-to-spectrum basis.  We used the same procedure
as in \citet{an:11} to decompose a wide CO$_2$ ice absorption band over $\sim15\
\mu$m--$16\ \mu$m using five laboratory spectral components of different CO$_2$
ice mixtures, and obtained an optical depth of the CO$_2$ ice from the
best-fitting model profile. The mean optical depth from all of the GC ISM
spectra is $\langle \Delta \tau \rangle=0.065$ at the position of the
[\ion{Ne}{3}] line. The relatively small optical depth at $15.56\ \mu$m is
because the [\ion{Ne}{3}] line is located in the long-wavelength wing of the
CO$_2$ ice absorption, in addition to the fact that the CO$_2$ ice band in the
ISM is weaker and narrower than those seen in massive YSOs \citep[see][]{an:11}.

For a given $\tau_{9.7}$, we corrected a line flux for interstellar extinction
at each wavelength using \begin{equation} A_\lambda = \left(
\frac{A_\lambda}{A_K} \right) \left( \frac{A_K}{A_V} \right) \left(
\frac{A_V}{\tau_{9.7}} \right) \tau_{9.7}, \end{equation} where $A_\lambda /
A_K$ is the GC extinction curve value in \citet{chiar:06}, normalized to the
extinction in the $K$ bandpass (see Table~\ref{tab:tab1}). We adopted $A_K /
A_V\approx0.11$ \citep{figer:99} and the GC relation of $A_V / \tau_{9.7} = 9$
as determined by \citet{roche:85}. The extinction-corrected line flux was then
obtained using \begin{equation} \log{f_0} = \log{f_{\rm obs}} + 0.4 A_\lambda.
\end{equation}

\subsection{Coadded GC ISM Spectra}\label{sec:coadd}

\begin{figure*}
\epsscale{0.95}
\plotone{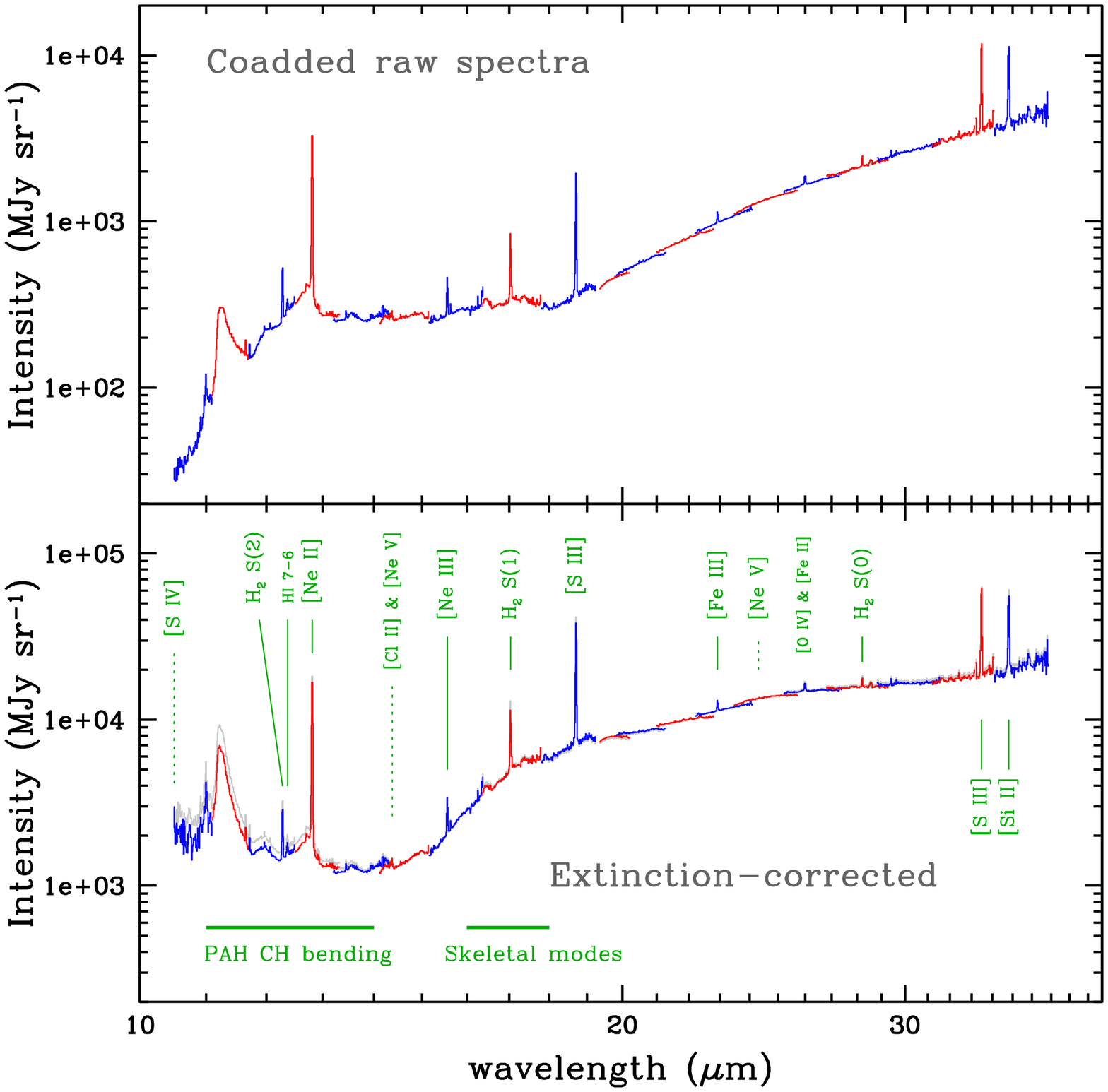}
\caption{Coadded GC ISM spectra constructed from 428 high-resolution IRS
spectra.  Locations of ionic forbidden emission lines and molecular hydrogen
lines are marked, where dotted lines indicate positions of weakly detected
emission lines.  {\it Top:} Coadded spectrum created by adding fluxes from
individual IRS spectra without foreground extinction corrections. {\it Bottom:}
Same as in the top panel, but coadded after correcting individual spectra for
extinction based on the \citet{simpson:07} method.  Underlying grey line shows
the same but with extinction corrections based on the \citet{schultheis:09}
extinction map. Different spectral orders are shown in alternating colors.
\label{fig:coadd}}
\end{figure*}

In addition to individual ISM spectra, we also created and analyzed a coadded GC
spectrum as shown in Figure~\ref{fig:coadd}. This allows us to compare the ISM
spectrum measured across $\sim200$~pc to the individual spectra with the
$\sim3.5$~pc resolution.  We constructed the coadded IRS spectrum by summing
fluxes from $428$ individual high-resolution spectra in the GC.  The top panel
in Figure~\ref{fig:coadd} shows the spectrum coadded without any foreground
extinction correction.  The bottom panel displays the result of correcting each
individual spectrum for extinction based on the \citet{simpson:07} method
(\S~\ref{sec:extinction}), then coadding the corrected spectra.  We also
illustrate the spectrum created by correcting each individual spectrum for
extinction with the \citet{schultheis:09} extinction map, then coadding.  The
coadded spectra from each extinction correction method are very similar.  We
present below line fluxes from the coadded spectrum together with those obtained
from individual spectra.

\section{Results}\label{sec:results}

\subsection{Panoramic Emission Line Mapping in the GC}\label{sec:mapping}

Panoramic emission line maps are displayed in Figure~\ref{fig:elines} in a
logarithmic flux scale for each line. Only those lines detected at more than a
$3\sigma$ level were included. Table~\ref{tab:tab2} provides the individual line
fluxes used in these maps, after correcting for dust extinction using the
\citet{simpson:07} technique. Only a portion of Table~\ref{tab:tab2} is shown
here to demonstrate its form and content, and a machine-readable version of the
full table is available in the electronic edition of the Journal.

\begin{figure*}
\centering
\includegraphics[scale=0.42]{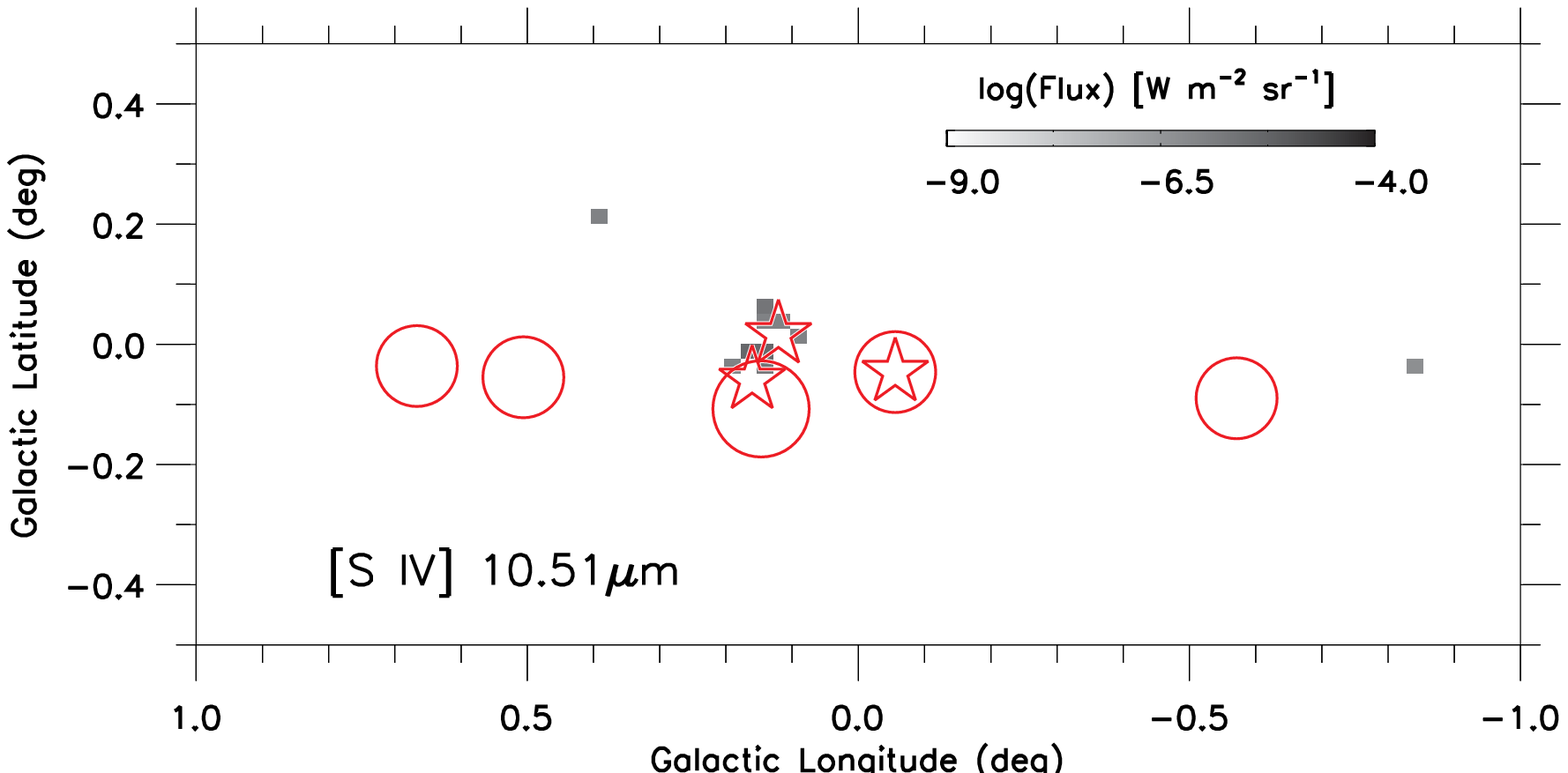}
\includegraphics[scale=0.42]{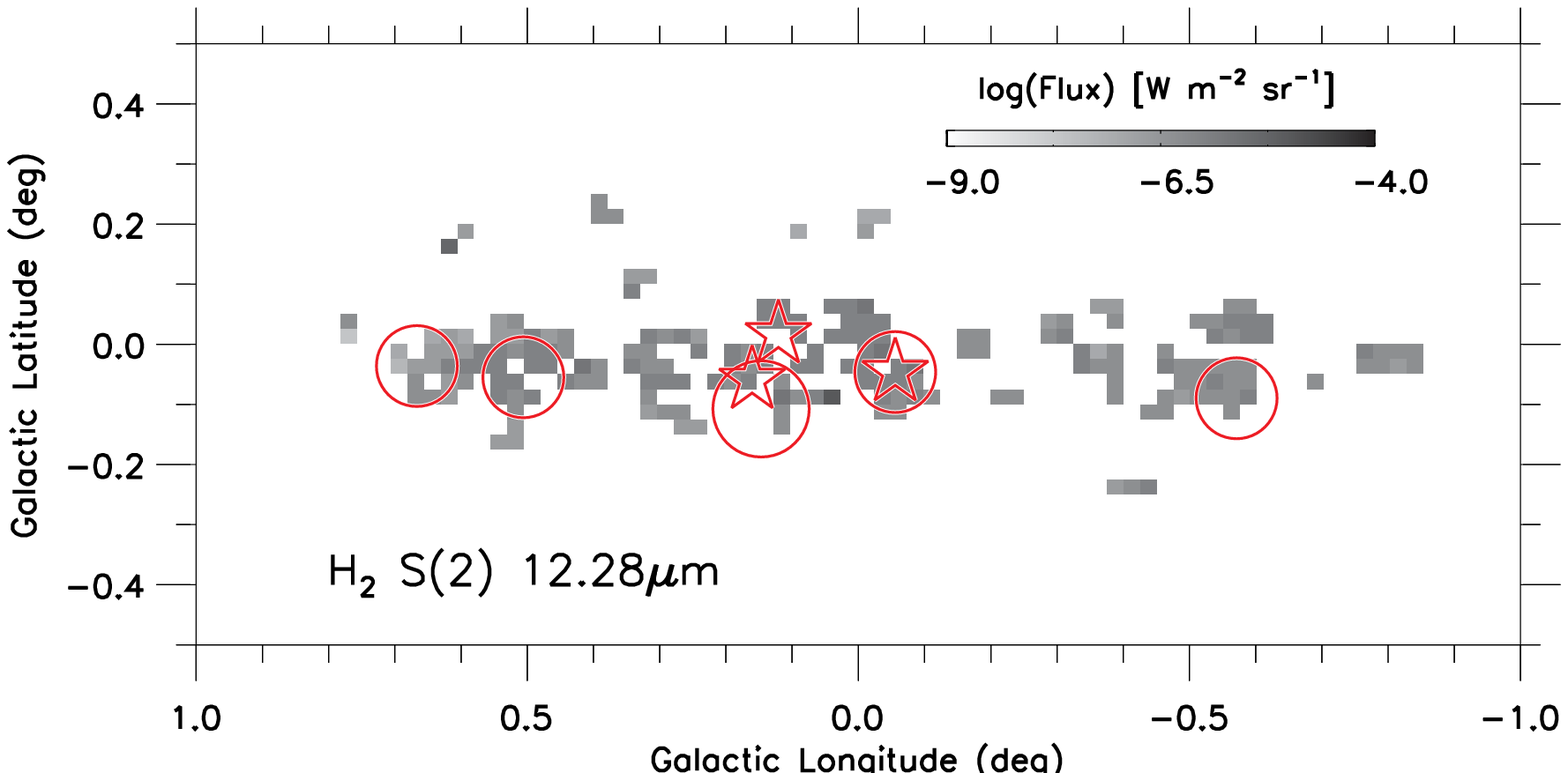}
\includegraphics[scale=0.42]{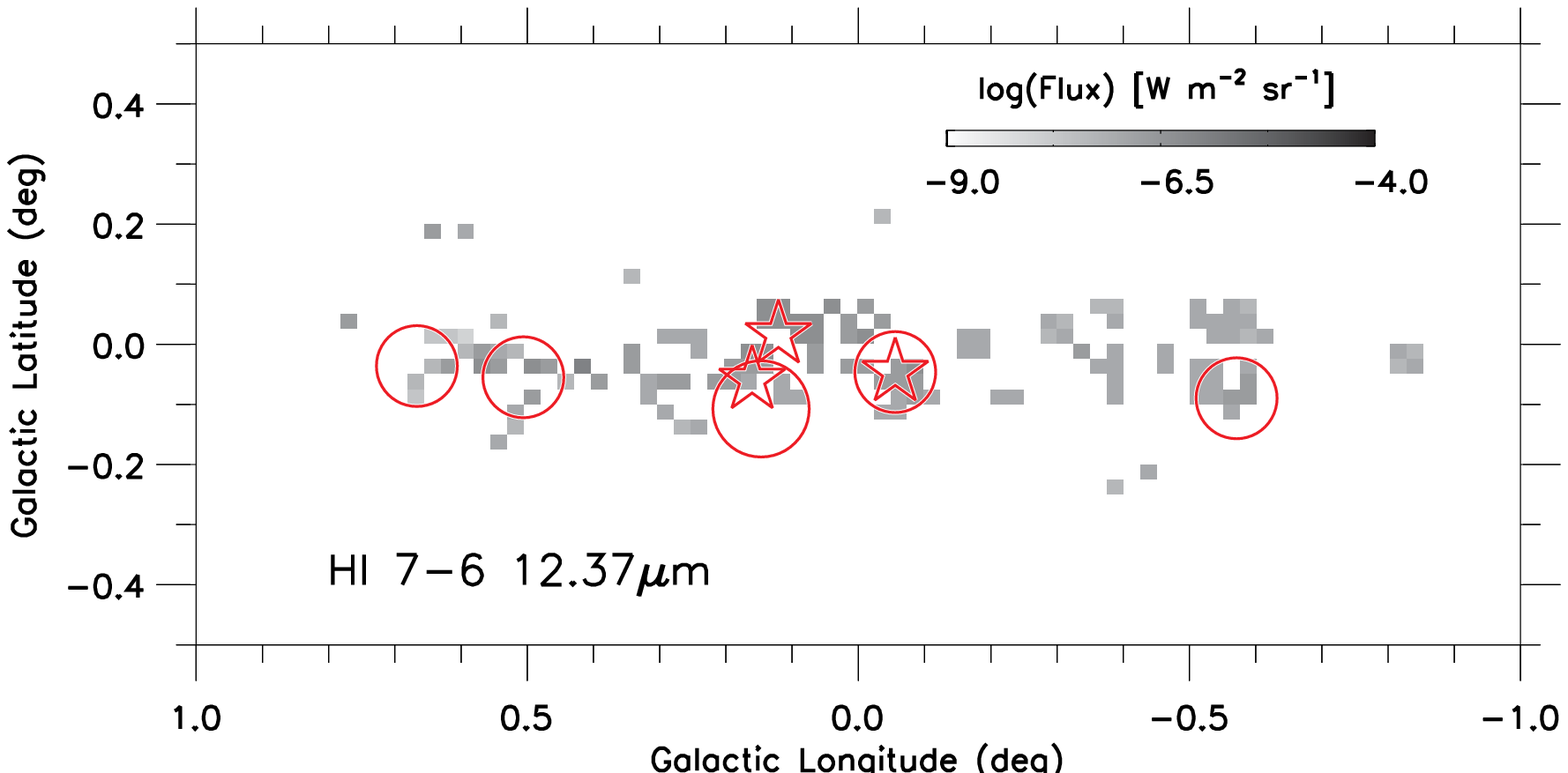}
\includegraphics[scale=0.42]{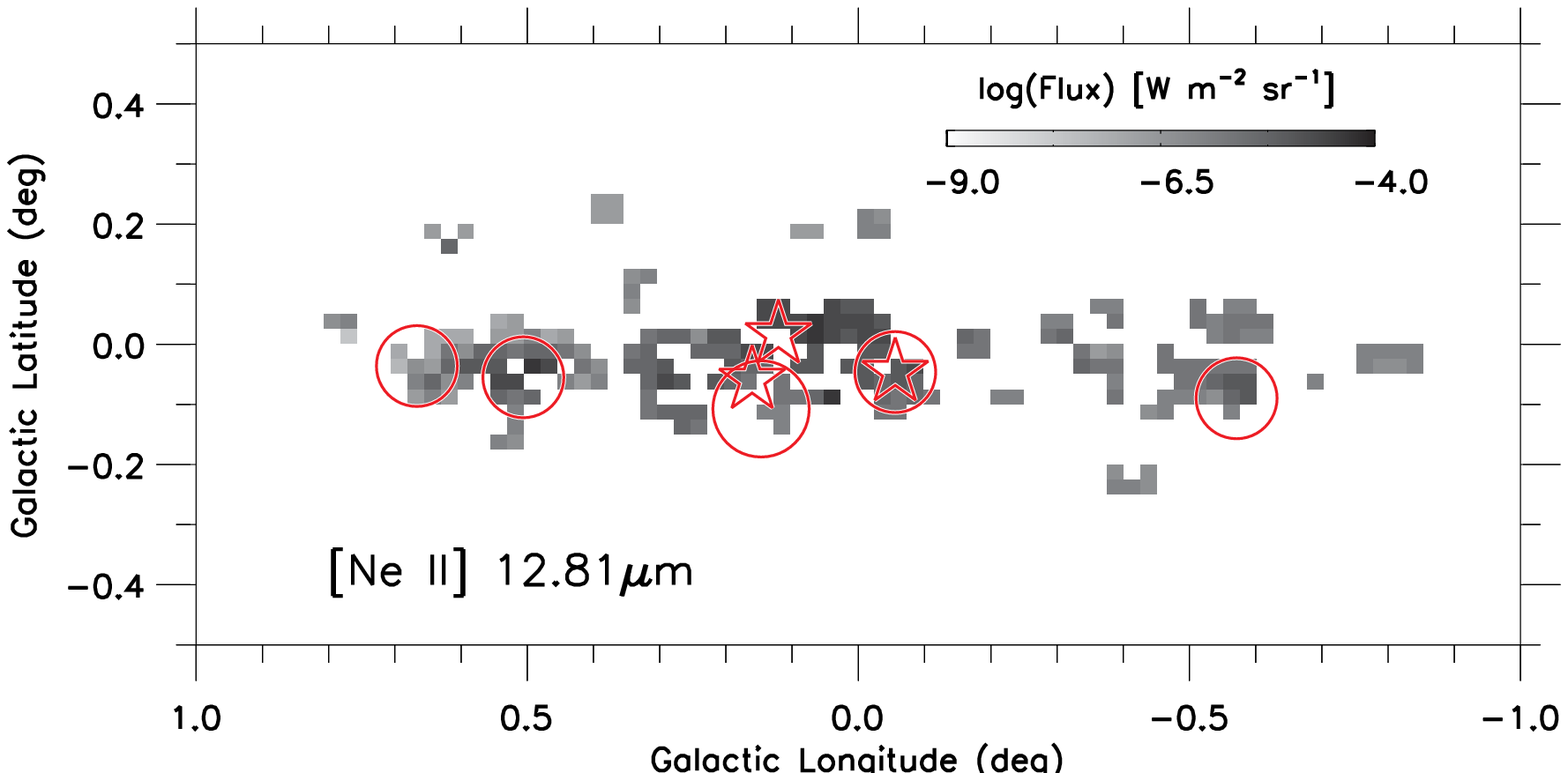}
\includegraphics[scale=0.42]{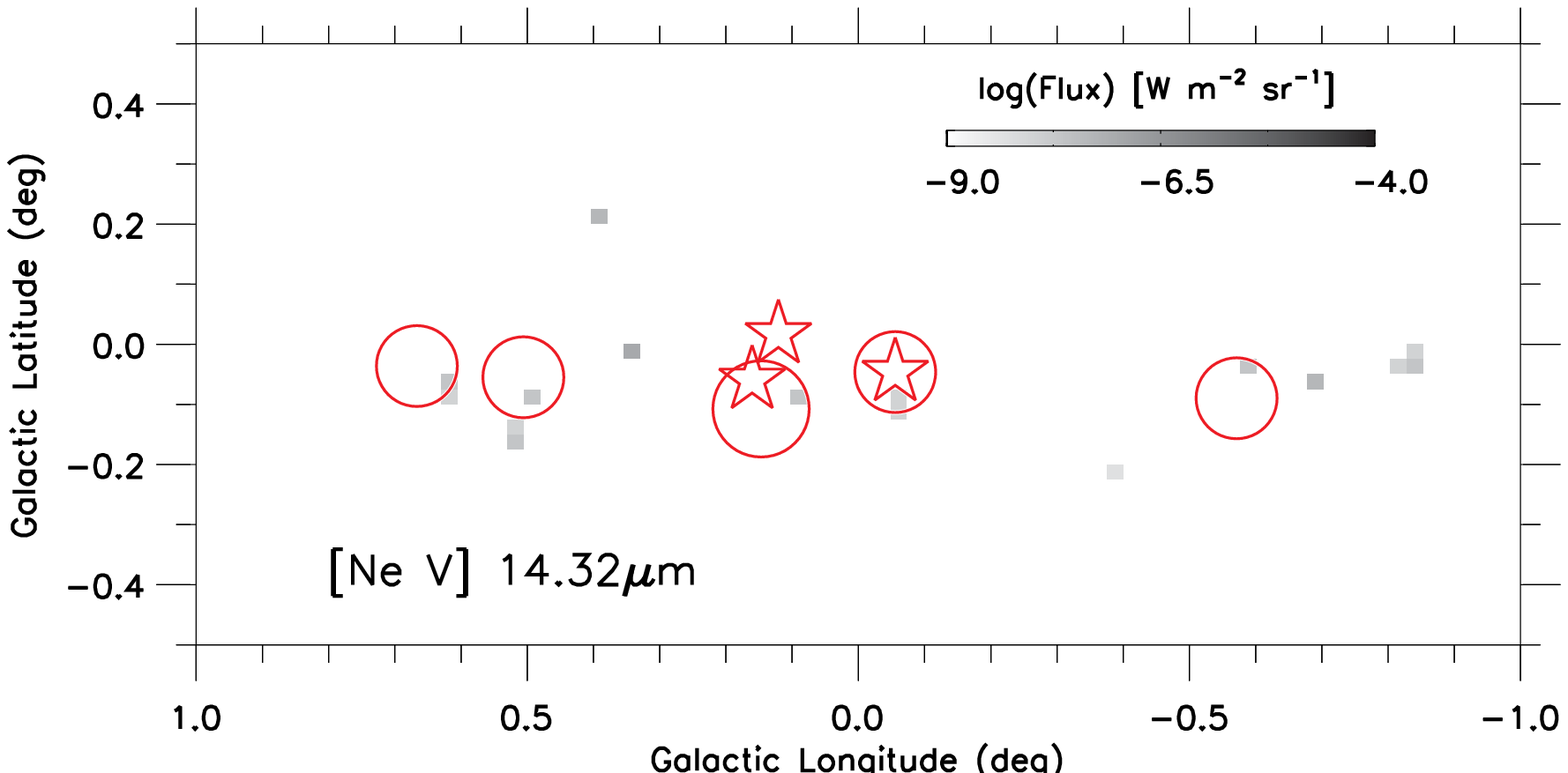}
\includegraphics[scale=0.42]{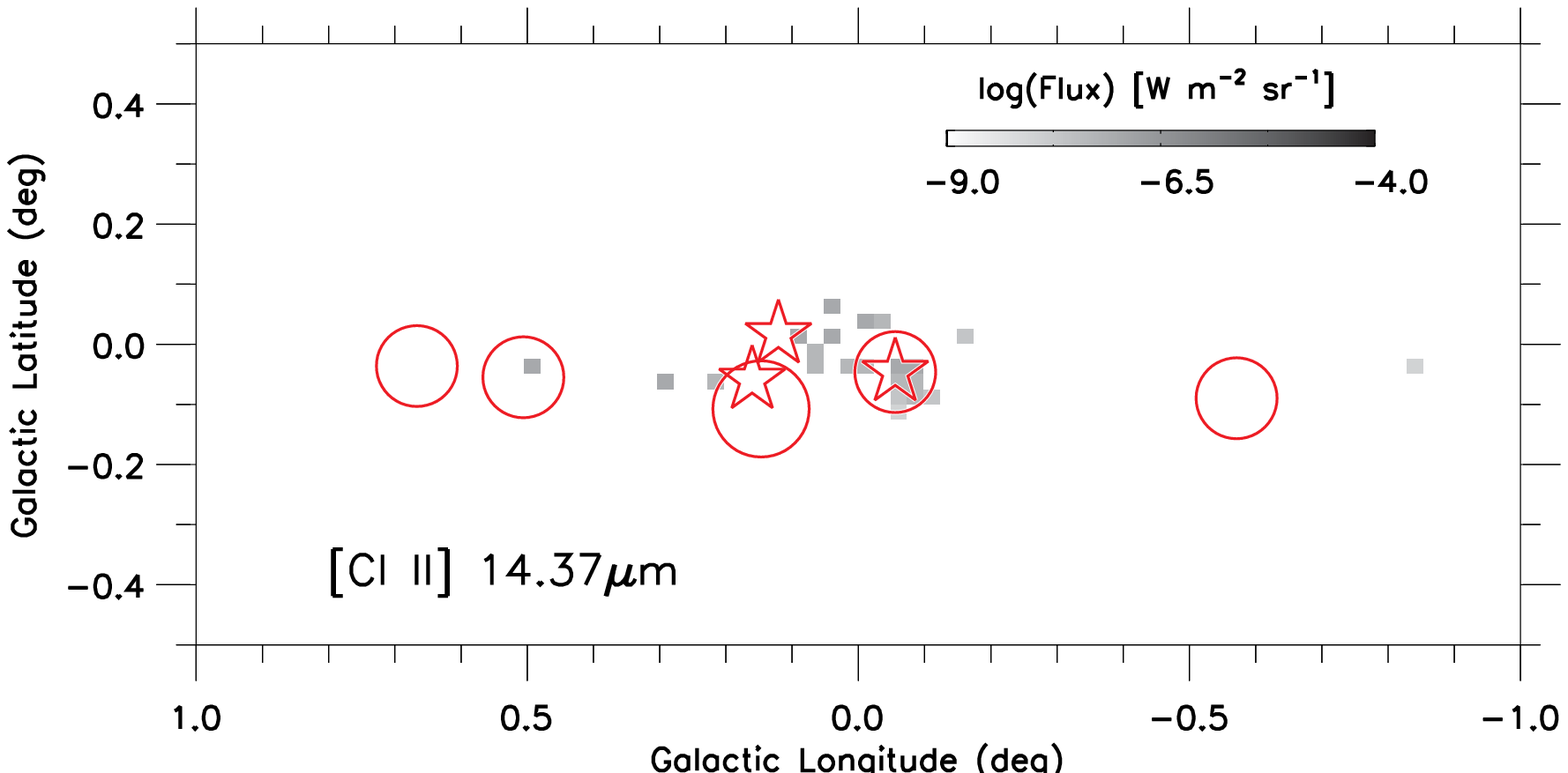}
\includegraphics[scale=0.42]{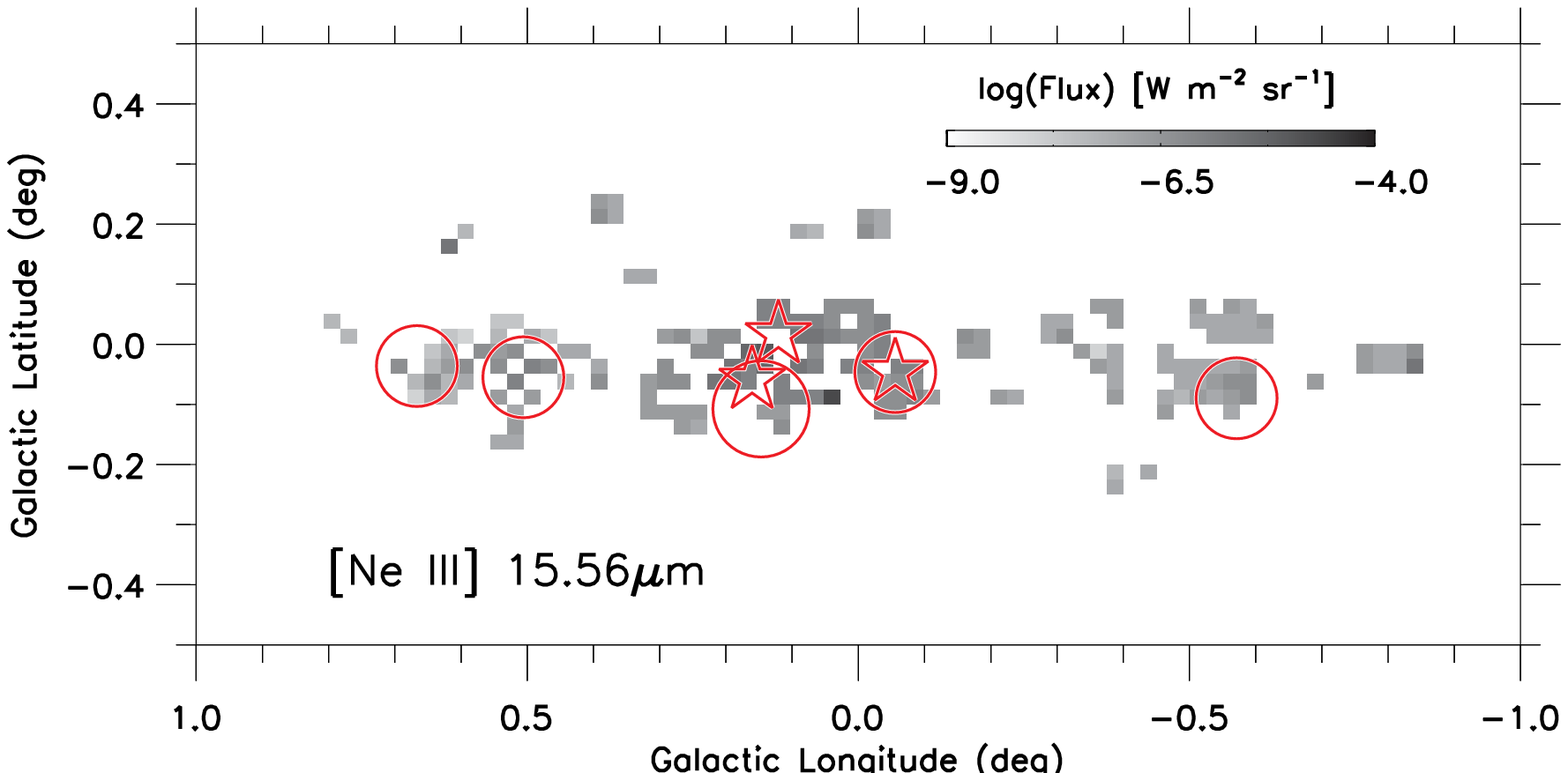}
\includegraphics[scale=0.42]{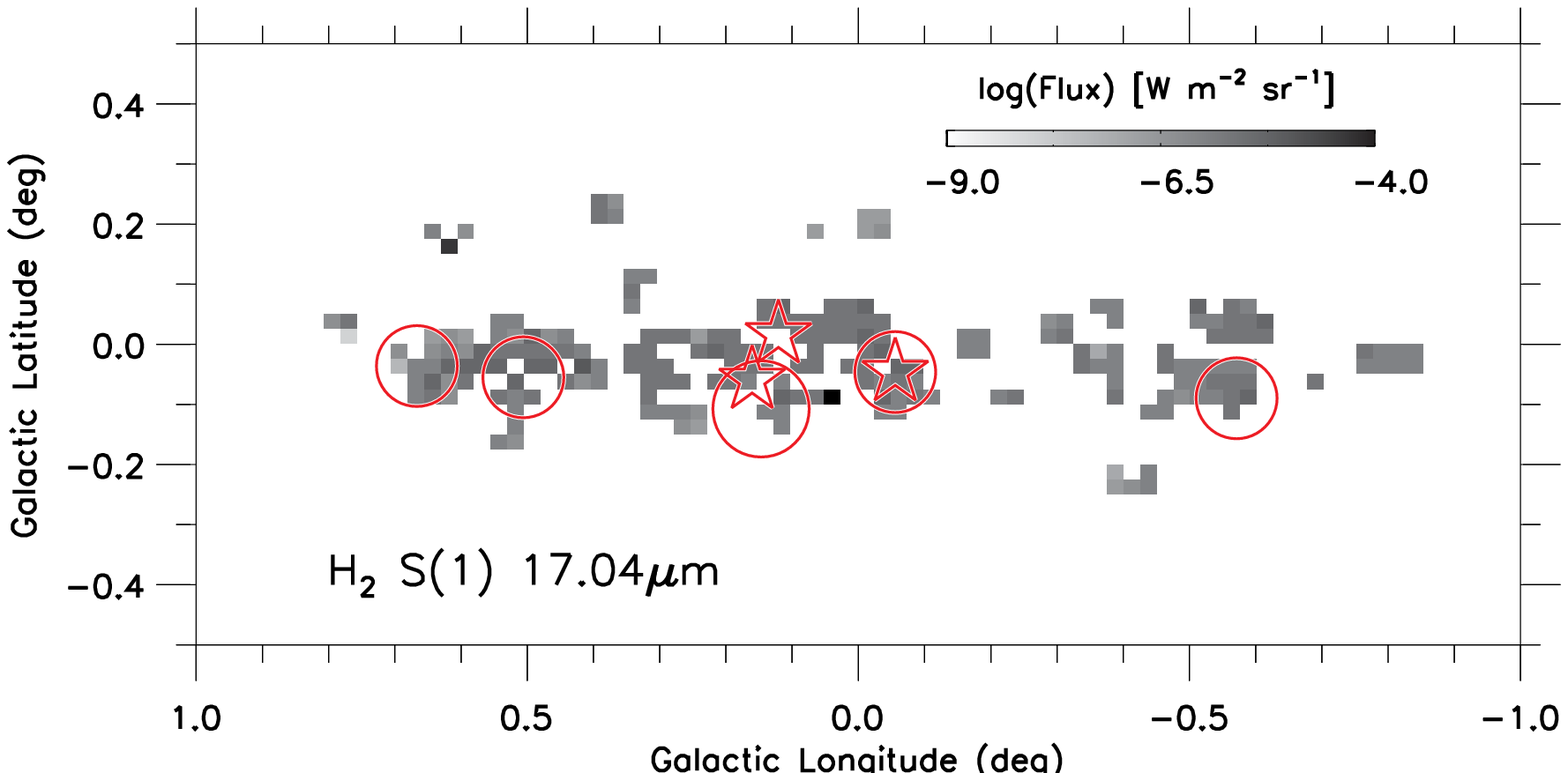}
\caption{Panoramic emission line maps in the GC. Line intensities from
individual spectra are averaged together within each
$1.5\arcmin\times1.5\arcmin$ ($\sim 3.5$~pc $\times 3.5$~pc) pixel if the lines
are detected at more than a $3\sigma$ level. Average line intensities are shown
in a logarithmic scale (W m$^{-2}$ sr$^{-1}$), after correcting for foreground
extinction ($\tau_{9.7}$) derived from the ratio between $10\ \mu$m and $14\
\mu$m continuum fluxes ($F_{14}/F_{10}$; see top panel in Figure~\ref{fig:tau}).
Key GC features are shown (see Figure~\ref{fig:map}).
\label{fig:elines}}
\end{figure*}

%Figure 7b
\setcounter{figure}{6}
\begin{figure*}
\centering
\includegraphics[scale=0.42]{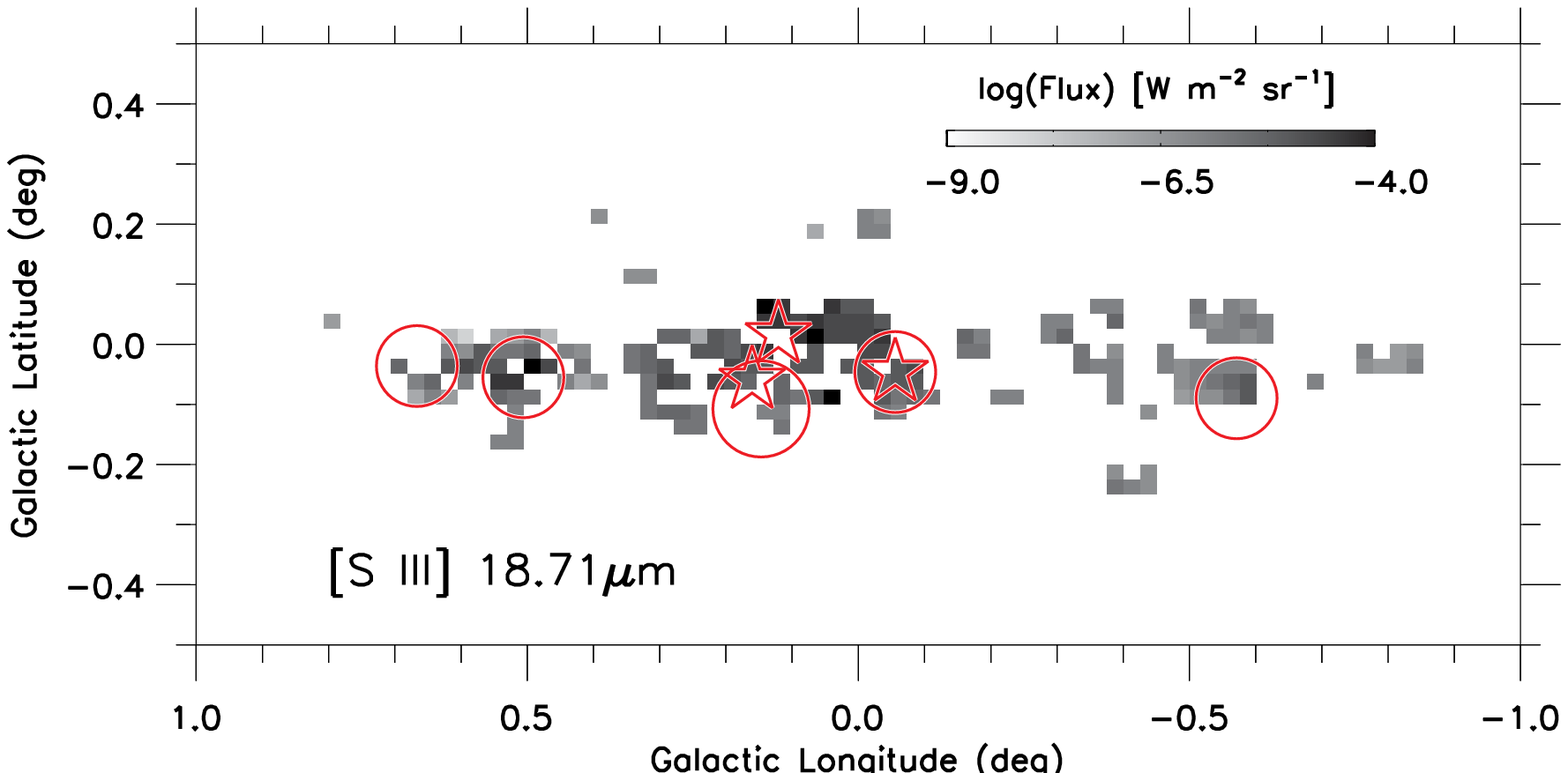}
\includegraphics[scale=0.42]{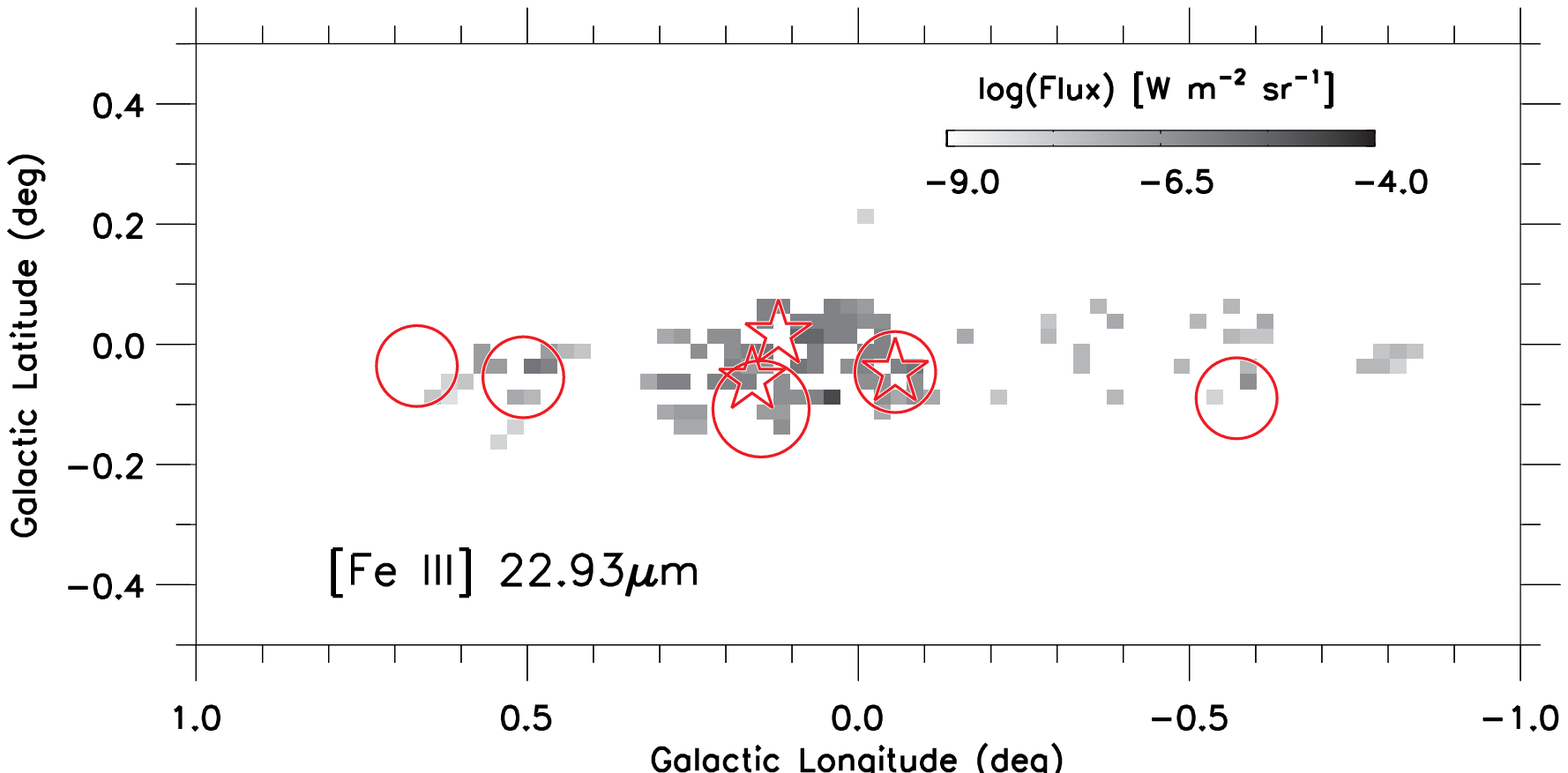}
\includegraphics[scale=0.42]{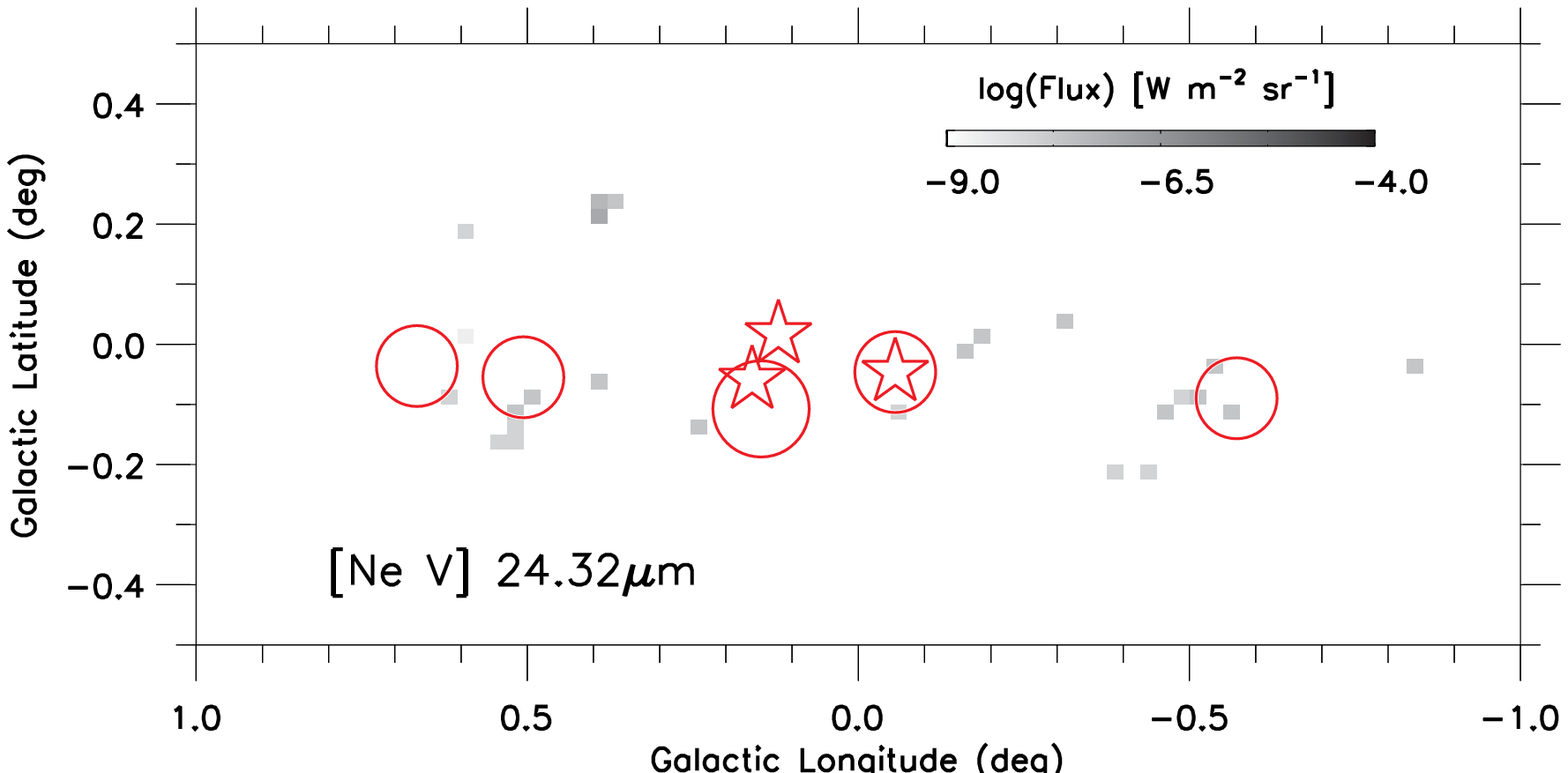}
\includegraphics[scale=0.42]{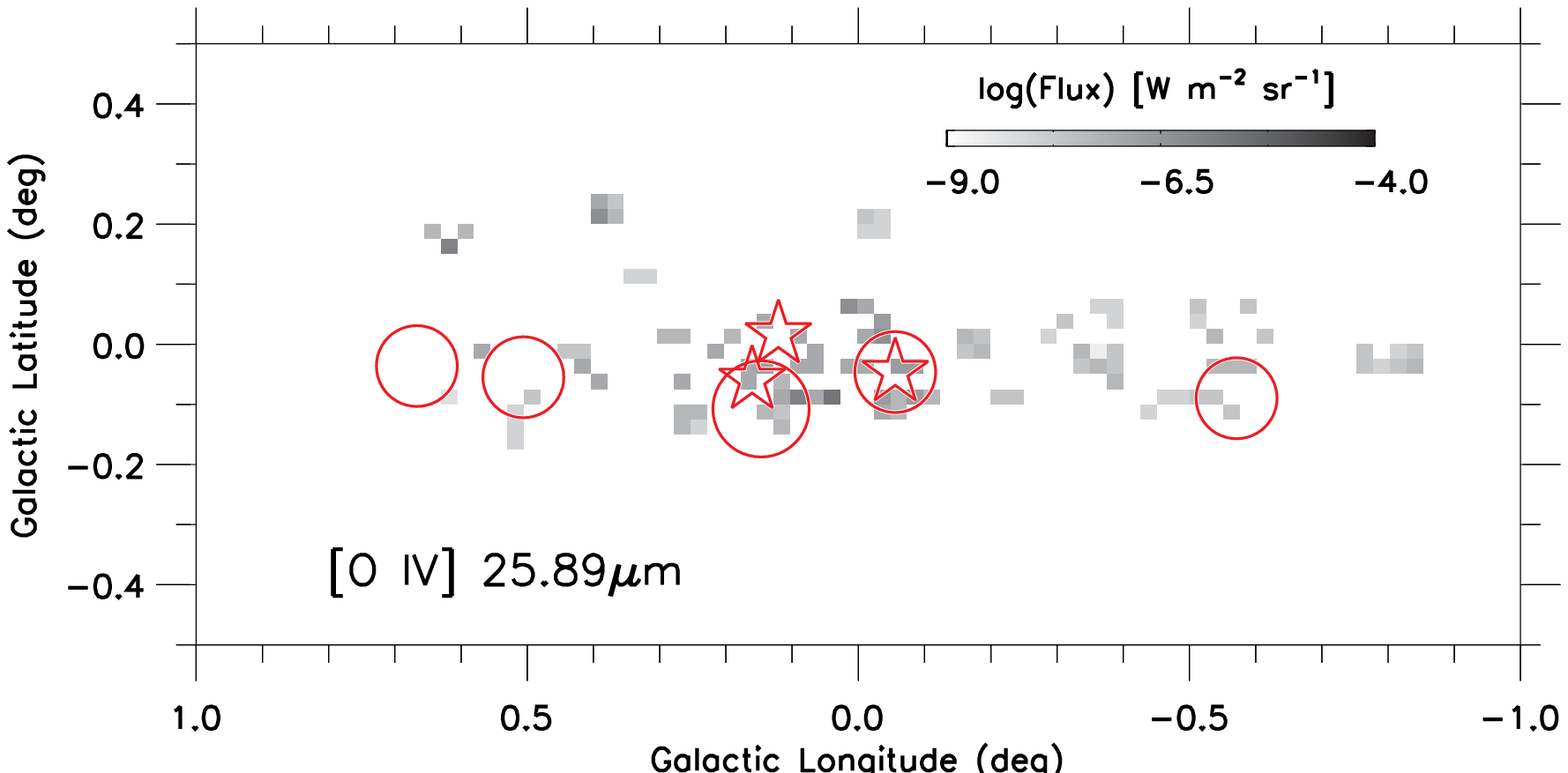}
\includegraphics[scale=0.42]{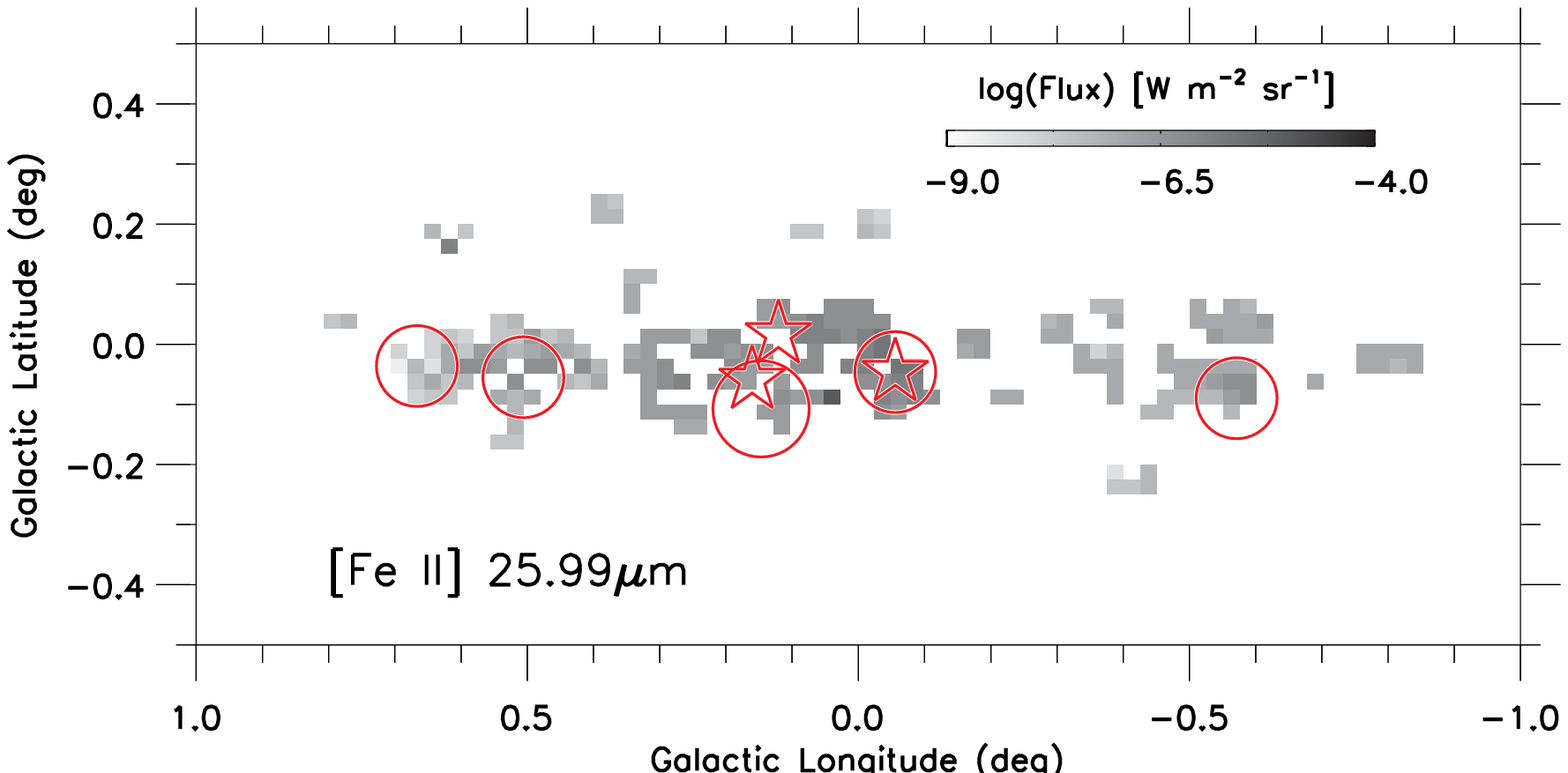}
\includegraphics[scale=0.42]{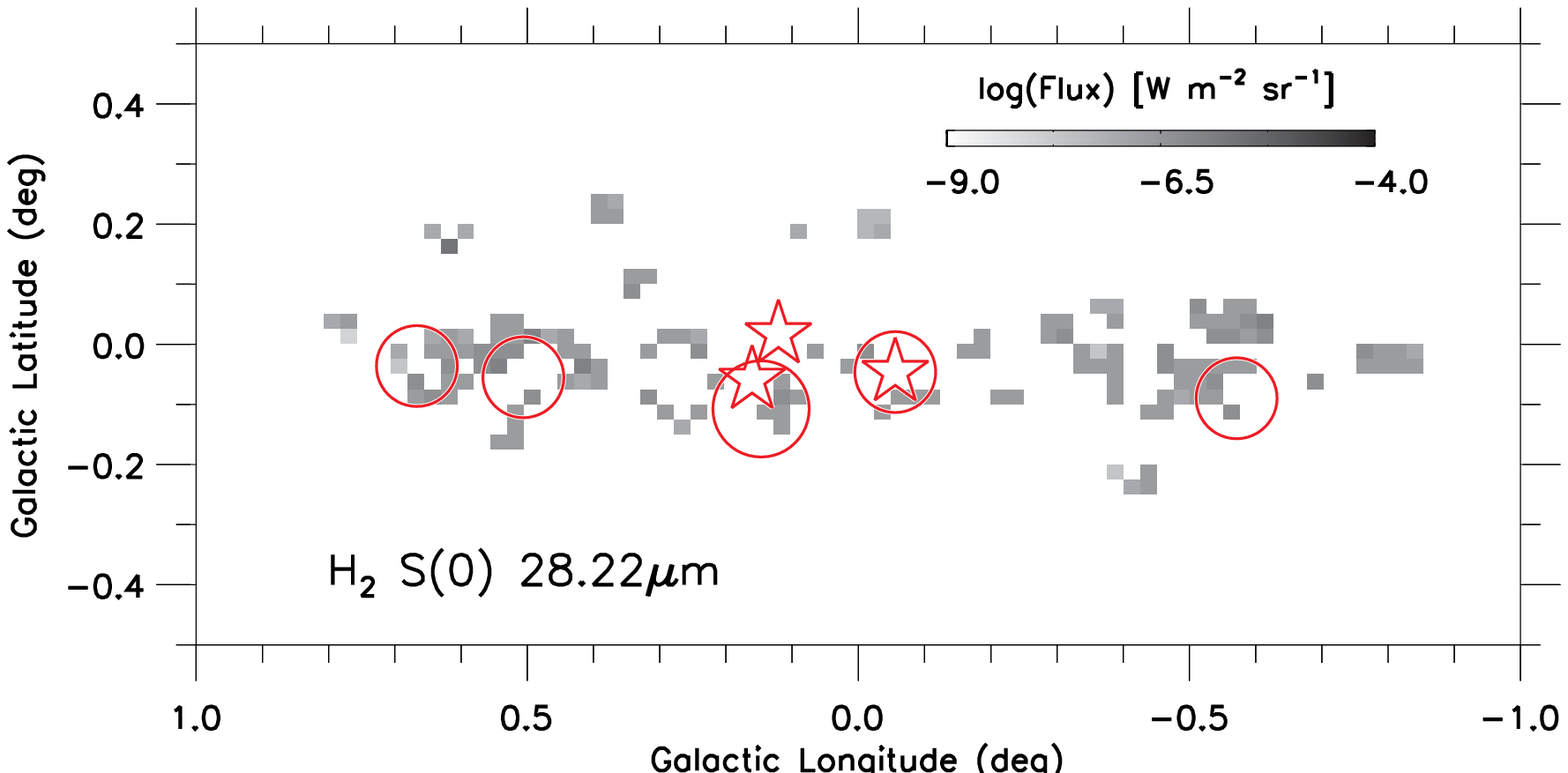}
\includegraphics[scale=0.42]{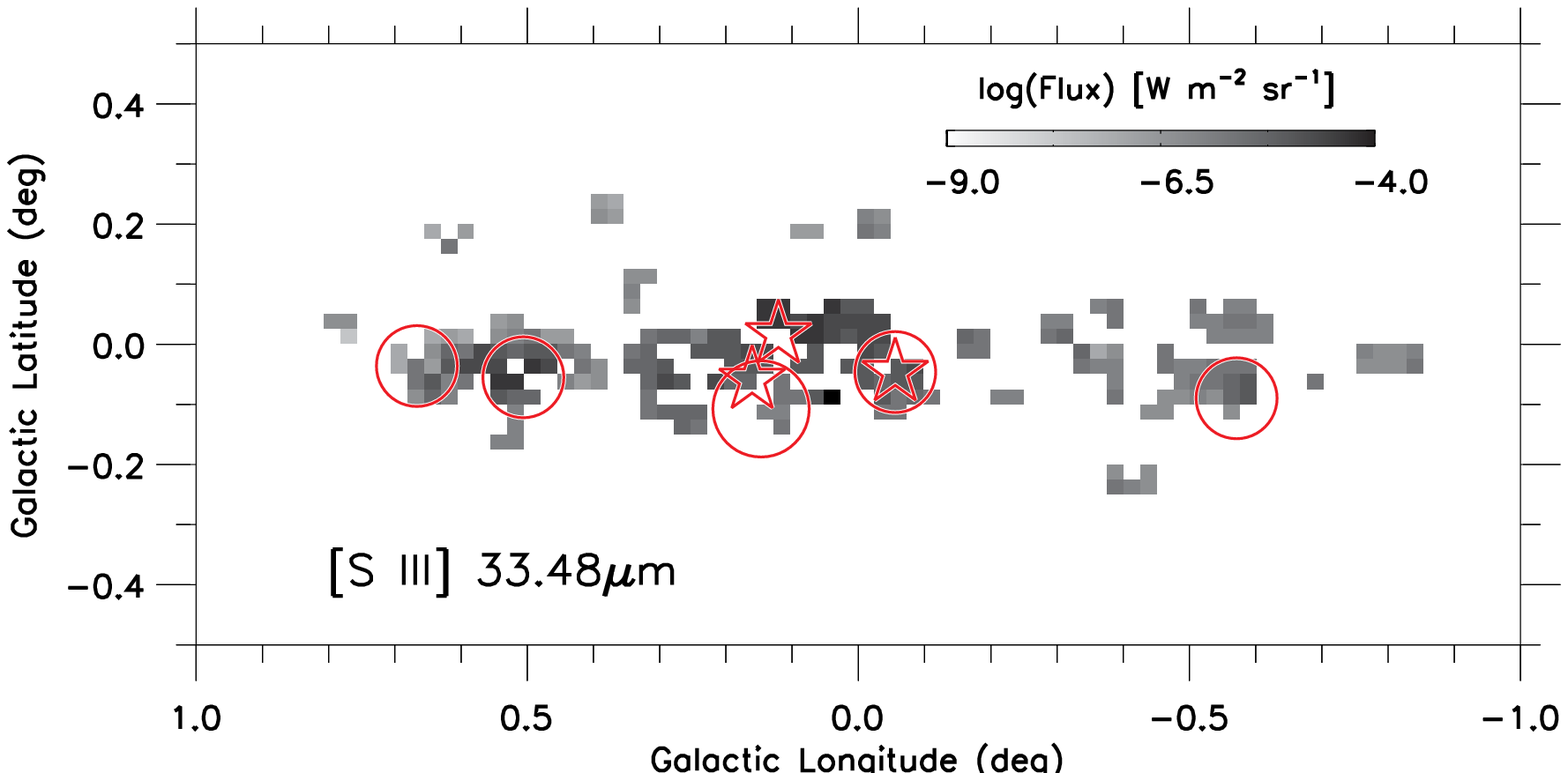}
\includegraphics[scale=0.42]{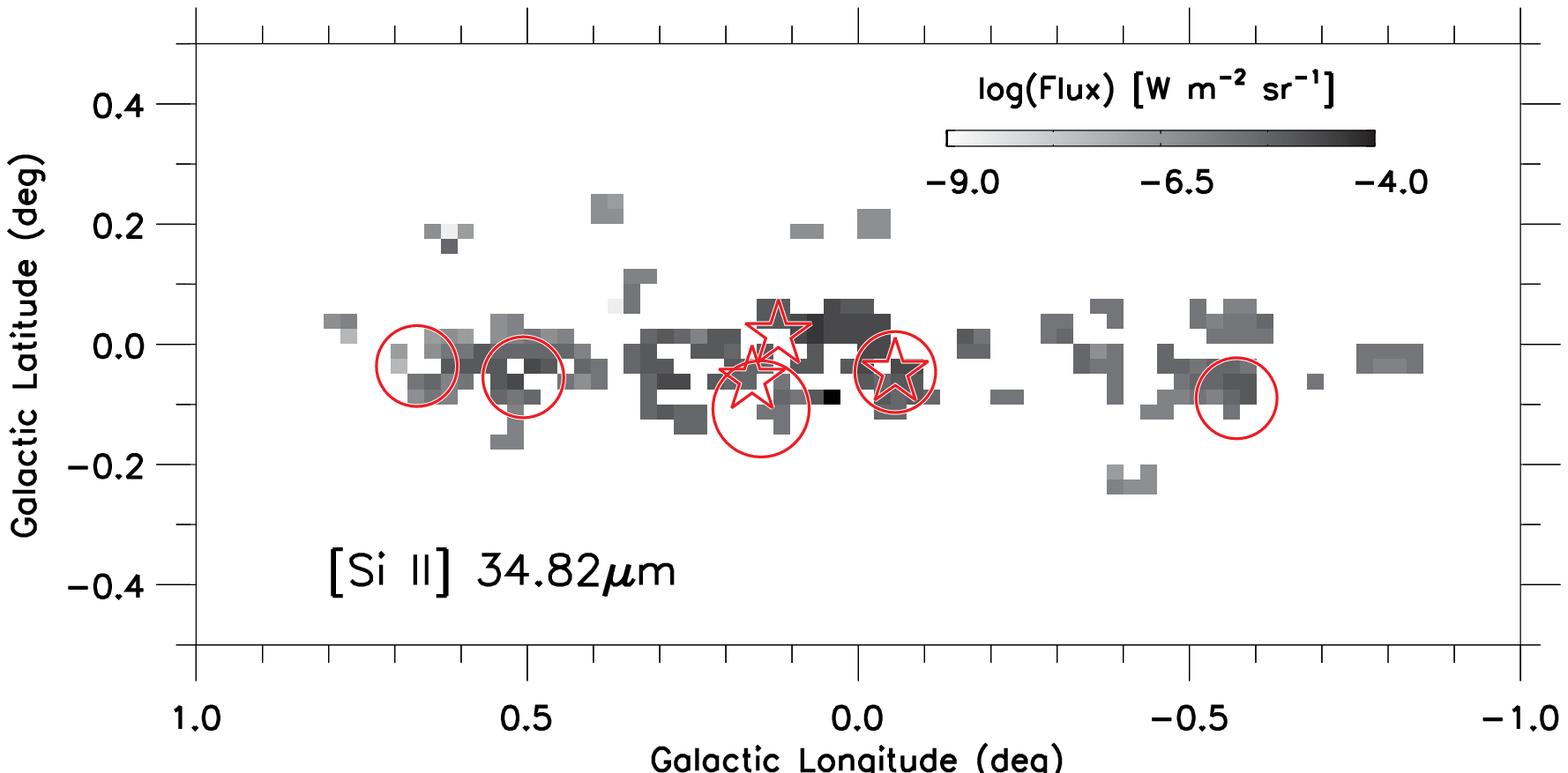}
\caption{Cont'd.
\label{fig:elines_b}}
\end{figure*}

\begin{deluxetable*}{rrrrrrrr}
\tablewidth{0pt}
\tabletypesize{\scriptsize}
\tablecaption{Line Flux Measurements from Individual Spectra\label{tab:tab2}}
\tablehead{
  \colhead{Galactic} &
  \colhead{Galactic} &
  \colhead{{\rm [\ion{Ne}{2}]} } &
  \colhead{{\rm H$_2$ S(1)}    } &
  \colhead{{\rm [\ion{S}{3}]}  } &
  \colhead{{\rm H$_2$ S(0)}    } &
  \colhead{{\rm [\ion{S}{3}]}  } &
  \colhead{{\rm [\ion{Si}{2}]} } \nl
  \colhead{longitude} &
  \colhead{latitude} &
  \colhead{$12.81\ \mu$m} &
  \colhead{$17.04\ \mu$m} &
  \colhead{$18.71\ \mu$m} &
  \colhead{$28.22\ \mu$m} &
  \colhead{$33.48\ \mu$m} &
  \colhead{$34.82\ \mu$m} \nl
  \colhead{(deg)} &
  \colhead{(deg)} &
  \colhead{} &
  \colhead{} &
  \colhead{} &
  \colhead{} &
  \colhead{} &
  \colhead{}
}
\startdata
$      -0.7954 $ & $      -0.0263 $ & $    7.32 \pm  0.11 $ & $    8.67 \pm  0.08 $ & $    2.87 \pm  0.15 $ & $    1.99 \pm  0.15 $ & $    4.25 \pm  0.14 $ & $   16.88 \pm  0.20 $ \nl 
$      -0.8254 $ & $      -0.0419 $ & $   10.94 \pm  0.13 $ & $    8.10 \pm  0.09 $ & $   10.21 \pm  0.17 $ & $    1.87 \pm  0.13 $ & $   16.04 \pm  0.35 $ & $   25.87 \pm  0.45 $ \nl 
$      -0.8037 $ & $      -0.0450 $ & $    4.99 \pm  0.11 $ & $    6.83 \pm  0.08 $ & $    2.01 \pm  0.15 $ & $    1.68 \pm  0.05 $ & $    3.42 \pm  0.11 $ & $   11.57 \pm  0.16 $ \nl 
$      -0.8198 $ & $      -0.0212 $ & $    8.10 \pm  0.11 $ & $    6.65 \pm  0.07 $ & $    3.99 \pm  0.22 $ & $    1.46 \pm  0.10 $ & $    6.64 \pm  0.11 $ & $   15.18 \pm  0.32 $ \nl 
$      -0.7488 $ & $      -0.0254 $ & $    4.43 \pm  0.13 $ & $    5.13 \pm  0.18 $ & $    1.91 \pm  0.23 $ & $    1.60 \pm  0.10 $ & $    3.68 \pm  0.18 $ & $    7.82 \pm  0.16 $ \nl 
$      -0.7814 $ & $      -0.0477 $ & $    5.51 \pm  0.20 $ & $    9.42 \pm  0.42 $ & $    2.93 \pm  0.34 $ & $    2.08 \pm  0.07 $ & $    3.54 \pm  0.31 $ & $   11.85 \pm  0.19 $ \nl 
$      -0.7594 $ & $      -0.0510 $ & $    6.50 \pm  0.18 $ & $    9.64 \pm  0.56 $ & $    3.41 \pm  0.54 $ & $    2.75 \pm  0.15 $ & $    4.65 \pm  0.40 $ & $   13.44 \pm  0.27 $ \nl 
$      -0.7720 $ & $      -0.0227 $ & $    7.00 \pm  0.18 $ & $   10.87 \pm  0.33 $ & $    0.00 \pm  0.00 $ & $    2.38 \pm  0.09 $ & $    5.50 \pm  0.34 $ & $   14.19 \pm  0.21 $ \nl 
$      -0.7529 $ & $      -0.0297 $ & $   25.84 \pm  0.70 $ & $   54.67 \pm  0.95 $ & $   44.04 \pm  2.32 $ & $   13.19 \pm  0.60 $ & $   28.87 \pm  1.07 $ & $   58.81 \pm  1.89 $ \nl 
$      -0.7836 $ & $      -0.0515 $ & $    6.07 \pm  0.10 $ & $    9.02 \pm  0.09 $ & $    2.38 \pm  0.15 $ & $    1.98 \pm  0.07 $ & $    3.50 \pm  0.59 $ & $   12.99 \pm  0.71 $ \nl 
\enddata
\tablecomments{Fluxes are corrected for dust extinction based on the
\citet{simpson:07} technique (see text), and are presented in units of $10^{-7}
$ W m$^{-2}\ {\rm sr}^{-1}$.  Only those detected at more than $3\sigma$ are
shown in this table.}
\tablecomments{Table~2 is published in its entirety in the electronic edition of
the {\it Astrophysical Journal Supplement Series}.  A portion is shown here for
guidance regarding its form and content.  The electronic edition of Table~2
contains the observed fluxes and errors for the emission lines listed in
Table~1.}
\end{deluxetable*}

Table~\ref{tab:tab3} lists the line fluxes measured by coadding all GC spectra,
with three different extinction correction methods. The line fluxes and ratios
listed in the second column are those measured from a spectrum coadded from
individual spectra that have first been corrected for extinction using the
\citet{simpson:07} $F_{14}/F_{10}$ technique. Similarly, values in the third
column are those measured in a spectrum coadded from individual spectra that
were first corrected using the \citet{schultheis:09} extinction map.  The last
column lists line fluxes and ratios measured by coadding all GC spectra, then
applying an extinction correction for $\tau_{9.7} = 3.439$ (the median value
obtained from the \citet{simpson:07} $F_{14}/F_{10}$ technique; see
Figure~\ref{fig:extinction}).

\begin{deluxetable*}{lrrr}
\tablewidth{0pt}
\tabletypesize{\scriptsize}
\tablecaption{Line Flux Measurements from Coadded Spectra\label{tab:tab3}}
\tablehead{
  \colhead{} &
  \multicolumn{3}{c}{Extinction Correction} \nl
  \cline{2-4}
  \colhead{Line Flux} &
  \colhead{\citet{simpson:07}\tablenotemark{a}} &
  \colhead{\citet{schultheis:09}\tablenotemark{b}} &
  \colhead{Posterior Correction\tablenotemark{c}} \nl
  \colhead{(Line Ratio)} &
  \colhead{($10^{-7} $ W m$^{-2}\ {\rm sr}^{-1}$)} &
  \colhead{($10^{-7} $ W m$^{-2}\ {\rm sr}^{-1}$)} &
  \colhead{($10^{-7} $ W m$^{-2}\ {\rm sr}^{-1}$)}
}
\startdata
{\rm [\ion{S}{4}]}   $10.51\ \mu$m & $ 2.30\pm3.22$ & $ 0.00\pm2.90$ & $ 0.54\pm4.38$\nl
{\rm H$_2$ S(2)}     $12.28\ \mu$m & $ 7.06\pm0.34$ & $ 8.45\pm0.34$ & $ 6.28\pm1.60$\nl
{\rm \ion{H}{1} 7-6} $12.37\ \mu$m & $ 1.67\pm0.17$ & $ 1.95\pm0.17$ & $ 1.36\pm0.75$\nl
{\rm [\ion{Ne}{2}]}  $12.81\ \mu$m & $58.09\pm0.30$ & $63.00\pm0.32$ & $45.94\pm1.23$\nl
{\rm [\ion{Ne}{5}]}  $14.32\ \mu$m & $ 0.00\pm0.08$ & $ 0.00\pm0.08$ & $ 0.06\pm0.32$\nl
{\rm [\ion{Cl}{2}]}  $14.37\ \mu$m & $ 0.32\pm0.20$ & $ 0.33\pm0.09$ & $ 0.22\pm0.37$\nl
{\rm [\ion{Ne}{3}]}  $15.56\ \mu$m & $ 4.74\pm0.26$ & $ 4.94\pm0.26$ & $ 3.80\pm1.48$\nl
{\rm H$_2$ S(1)}     $17.04\ \mu$m & $20.57\pm0.25$ & $24.90\pm0.30$ & $15.43\pm2.28$\nl
{\rm [\ion{S}{3}]}   $18.71\ \mu$m & $89.87\pm0.56$ & $99.51\pm0.67$ & $52.91\pm2.54$\nl
{\rm [\ion{Fe}{3}]}  $22.93\ \mu$m & $ 4.12\pm0.19$ & $ 4.05\pm0.18$ & $ 2.69\pm1.39$\nl
{\rm [\ion{Ne}{5}]}  $24.32\ \mu$m & $ 0.23\pm0.19$ & $ 0.23\pm0.25$ & $ 0.16\pm1.80$\nl
{\rm [\ion{O}{4}]}   $25.89\ \mu$m & $ 0.70\pm0.12$ & $ 0.86\pm0.13$ & $ 0.60\pm0.45$\nl
{\rm [\ion{Fe}{2}]}  $25.99\ \mu$m & $ 2.97\pm0.12$ & $ 3.48\pm0.14$ & $ 2.26\pm0.83$\nl
{\rm H$_2$ S(0)}     $28.22\ \mu$m & $ 3.25\pm0.59$ & $ 3.85\pm0.61$ & $ 2.73\pm1.27$\nl
{\rm [\ion{S}{3}]}   $33.48\ \mu$m & $75.30\pm0.98$ & $77.48\pm1.03$ & $58.07\pm0.77$\nl
{\rm [\ion{Si}{2}]}  $34.82\ \mu$m & $61.76\pm1.34$ & $70.83\pm1.41$ & $51.29\pm1.04$\nl

\\
\hline
\\
{\rm [\ion{Ne}{3}] / [\ion{Ne}{2}]} & $0.082\pm0.005$ & $0.078\pm0.004$ & $0.083\pm0.032$\nl
{\rm [\ion{Si}{2}] / [\ion{S}{3}]}  & $0.820\pm0.021$ & $0.914\pm0.022$ & $0.883\pm0.021$\nl
{\rm [\ion{Fe}{2}] / [\ion{Ne}{2}]} & $0.051\pm0.002$ & $0.055\pm0.002$ & $0.049\pm0.018$\nl
{\rm [\ion{O}{4}]  / [\ion{Ne}{2}]} & $0.012\pm0.002$ & $0.014\pm0.002$ & $0.013\pm0.010$\nl
{\rm [\ion{Fe}{2}] / [\ion{O}{4}]}  & $4.228\pm0.766$ & $4.071\pm0.624$ & $3.766\pm3.120$\nl
\enddata
\tablenotetext{a}{Fluxes corrected for dust extinction using the \citet{simpson:07} technique (see text).}
\tablenotetext{b}{Fluxes corrected for dust extinction using the \citet{schultheis:09} map (see text).}
\tablenotetext{c}{Extinction corrections applied to the coadded spectrum assuming
$\langle \tau_{9.7} \rangle = 3.439$ (see text).}
\end{deluxetable*}

Mapping results in Figure~\ref{fig:elines} are shown in the same order of
increasing wavelength as in Table~\ref{tab:tab1}. Each pixel in the map covers a
$1.5\arcmin\times1.5\arcmin$ region of the sky ($\sim 3.5$~pc $\times 3.5$~pc),
which is an order of magnitude larger than the area covered by a slit entrance
of either {\tt SH} or {\tt LH}. We divided each extracted line flux by the areal
coverage of the corresponding slit entrance ($53\ {\rm arcsec}^2$ for {\tt SH}
and $248\ {\rm arcsec}^2$ for {\tt LH}), and computed unweighted mean
intensities (W~m$^{-2}$~sr$^{-1}$) in each pixel of the map in
Figure~\ref{fig:elines}. Each pixel includes two background IRS pointings on
average.

Average line intensities in Figure~\ref{fig:elines} were corrected for
foreground dust extinction derived using the \citet{simpson:07} method
(\S~\ref{sec:extinction}), which relies on the continuum flux ratio between $10\
\mu$m and $14\ \mu$m (see the top panel in Figure~\ref{fig:tau}).  Mid-IR
forbidden emission lines, such as [\ion{Ne}{2}] $12.81\ \mu$m, [\ion{Ne}{3}]
$15.56\ \mu$m, [\ion{S}{3}] $18.71\ \mu$m, [\ion{S}{3}] $33.48\ \mu$m, and
[\ion{Si}{2}] $34.82\ \mu$m, are strong in the CMZ.  Their line strengths vary
across the region, with strongest emission observed near the Arches cluster and
Sgr~B1. Weaker emission is observed in the CMZ for H I 7--6 12.37 $\mu$m,
[\ion{Cl}{2}] 14.37 $\mu$m, [\ion{Fe}{3}] 22.93 $\mu$m, and [\ion{Fe}{2}] 25.99
$\mu$m.  Emission from pure rotational H$_2$ lines, 0--0 S(0), 0--0 S(1), and
0--0 S(2) at $28.22\ \mu$m, $17.04\ \mu$m, and $12.28\ \mu$m, respectively, is
also strong, but these H$_2$ lines are relatively constant over the CMZ. We will
discuss the uniform H$_2$ emission in \S~\ref{sec:hydrogen}, further utilizing
radial velocities measured from the IRS spectra.  Fine structure lines from
highly ionized species such as [\ion{O}{4}] $25.89\ \mu$m are found throughout
the CMZ, but [\ion{S}{4}] 10.51 $\mu$m, [\ion{Ne}{5}] $14.32\ \mu$m and
[\ion{Ne}{5}] $24.32\ \mu$m were detected only in a few lines of sight to the
GC. Our mapping results and interpretation remain qualitatively unchanged if the
\citet{schultheis:09} map is used for the foreground extinction correction.

\subsection{Molecular Hydrogen Line Emission and Radial Velocity
Mapping}\label{sec:hydrogen}

Figure~\ref{fig:elines} shows line intensity maps for pure rotational
transitions from the lowest three levels of molecular hydrogen, H$_2$ S(2)
$12.28\ \mu$m, S(1) $17.04\ \mu$m, and S(0) $28.22\ \mu$m. We detected these
lines in almost all lines of sight in the GC at more than a $3\sigma$ level.
Their intensity distributions are rather uniform in the CMZ, compared to the
spatial structures seen for ionic forbidden emission lines. We also find weak
correlations in line intensities between the pure rotational hydrogen emission
and ionic forbidden lines ($0.3 \la p \la 0.6$); correlations are stronger ($p
\ga 0.9$) among strong forbidden emission lines.

The pure rotational H$_2$ lines from warm molecular gas are important tools for
studying heating mechanisms in the GC \citep[e.g., see][for extragalactic
H$_2$ line measurement]{roussel:07}. \citet{rf:04} suggested that a combination
of PDRs and diffuse ionized gas can be used to explain the observed dust, H$_2$,
neutral gas, and ionized gas emission in the GC. \citet{pak:96} also found that
the most likely cause of the large-scale ro-vibrational H$_2$
$\nu=1\rightarrow0$ S(1) emission at $2.12\ \mu$m in the GC is UV excitation by
hot massive stars.  However, \citet{rf:05} showed that while PDR models can
reproduce the observed pure rotational H$_2$ lines from excited levels in the
GC, they are not enough to explain all the emission from the lowest levels, S(0)
and S(1). They argued that low velocity shocks or turbulent motions are needed
as an additional heating mechanism to reproduce excess emission from low
excitation H$_2$ lines.  On the other hand, \citet{simpson:07} concluded, based
on their H$_2$ rotational line measurements, that multi-component models of warm
molecular gas in PDRs along the line of sight to the GC can fully explain the
observed H$_2$ line ratios, without requiring shocks.

\begin{figure*}
\centering
\includegraphics[scale=0.42]{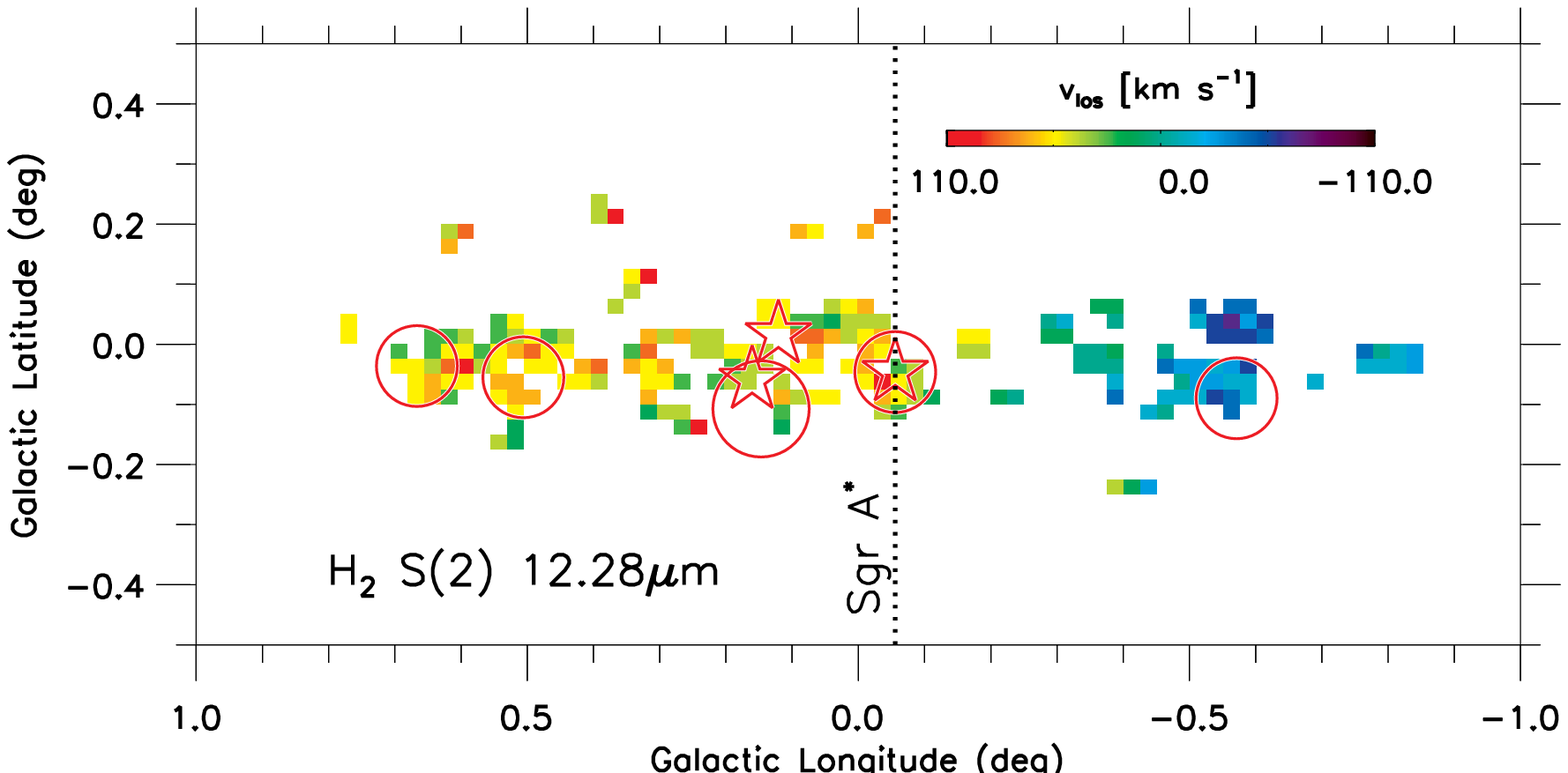}
\includegraphics[scale=0.42]{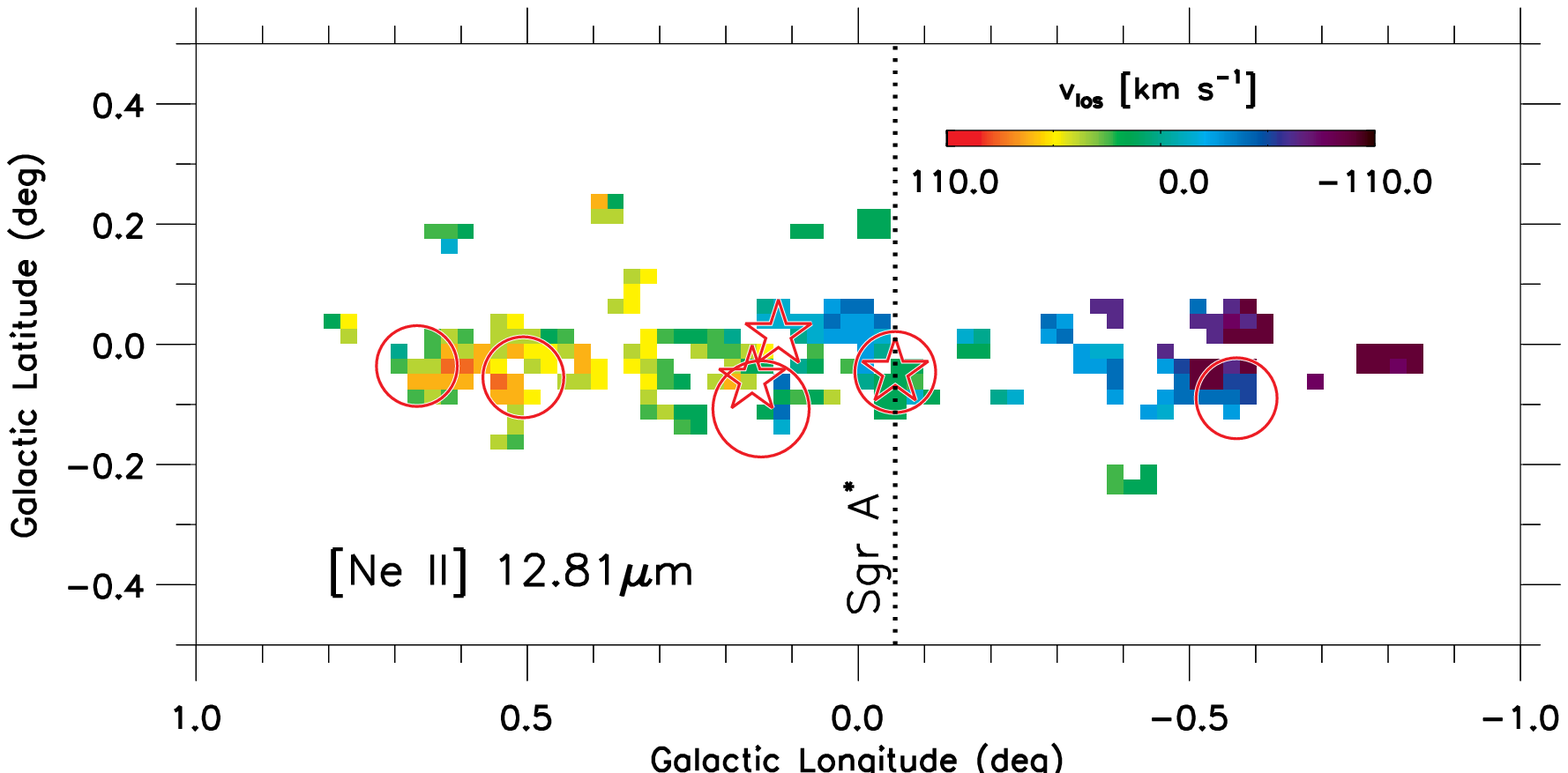}
\includegraphics[scale=0.42]{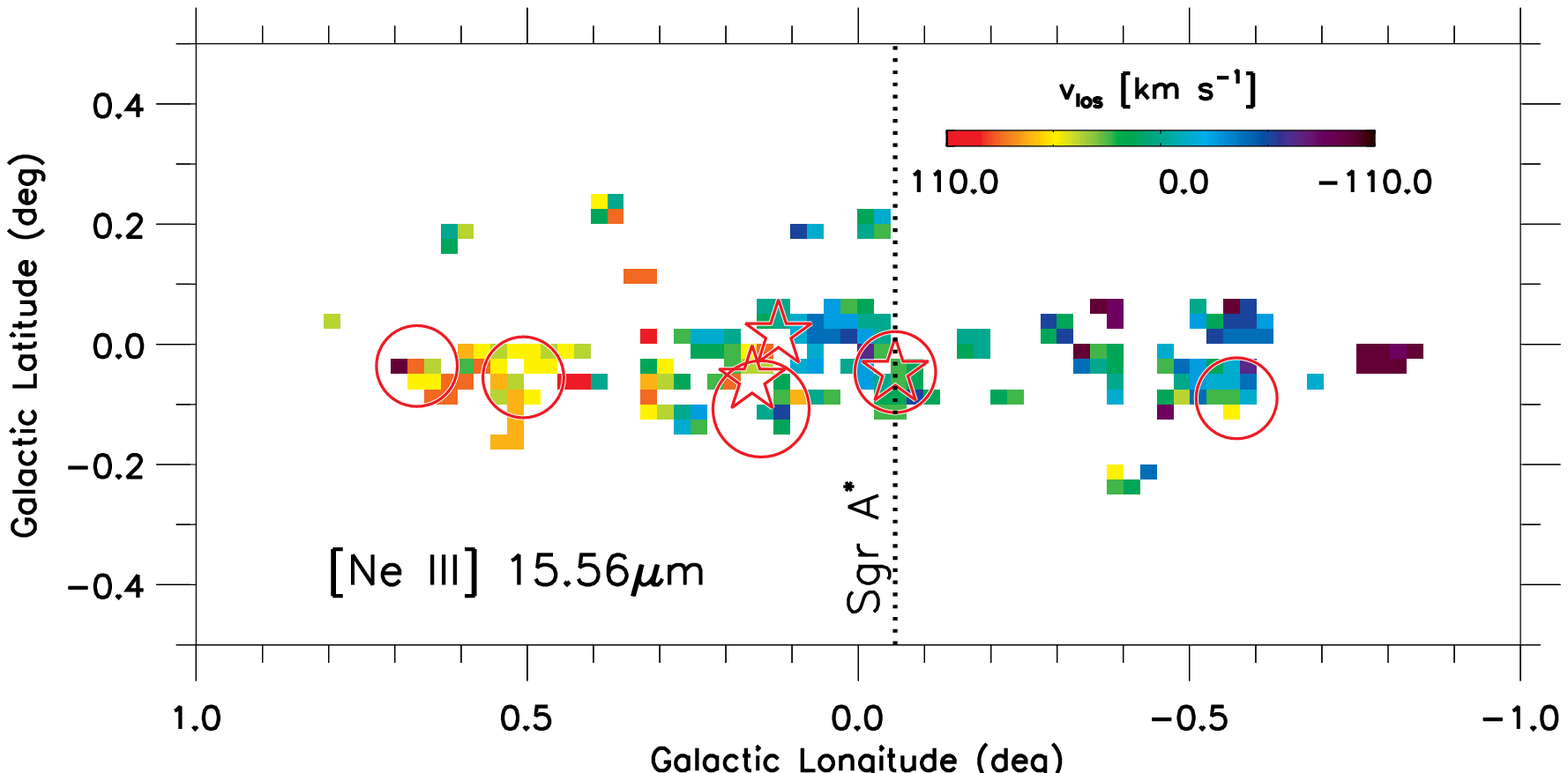}
\includegraphics[scale=0.42]{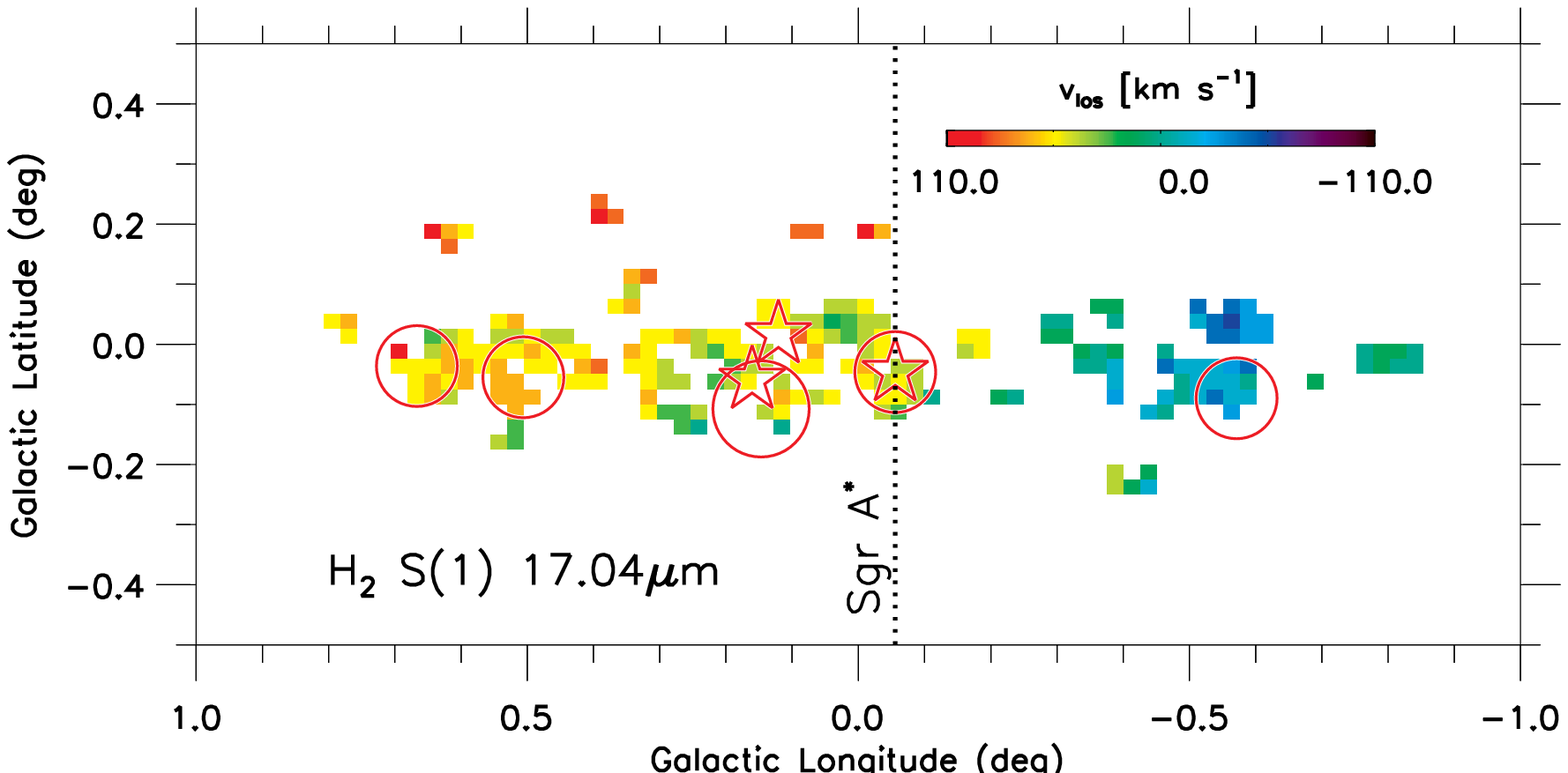}
\includegraphics[scale=0.42]{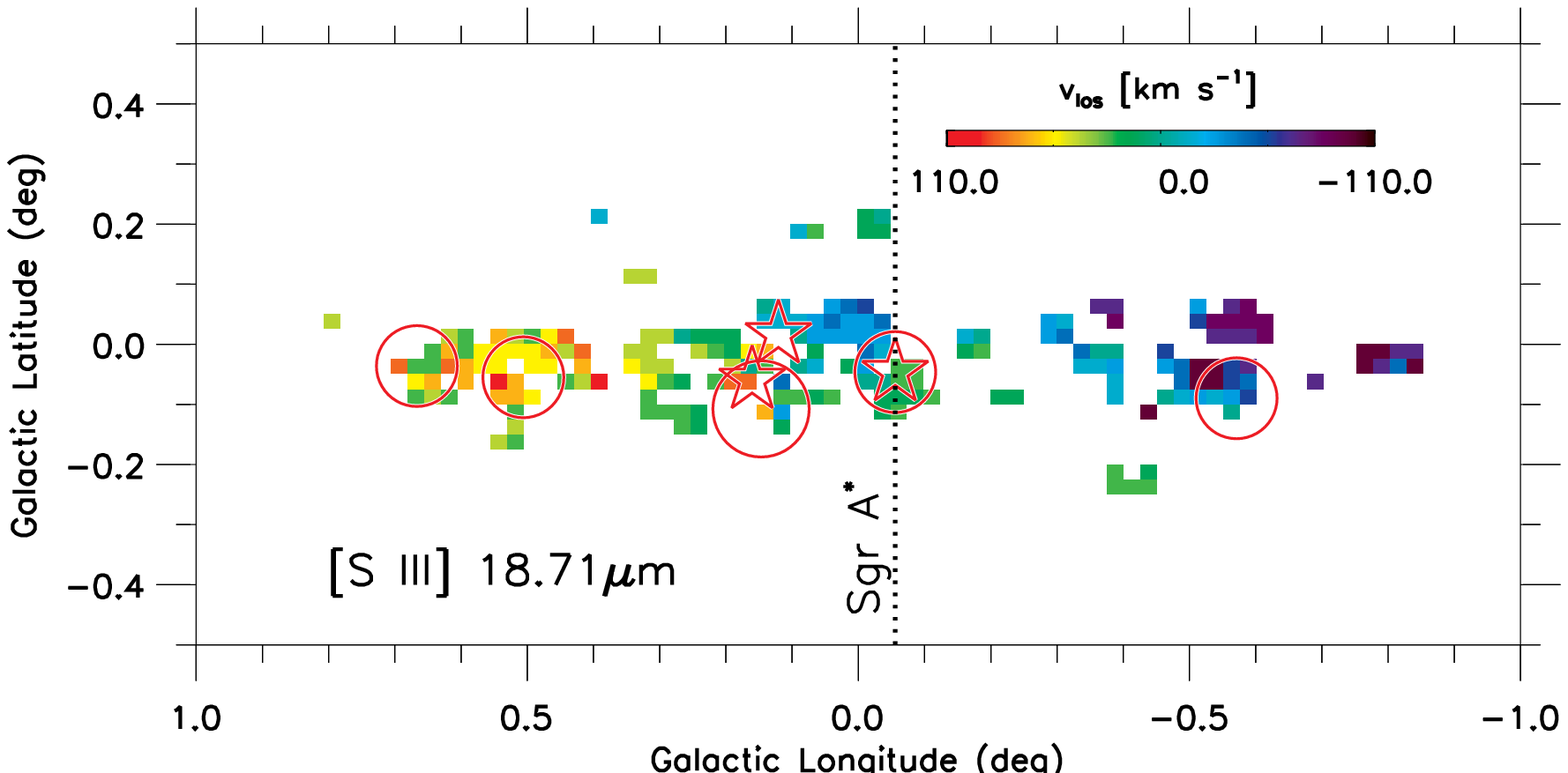}
\includegraphics[scale=0.42]{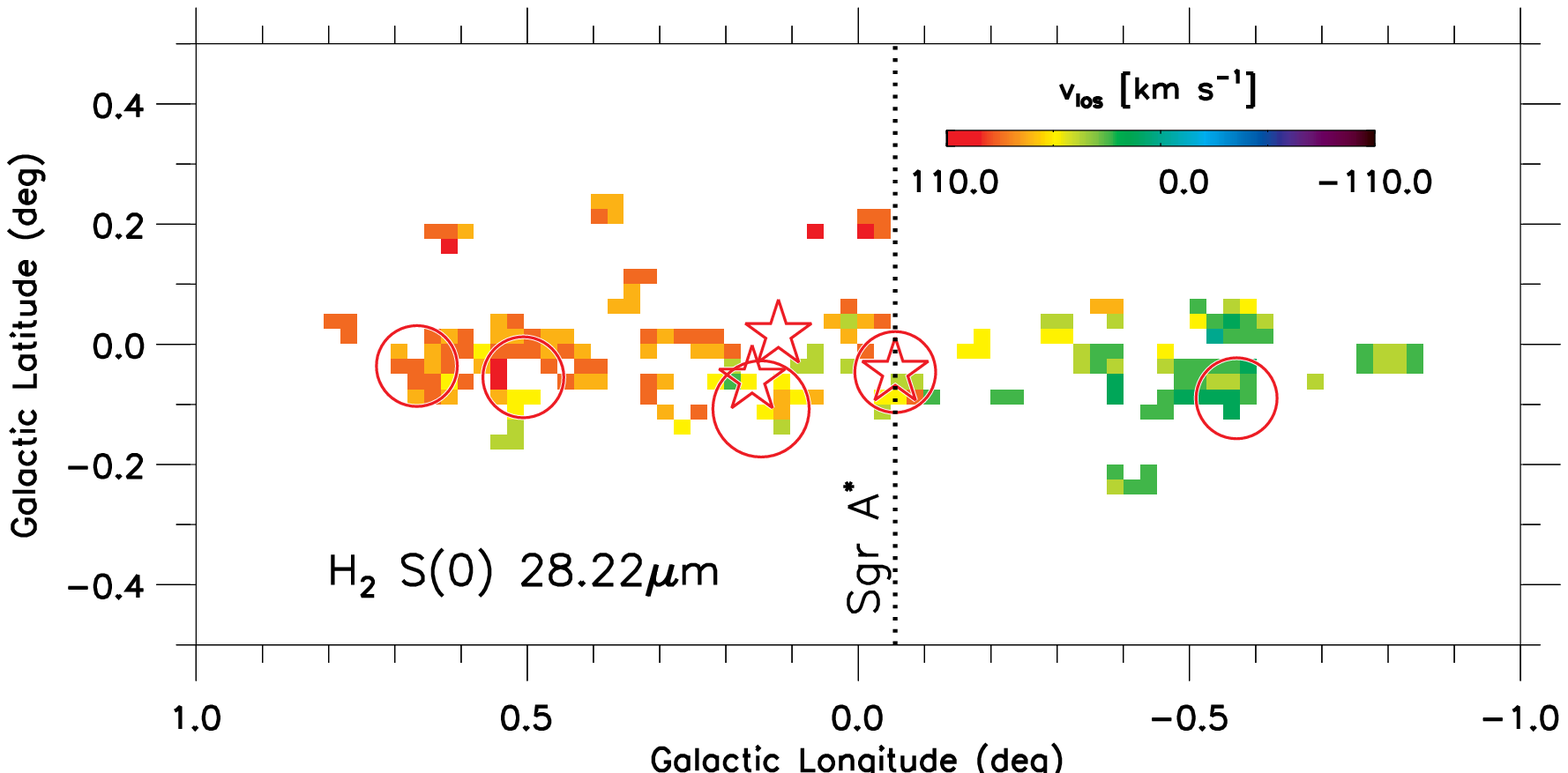}
\includegraphics[scale=0.42]{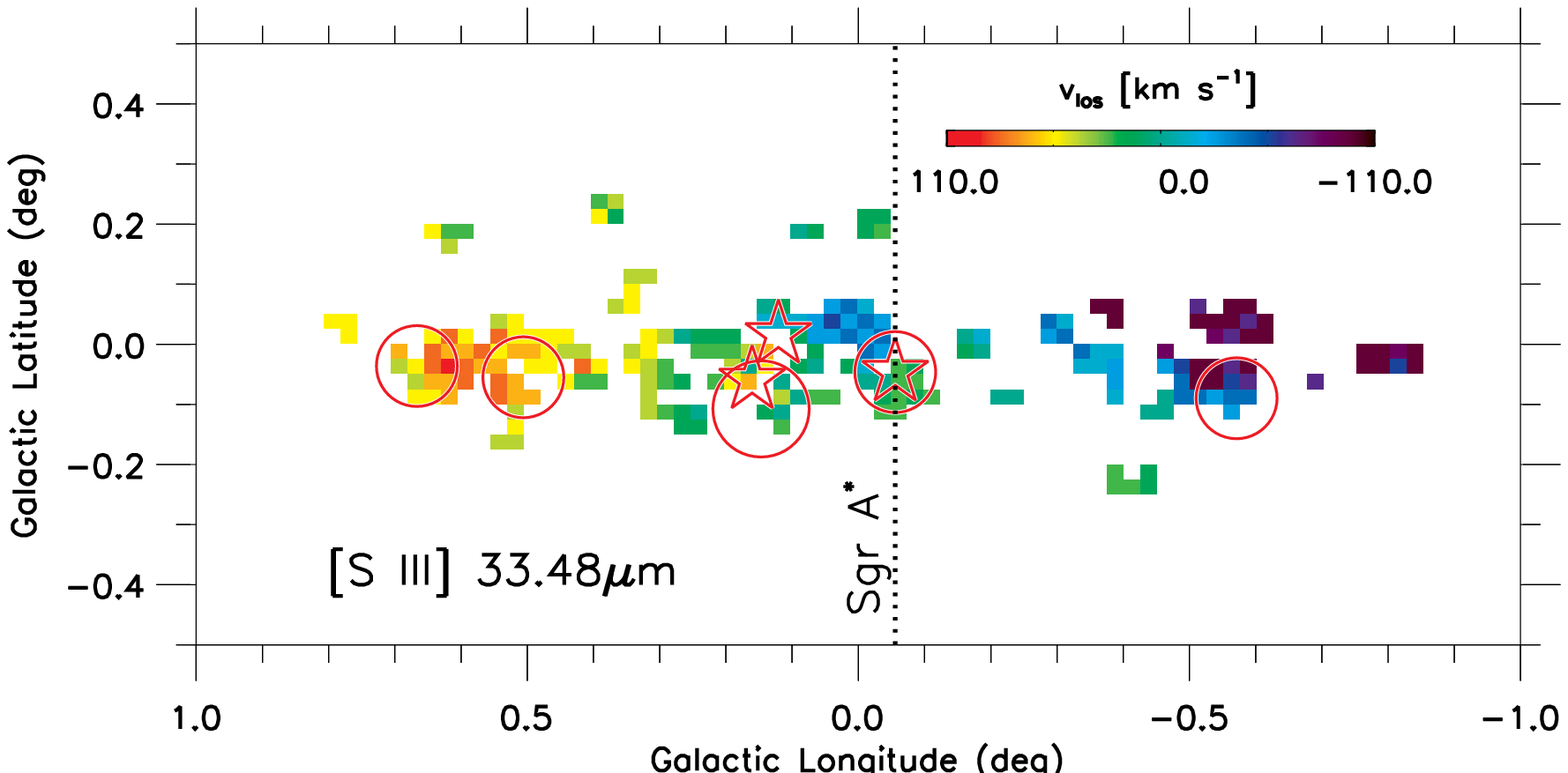}
\includegraphics[scale=0.42]{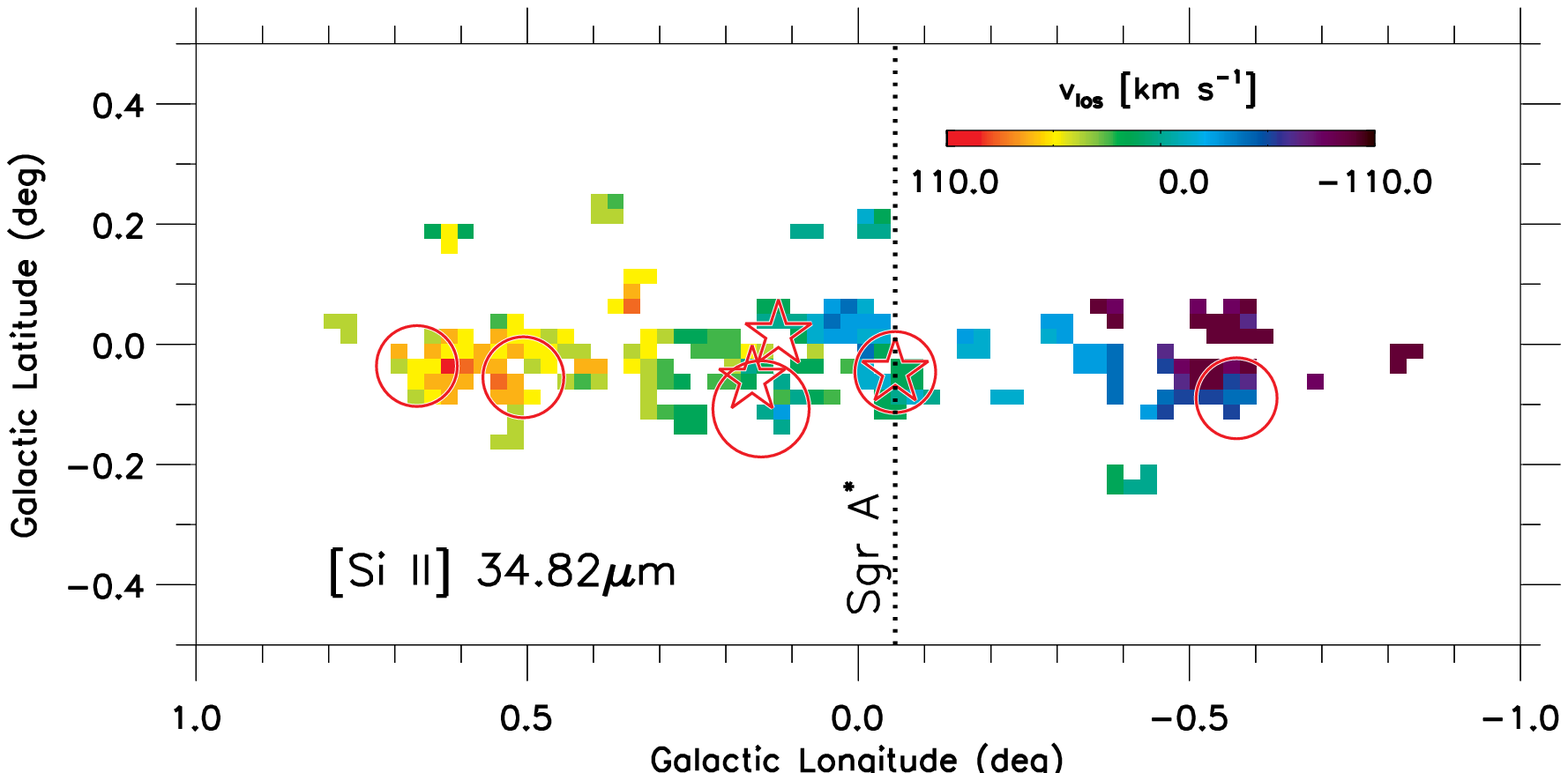}
\caption{Panoramic radial velocity maps of the GC, constructed from strong ionic
and molecular hydrogen emission lines. The vertical line indicates the Galactic
longitude of Sgr~A* at the dynamical center of the Galaxy. The sense of the
rotation is that the eastern part of the GC, including the Sgr~B complex, is
systematically receding from the Sun (positive $v_r$), consistent with the
rotation of the Galactic disk.  Each pixel covers $1.5\arcmin\times1.5\arcmin$
($\sim 3.5$~pc $\times 3.5$~pc).  Key features of the GC are overlaid (see
Figure~\ref{fig:map}).
\label{fig:vlos}}
\end{figure*}

To further investigate the source of low-level rotational H$_2$ lines, we
compared radial velocity distributions of various mid-IR emission lines.
Figure~\ref{fig:vlos} shows panoramic radial velocity ($v_r$) maps of the CMZ
for several strong ionic and molecular hydrogen emission lines.  Radial
velocities from strong ionic lines exhibit a systematic rotation of the CMZ,
consistent with the rotation of the Galactic disk.  The sense of the rotation
is that the eastern part of the GC, including the Sgr~B complex, is
systematically receding from the Sun (positive $v_r$), and the Sgr~C complex is
systematically approaching the Sun (negative $v_r$), with respect to the
dynamical center of the Galaxy, indicated by a vertical dotted line at Sgr~A*.

\begin{figure*}
\centering
\includegraphics[scale=0.45]{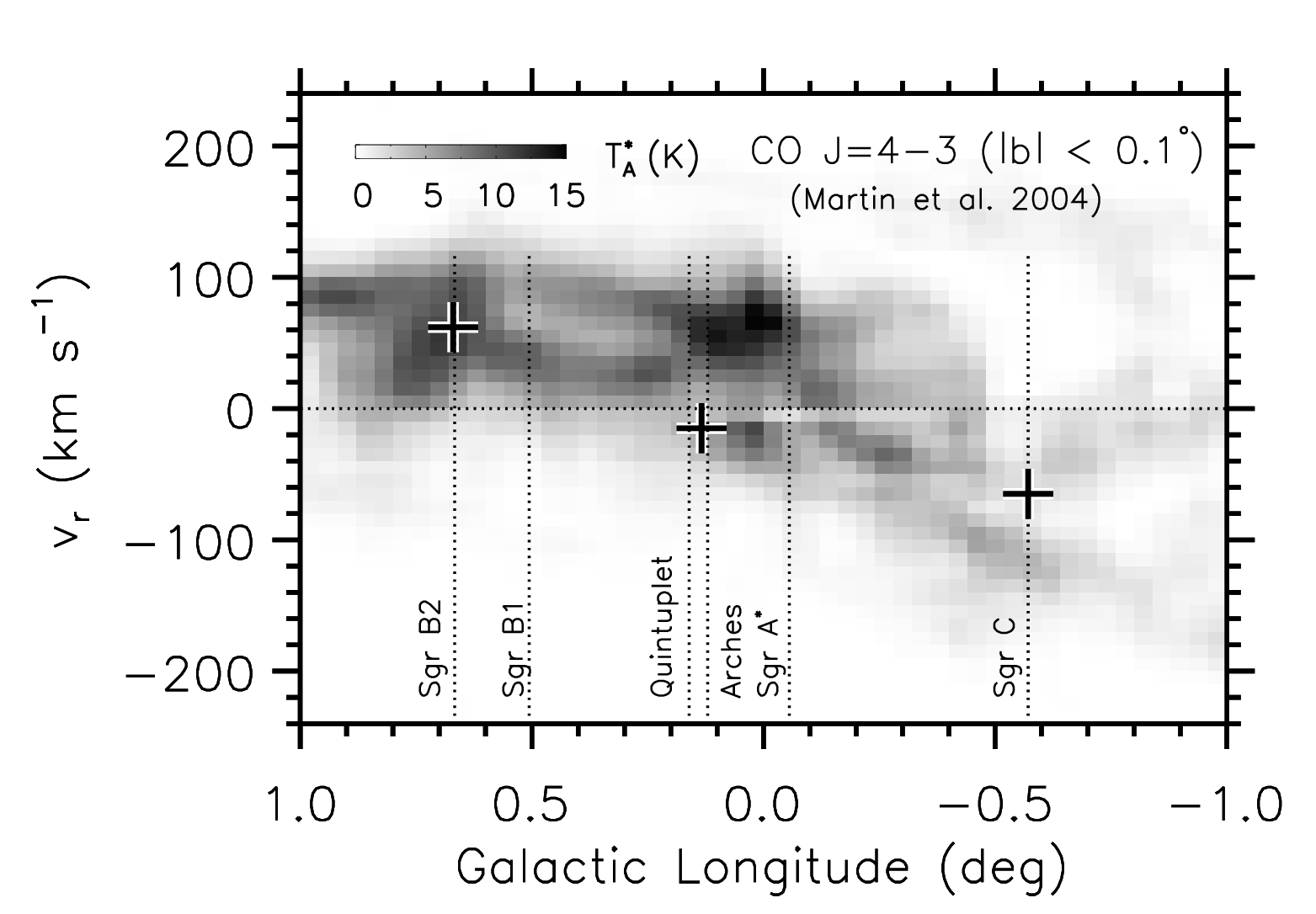}
\includegraphics[scale=0.45]{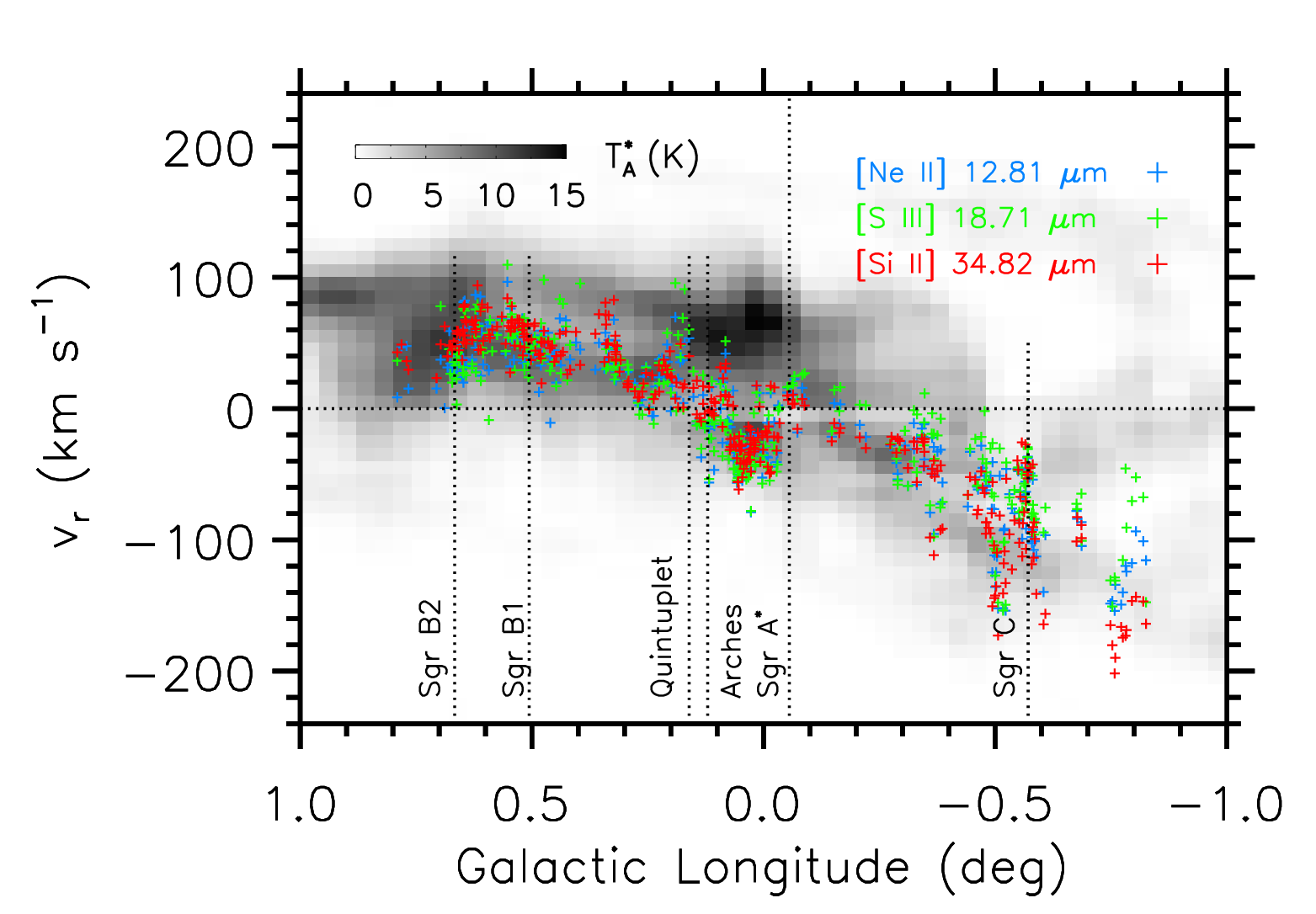}
\includegraphics[scale=0.45]{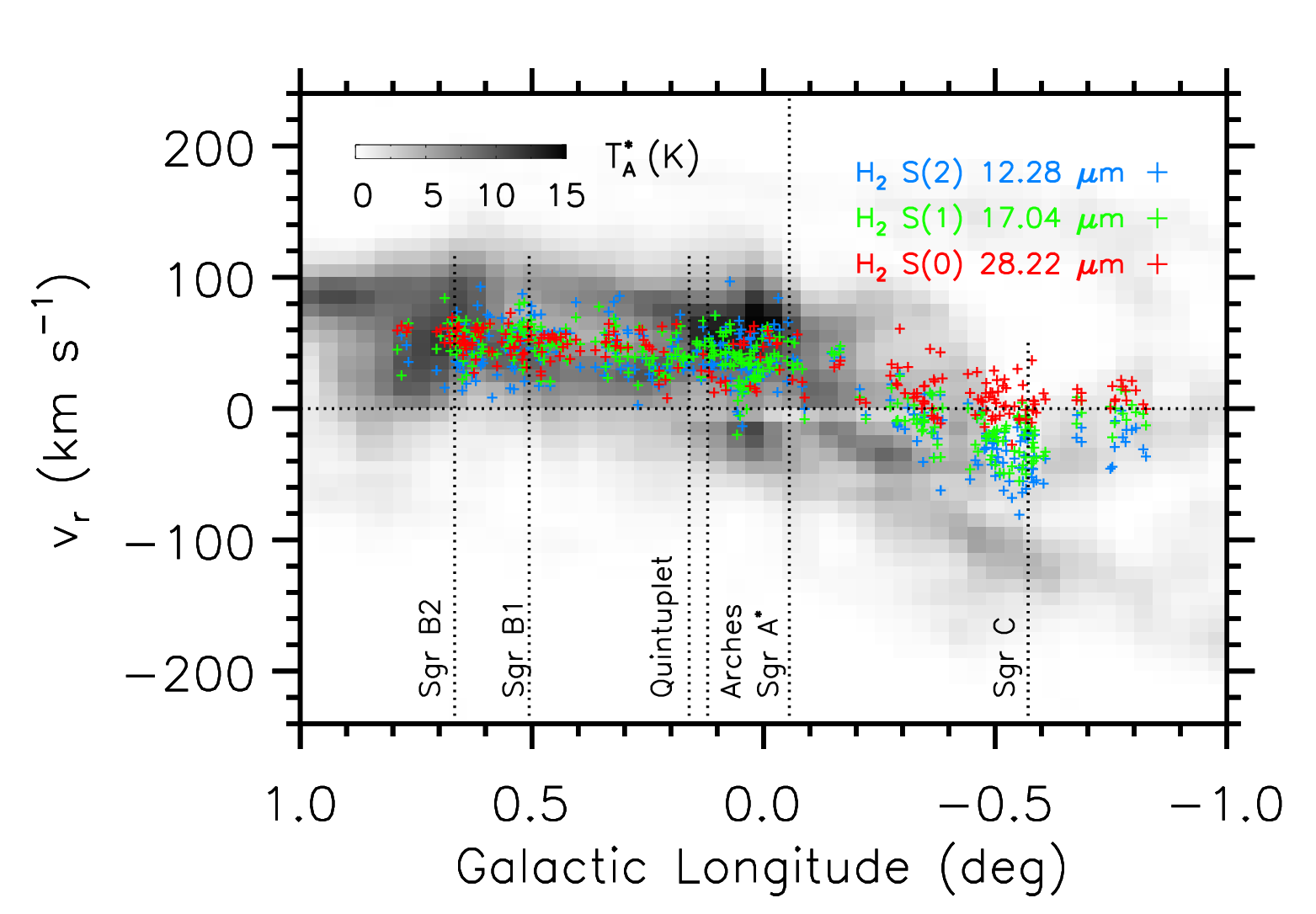}
\caption{{\it Top:} Average antenna temperatures (T$^*_A$) at $|b| <
0.1\arcdeg$ from CO $J=4\rightarrow3$ survey \citep{martin:04}. Large cross
signs indicate radial velocities from radio recombination-line studies:
\citet[][Sgr~B2]{depree:96}, \citet[][Arches cluster]{lang:01}, and
\citet[][Sgr~C]{liszt:95}.  {\it Middle:} Radial velocities from strong
forbidden emission lines, at $|b| < 0.1\arcdeg$, on top of the CO $l$-$v$
diagram.  {\it Bottom:} Radial velocities from molecular hydrogen lines, on top
of the CO $l$-$v$ diagram.  Vertical dotted lines indicate Galactic longitudes
of several landmarks in the CMZ.
\label{fig:vlos.corr}}
\end{figure*}

Radial velocity distributions as a function of Galactic longitude are displayed
in Figure~\ref{fig:vlos.corr}, on top of the $l$-$v_r$ diagram of CO
$J=4\rightarrow3$ \citep{martin:04}; see \S~\ref{sec:fitting} for details on
the $v_r$ determination.  Only those $v_r$ measurements near the Galatic plane,
$|b| < 0.1\arcdeg$, are included in the $l$-$v_r$ diagrams, and averaged
antenna temperatures ($T^*_A$) from the CO survey are shown in the same
latitude range.

Three large cross signs in the upper panel of Figure~\ref{fig:vlos.corr}
indicate radial velocities from radio recombination-line studies
\citep{depree:96,lang:01, liszt:95}, independent of the $v_r$ measurement in
\citet{mehringer:92}. These show an excellent agreement with our $v_r$ values
from ionic lines (middle panel).  The IRS ionic lines generally follow a lower
branch (smaller $v_r$) in the CO $l$-$v_r$ diagram, and the highly negative
radial velocities observed in the northern rim of Sgr~A are associated with the
Arched Filaments and the Radio Bubble \citep[e.g.,][]{simpson:07}. The IRS
spectral resolution is not high enough to resolve individual radial velocity
structures along the line of sight as in the CO survey, and our $v_r$
measurements refer to those that belong to the highest peak of the emission.

Radial velocities from the molecular hydrogen lines S(0), S(1), and S(2) are
displayed in the bottom panel of Figure~\ref{fig:vlos.corr}.  As shown in this
panel, H$_2$ lines show a systematically flatter $v_r$ curve than those from
forbidden lines (middle panel). The flatness of the $v_r$ curve depends on
H$_2$ excitation, with S(0) showing the most difference from [\ion{Ne}{2}] and
S(2) showing the least.  The H$_2$ velocities and [\ion{Ne}{2}] velocities
differ most in the region between the Arches cluster and Sgr A, and toward Sgr
C.  \citet{simpson:07} also found a systematically different $v_r$ between
ionized species and molecular hydrogen.  Our mapping results show that $v_r$
from the lowest H$_2$ energy level, S(0), even exhibits flat rotation near
Sgr~A*. A flat $v_r$ distribution in the longitude vs.\ radial velocity diagram
toward the GC is typically attributed to gas clouds in the disk along the line
of sight to the GC \citep[e.g.,][]{binney:91,rf:06}. This suggests that the
H$_2$ line-emitting clouds are a superposition of several dense clouds along the
line of sight to the GC. The $v_r$ offsets for H$_2$ emission lines are still
uncertain, because we opted to use the CO $J=4\rightarrow3$ data cube
\citep{martin:04} at a particular longitude range, $-0.2\arcdeg < l <
+1.0\arcdeg$ (see \S~\ref{sec:fitting}).

Both the uniform H$_2$ intensity distributions and distinct H$_2$ radial
velocity structures suggest that a significant fraction of the S(0) emission,
and a smaller fraction of S(1) and S(2) emission, arise in clouds that are
probably not associated with the mid-IR forbidden line emitting cloud complexes
in the CMZ.

\subsection{Emission Line Ratio Mapping in the GC}

Mapping results for forbidden line ratios are presented in
Figure~\ref{fig:lineratio}.  Line fluxes with extinction corrections from the
\citet{simpson:07} $F_{14}/F_{10}$ technique were used in the computation of
these line ratios, and only those detected at more than a $3\sigma$ level were
included in Figure~\ref{fig:lineratio}. A moving boxcar average for these line
ratios is shown on the right panels as a function of the Galactic longitude,
where the error bars indicate the standard deviation of data points in each
longitude bin.  We also present moving boxcar-averaged line ratios with
extinction corrections from the \citet{schultheis:09} map (dashed green line)
and without any correction for extinction (dashed grey line). The [\ion{S}{3}]
18.71 $\mu$m / [\ion{S}{3}] 33.48 $\mu$m line ratio is strongly affected by
extinction, as \citet{simpson:07} also found, and we therefore move the
discussion of this line ratio to the Appendix.  All of the other line ratios in
Figure~\ref{fig:lineratio} are insensitive to the extinction correction, and
will be discussed in the following sections.

\begin{figure*}
\centering
\includegraphics[scale=0.42]{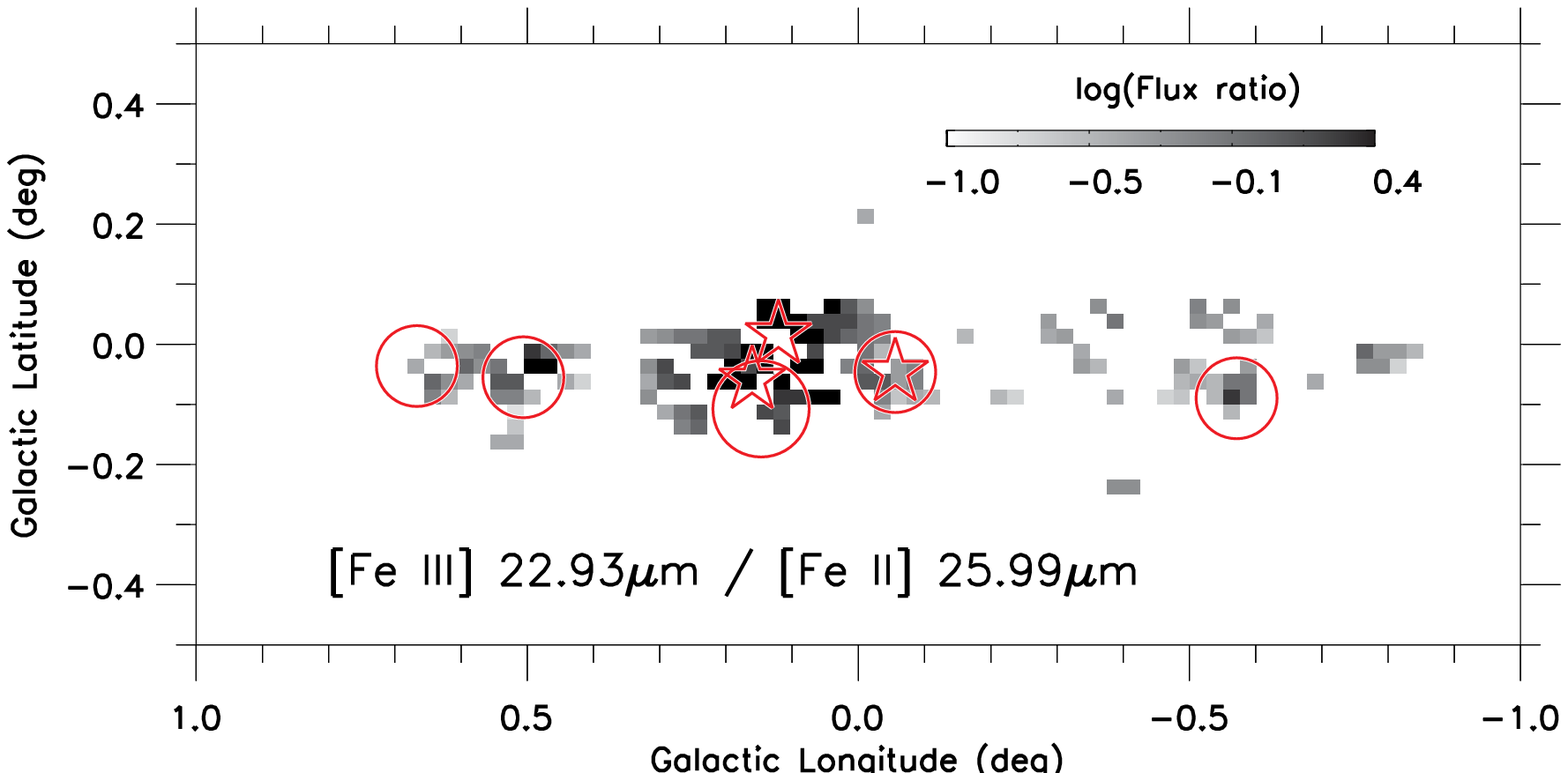}
\includegraphics[scale=0.40]{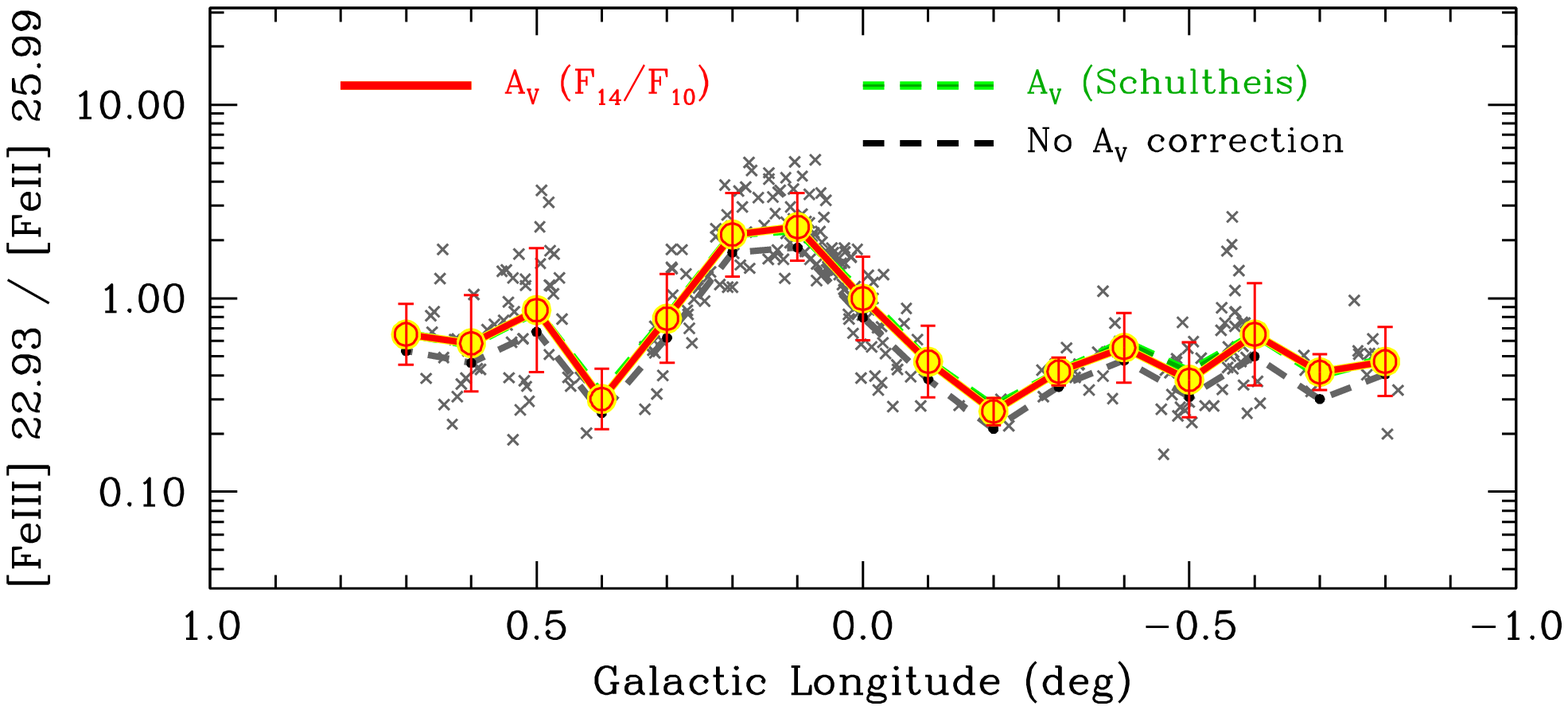}
\includegraphics[scale=0.42]{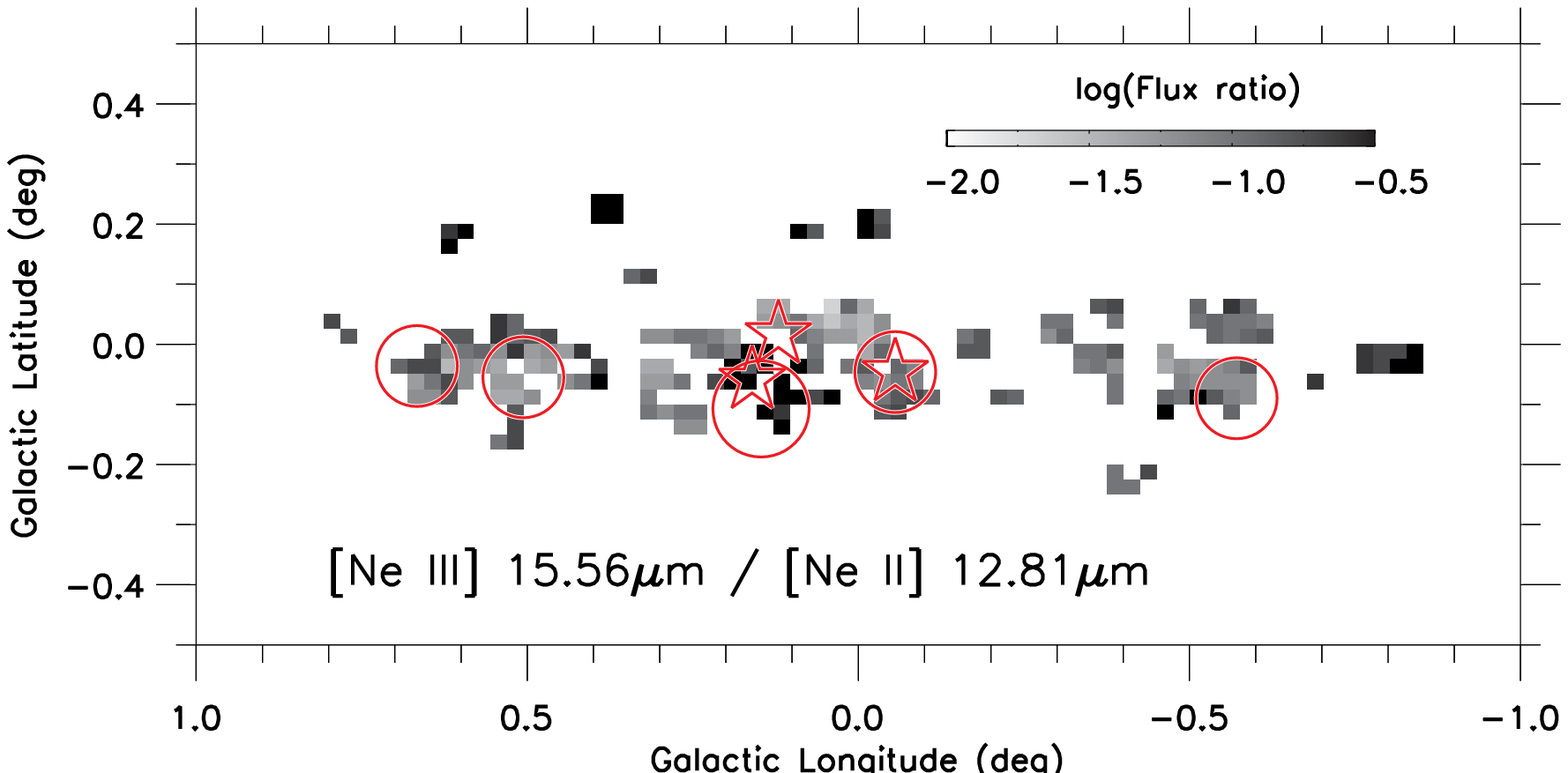}
\includegraphics[scale=0.40]{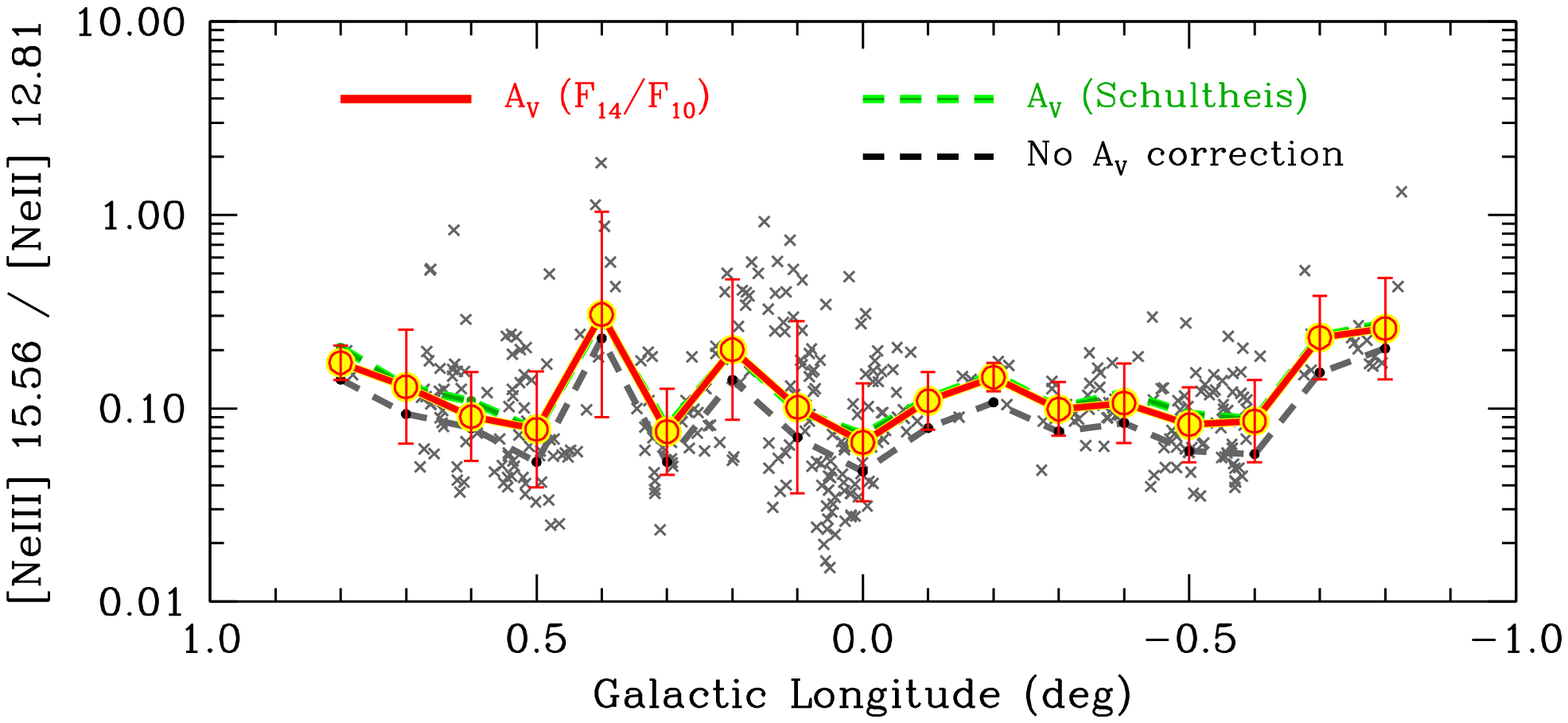}
\includegraphics[scale=0.42]{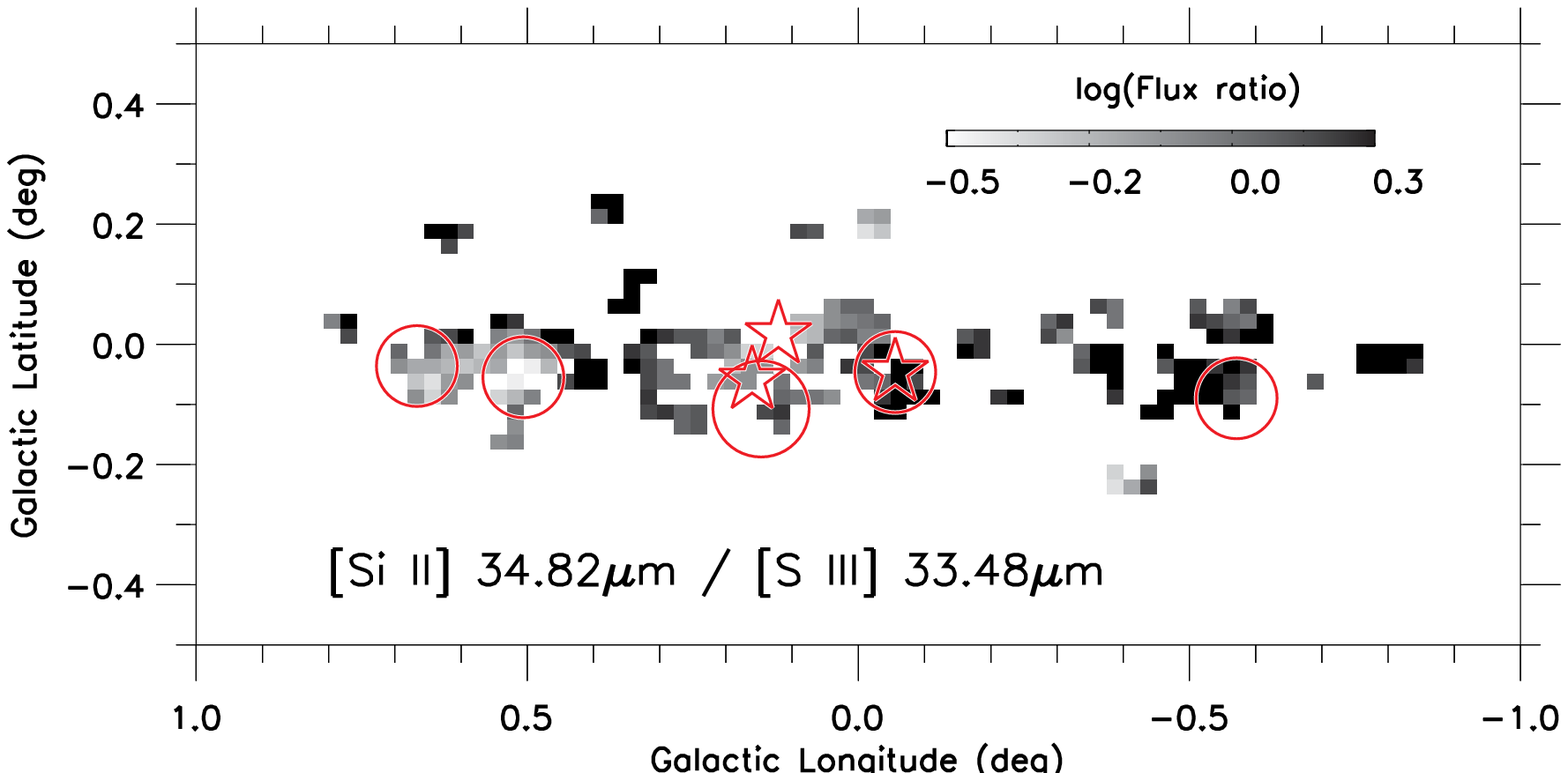}
\includegraphics[scale=0.40]{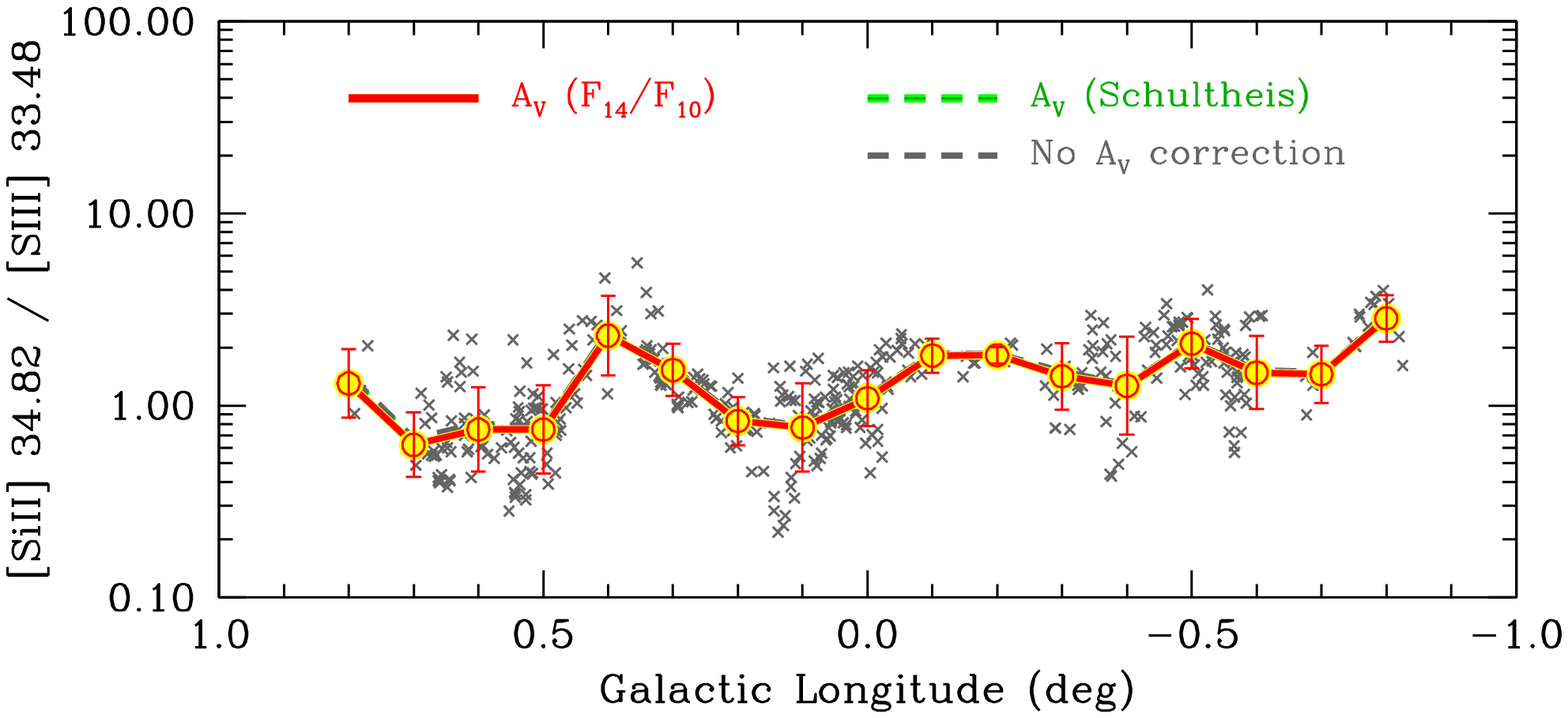}
\caption{Panoramic maps of line intensity ratios in the GC ({\it left}) and
corresponding line ratios as a function of Galactic longitude ({\it right}).
Only those lines detected at more than a $3\sigma$ level are included in the
above mapping of average line ratios within each $1.5\arcmin\times1.5\arcmin$
($\sim 3.5$~pc $\times 3.5$~pc) pixel. Line ratios are corrected for foreground
extinction based on the ratio between $10\ \mu$m and $14\ \mu$m continuum fluxes
($F_{14}/F_{10}$; see the top panel in Figure~\ref{fig:tau}). Moving averaged
points are connected with a solid red line, where error bars indicate a
$1\sigma$ dispersion in each moving average box.  Green dashed lines are moving
averaged flux ratios assuming the \citet{schultheis:09} extinction map, and the
grey dashed lines are the data without extinction corrections. Key features of
the GC are overlaid in the left panels (see Figure~\ref{fig:map}).  Horizontal
dotted line at [\ion{S}{3}] $18.71\ \mu$m / [\ion{S}{3}] $33.48\ \mu$m $=0.4$
represents the theoretical lower limit, where the gas electron density ($n_e$)
approaches zero \citep{rubin:89}.
\label{fig:lineratio}}
\end{figure*}

%figure 10b
\setcounter{figure}{9}
\begin{figure*}
\centering
\includegraphics[scale=0.40]{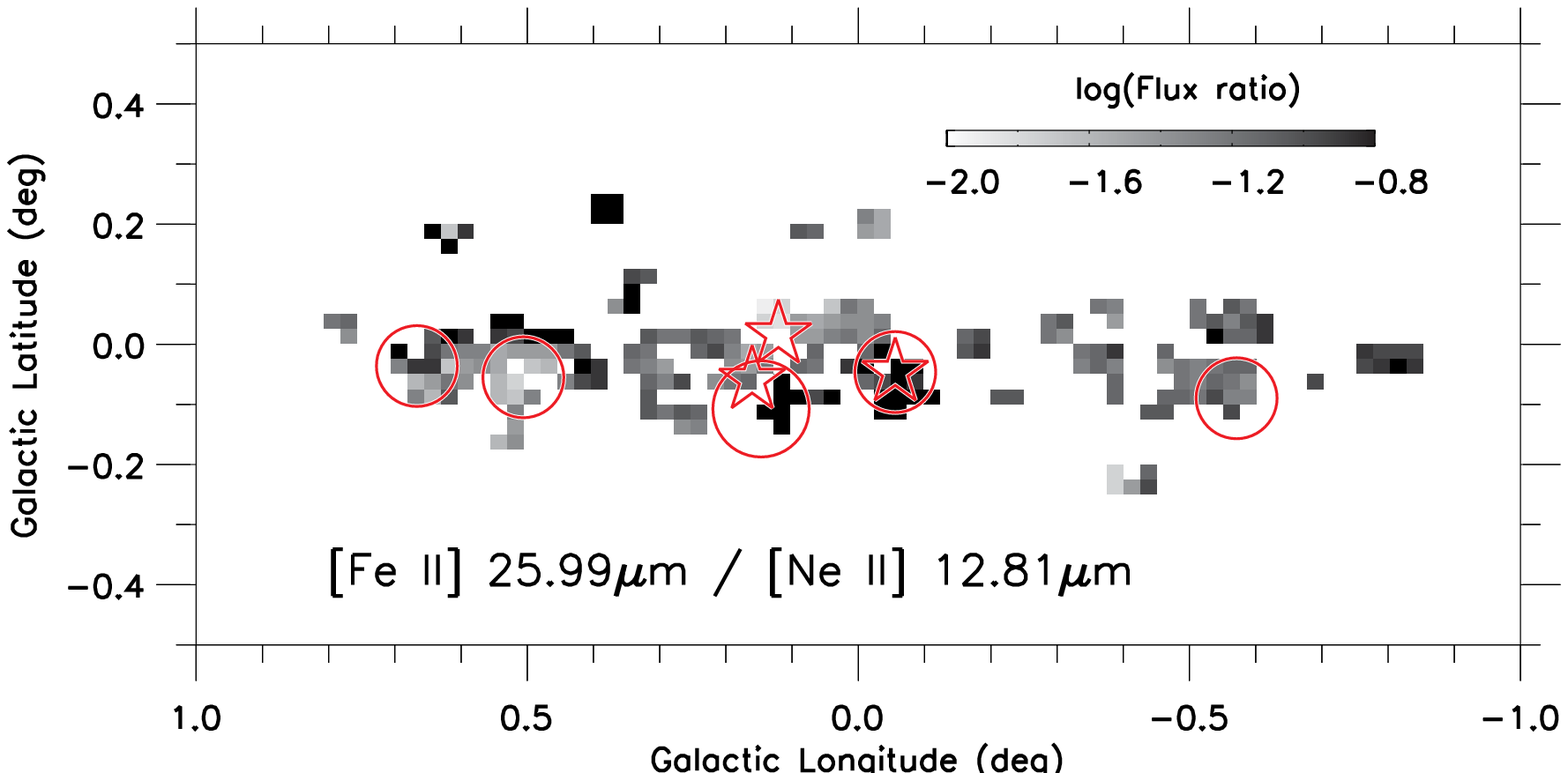}
\includegraphics[scale=0.40]{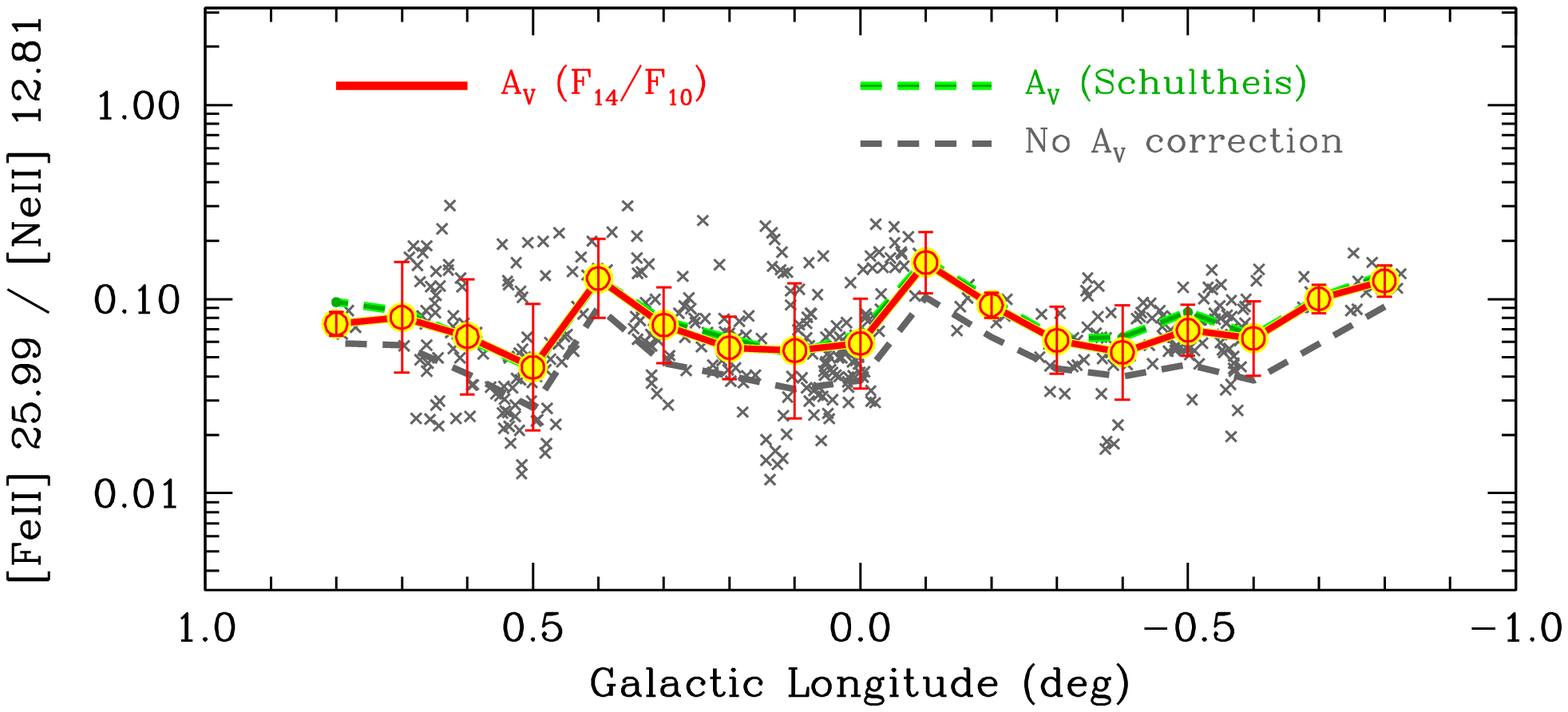}
\includegraphics[scale=0.40]{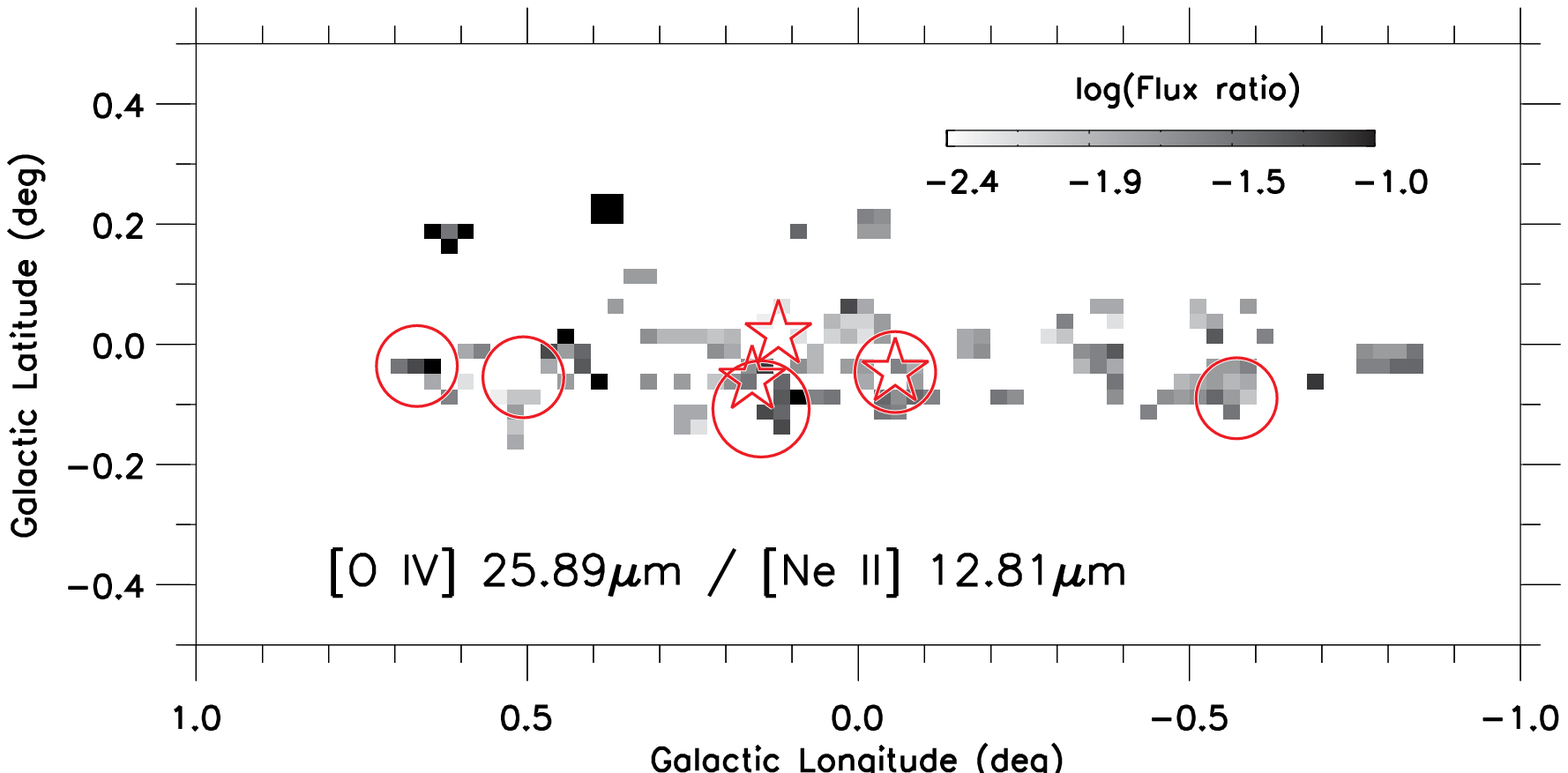}
\includegraphics[scale=0.40]{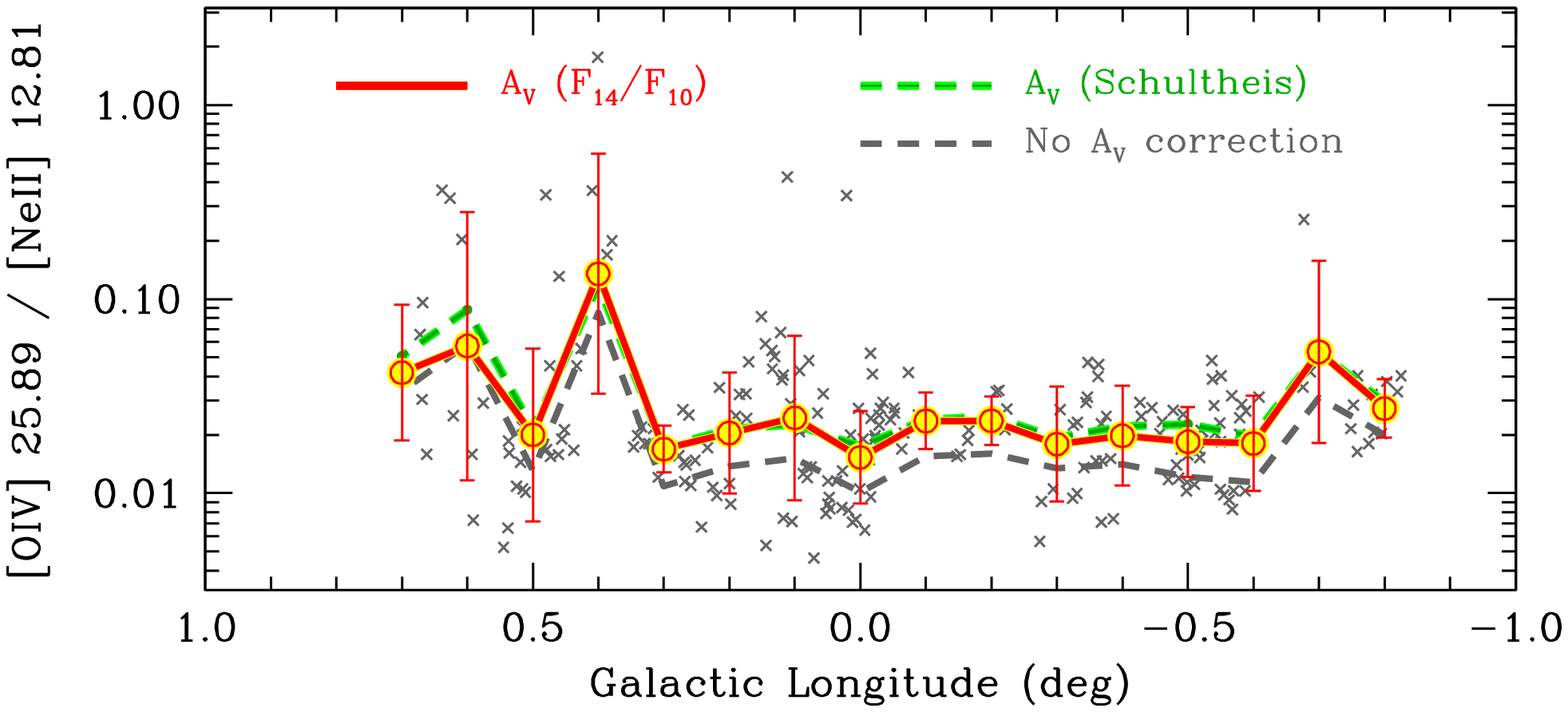}
\includegraphics[scale=0.40]{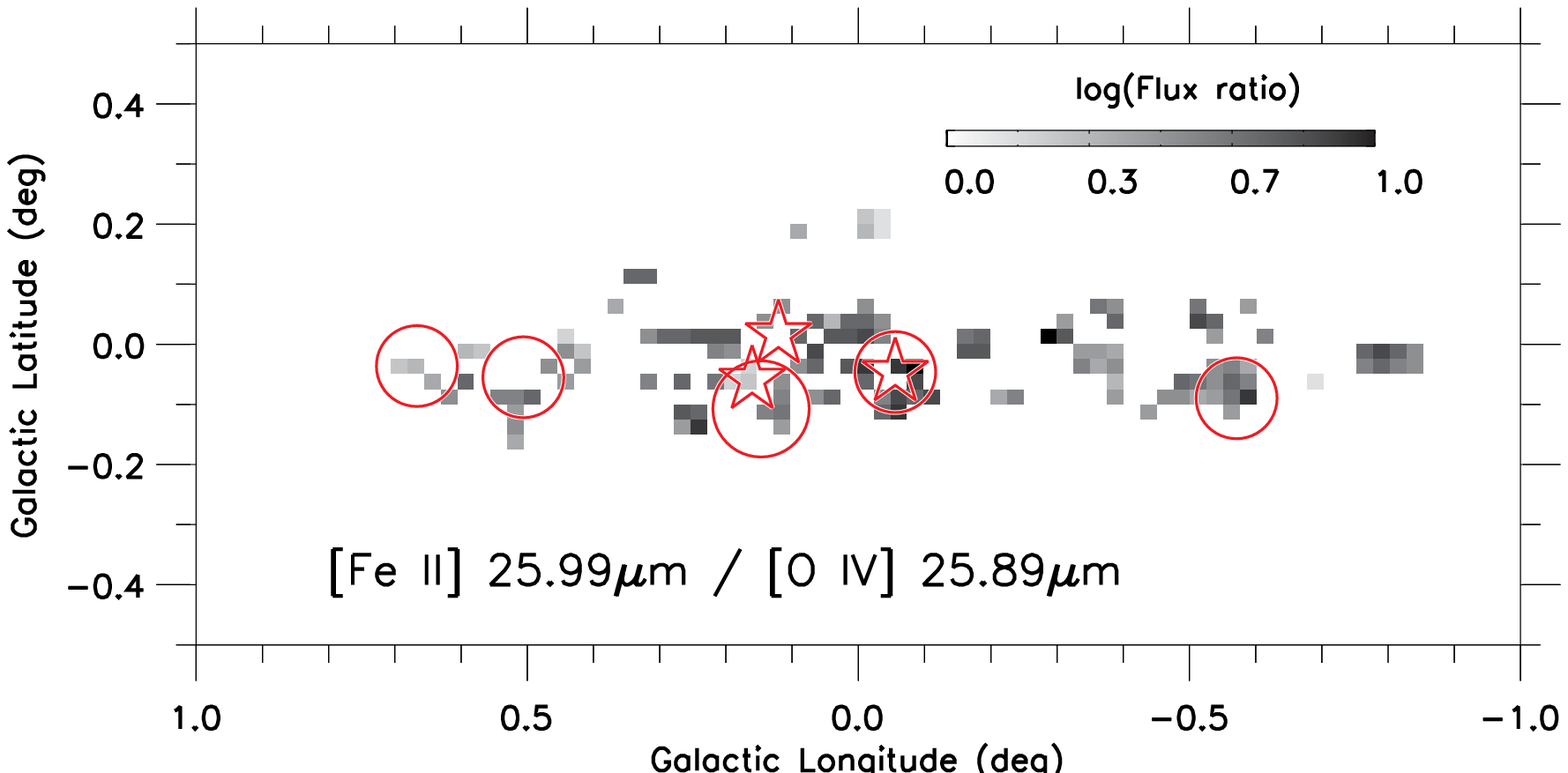}
\includegraphics[scale=0.40]{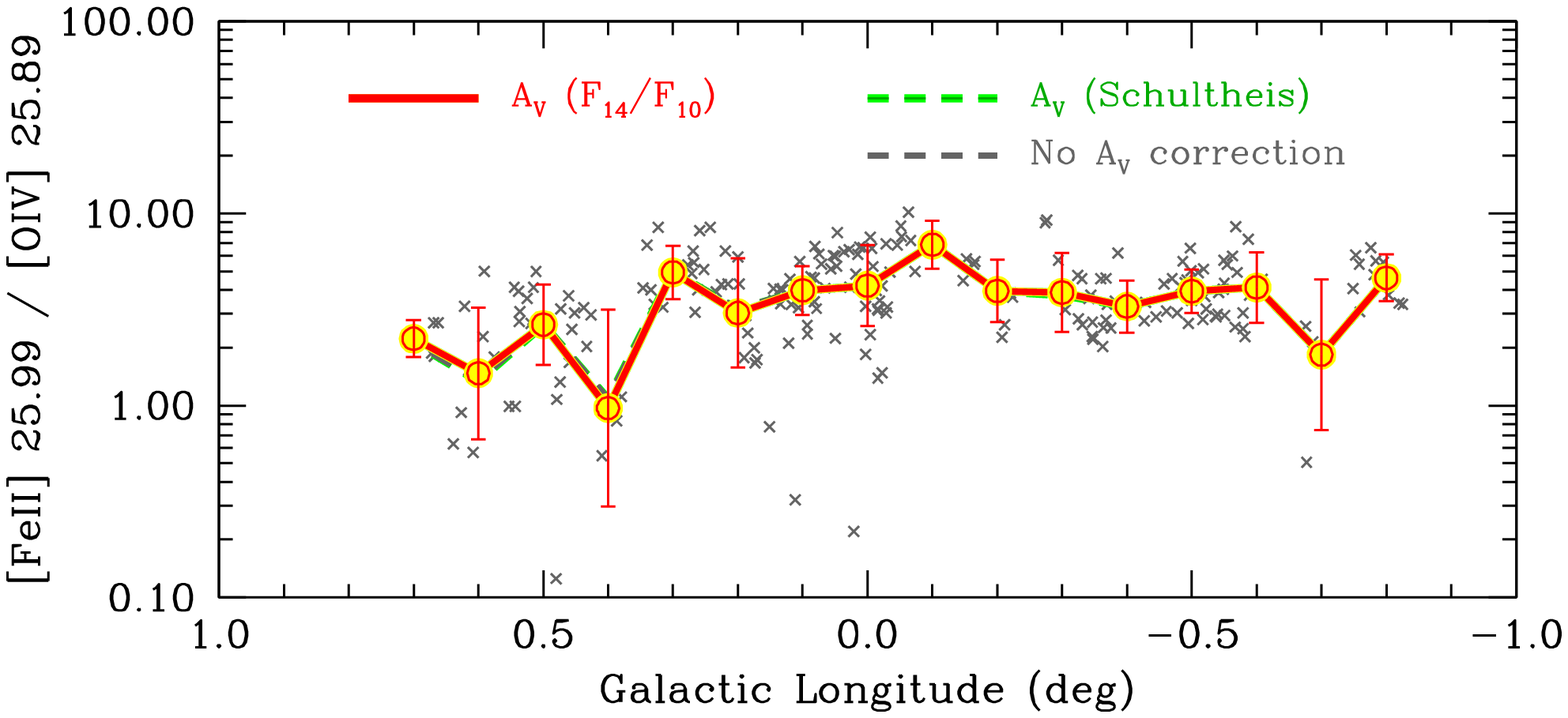}
\includegraphics[scale=0.40]{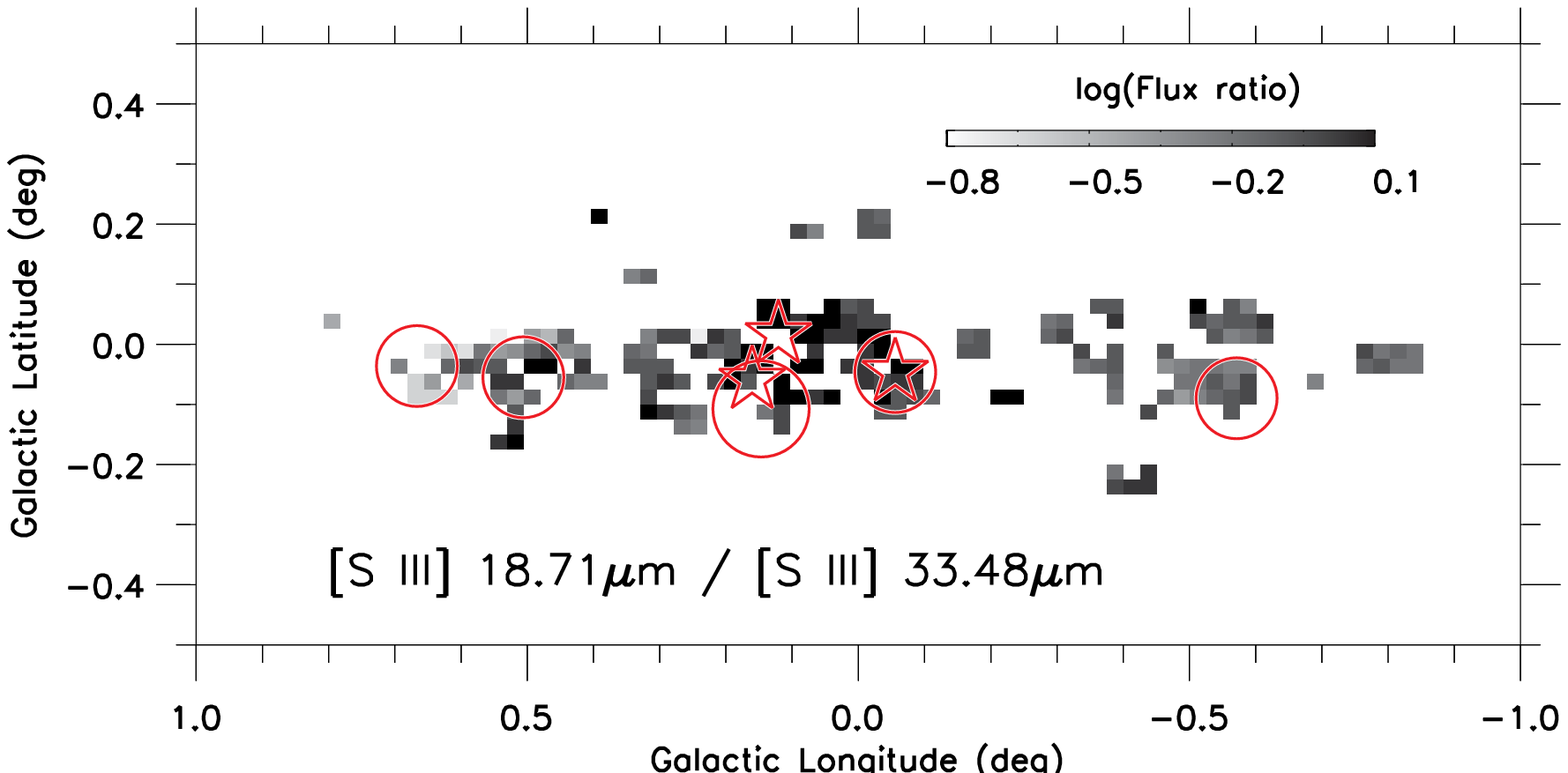}
\includegraphics[scale=0.40]{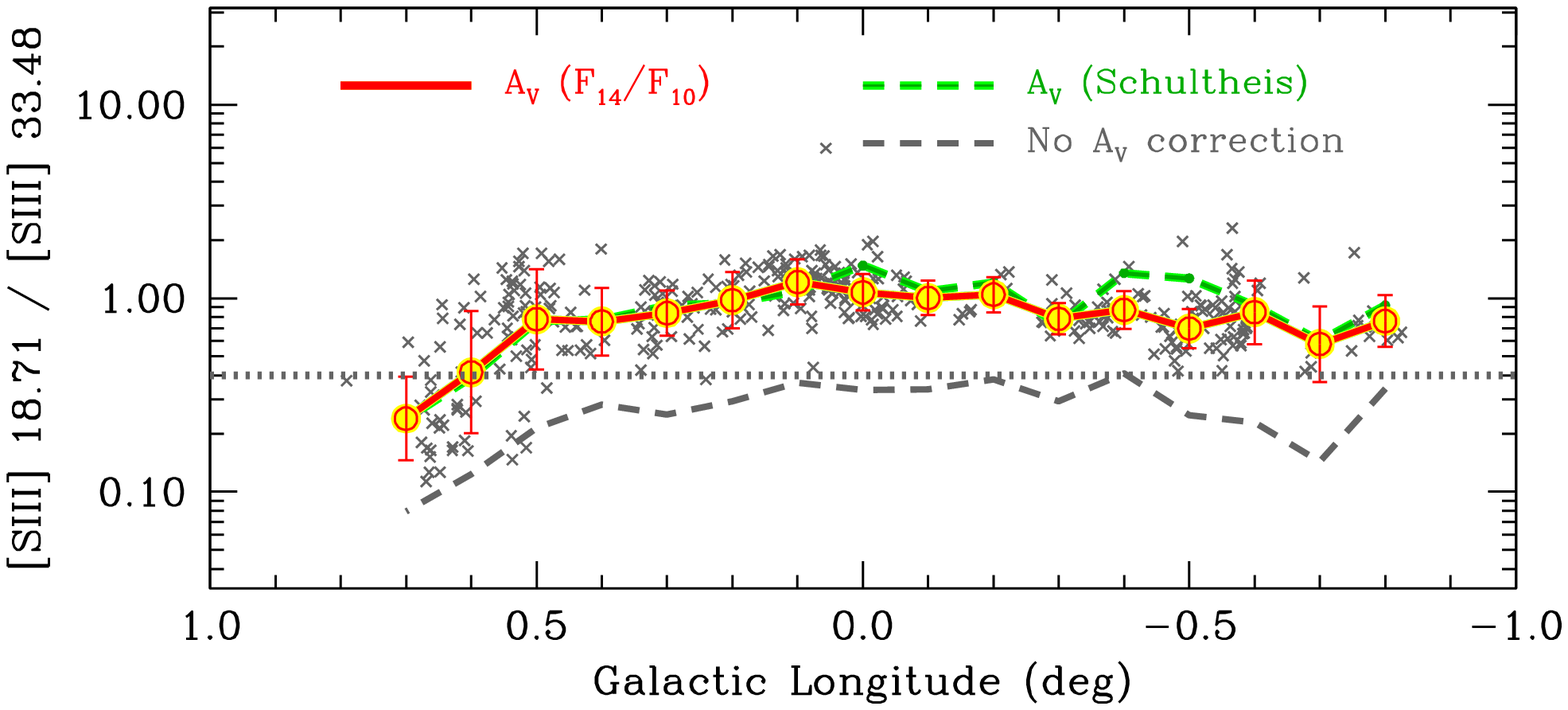}
\caption{Cont'd.
\label{fig:lineratio_b}}
\end{figure*}

We illustrate the distribution of [\ion{Fe}{3}] $22.93\ \mu$m / [\ion{Fe}{2}]
$25.99\ \mu$m as a function of Galactic longitude in Figure~\ref{fig:lineratio}.
This line ratio shows a broad peak centered at $l \sim +0.15\arcdeg$, near the
Quintuplet and Arches clusters of massive stars and the Radio Bubble.  The
[\ion{Fe}{3}] $22.93\ \mu$m / [\ion{Fe}{2}] $25.99\ \mu$m ratio is not sensitive
to the hardness of the ionizing radiation field, as [\ion{Ne}{3}] $15.56\ \mu$m
/ [\ion{Ne}{2}] $12.81\ \mu$m is (see below).  Instead, it is sensitive to the
ionization parameter, $U$, which is the ratio of the photoionization rate to the
recombination rate \citep{contini:09}.  This is because [\ion{Fe}{3}] $22.93\
\mu$m, with an ionization potential of $16.2$~eV, only arises in ionized gas,
while [\ion{Fe}{2}] $25.99\ \mu$m, with an ionization potential of $7.9$~eV,
can arise both from neutral and ionized gas \citep{kaufman:06}.
Comparisons of models with previous GC ISM observations have found $-3 \leq
\log(U) \leq -1$ \citep{rf:05,contini:09}.

\section{Comparison with Nearby Galaxies}\label{sec:discussion}

In this section, we compare ratios of ionic lines mapped in the GC to ratios
measured in the {\it Spitzer} Infrared Nearby Galaxies Survey (SINGS)
\citep{dale:09}.  The SINGS sample \citep{kennicutt:03} contains 75 galaxies,
evenly distributed among elliptical, spiral, and irregular galaxies, and
including a range of nuclear activity (quiescent, starburst, LINER, Seyfert).
The SINGS galaxies have a median distance of $9.5$~Mpc, and the closest is at
$0.6$~Mpc.  The IRS {\tt SH} and {\tt LH} slits correspond to $14$~pc $\times
33$~pc and $32$~pc $\times 65$~pc, respectively, for a galaxy at $0.6$~Mpc; they
cover $220$~pc $\times 520$~pc and $510$~pc $\times 1000$~pc, respectively, for
a galaxy at $9.5$~Mpc.  By comparison, we bin our GC line data into $1.5\arcmin
\times 1.5\arcmin$ pixels ($3.5$~pc $\times 3.5$~pc).  Our coadded spectrum of
the CMZ is constructed from spectra across a $210$~pc $\times 60$~pc region,
which is a good  match to the SINGS spatial resolution.

\subsection{Radiation Field Hardness and Oxygen Abundance}\label{sec:hardness}

The [\ion{Ne}{3}] $15.56\ \mu$m / [\ion{Ne}{2}] $12.81\ \mu$m line ratio is a
useful indicator of the radiation field hardness for star-forming regions, as
this ratio is higher when stars are hotter.  The mapping results for this line
ratio are included in Figure~\ref{fig:lineratio}. As shown in this map, the
highest excitation gas traced by [\ion{Ne}{3}]/[\ion{Ne}{2}] ratio is peaked in
the Radio Bubble and Quintuplet cluster, where ionizing photons from newly born
massive stars in the Quintuplet cluster are most likely responsible for the hard
radiation field.  \citet{simpson:07} also found the same result along the
$l \approx 0.1\arcdeg$ stripe \citep[see also][]{rf:05}.  The mean line ratio
($\approx0.1$--$1$) is consistent with recent bursts of massive star formation
in this region in the last few million years \citep[e.g.,][]{thornley:00,rf:05},
and is insensitive to a choice of foreground extinction corrections (see
dashed lines in Figure~\ref{fig:lineratio}).

\begin{figure*}
\epsscale{0.72}
\plotone{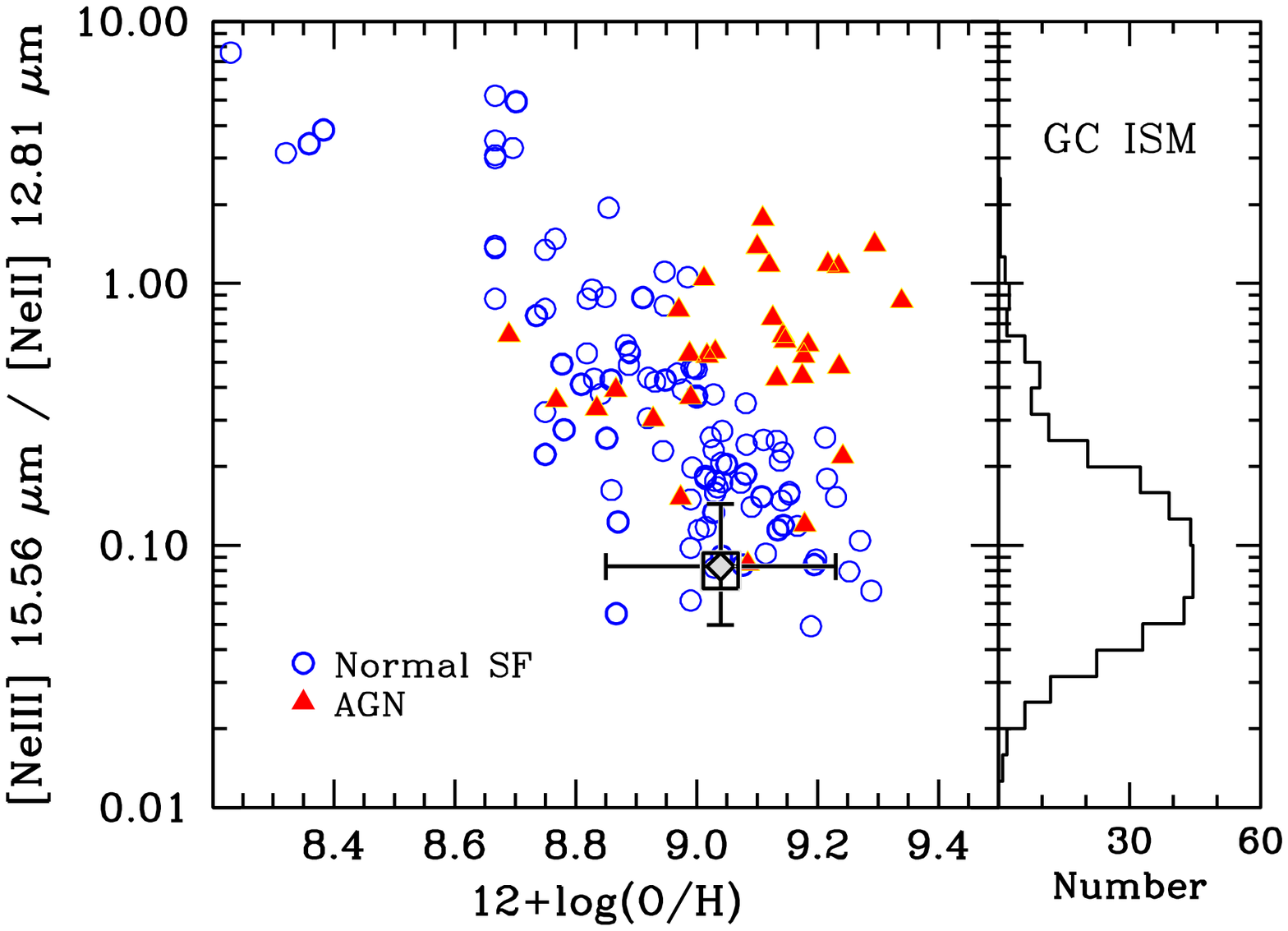}
\caption{Modification of Figure~$4$ in \citet{dale:09}, displaying [\ion{Ne}{3}]
$15.56\ \mu$m / [\ion{Ne}{2}] $12.81\ \mu$m as a function of oxygen abundance
for nuclear and extra-nuclear star-forming regions (open blue circles) in the
{\it Spitzer} Infrared Nearby Galaxies Survey (SINGS).  AGNs are shown as red
filled triangles.  The number distribution of GC spectra in
[\ion{Ne}{3}]/[\ion{Ne}{2}] is shown in the right panel. Filled diamond point
represents a median line ratio at the GC, plotted at the oxygen abundance, $[12
+ \log{({\rm O/H})}] = 9.04\pm0.19$, measured in GC stellar photospheres
\citep{cunha:07}. The vertical error bars indicate the interquartile range of
line ratios measured in the GC. Open box shows the line flux ratio from the
coadded GC spectrum.
\label{fig:lineratioOH}}
\end{figure*}

The [\ion{Ne}{3}] / [\ion{Ne}{2}] ratio also depends on the nebular oxygen
abundance, [O/H]. Figure~\ref{fig:lineratioOH} is a modification of Figure~$4$
in \citet{dale:09}, which shows [\ion{Ne}{3}]/[\ion{Ne}{2}] measured from a
sample of normal star-forming regions (both nuclear and extra-nuclear; open blue
circles) in SINGS.  As shown in Figure~\ref{fig:lineratioOH}, the line
ratio decreases as the nebular oxygen abundance increases in these normal
extragalactic star-forming regions, because the UV stellar spectrum of ionizing
hot stars depends on the (photospheric) metal abundance.  Active galactic
nuclei (AGNs) are also displayed as red filled triangles, but they do not follow
the line ratio vs.\ abundance trend observed among extra-galactic star-forming
regions.

The median [\ion{Ne}{3}] / [\ion{Ne}{2}] ratio ($\approx0.08$) measured in the
GC is shown as a filled diamond point in Figure~\ref{fig:lineratioOH}, and the
number distribution of GC spectra is displayed in the right panel. The vertical
error bars indicate the interquartile range of line ratios measured in the GC
($=0.05$--$0.14$). The line ratios from the coadded spectrum
(Table~\ref{tab:tab3}), shown as an open box symbol, are consistent with the
median value within the interquartile ranges.  The [\ion{Ne}{3}] / [\ion{Ne}{2}]
line ratio of the CMZ in Figure~\ref{fig:lineratioOH} is plotted at the stellar
oxygen abundance, $[12 + \log{({\rm O/H})}] = 9.04\pm0.19$ \citep{cunha:07},
which was derived from high-resolution infrared spectroscopy of five luminous
cool stars within $30$~pc of the GC. Their abundance measurement is consistent
with \citet{davies:09}, who measured $[12 + \log{({\rm O/H})}] = 9.09\pm0.11$
from a star that was not included in \citet{cunha:07}.
Figure~\ref{fig:lineratioOH} demonstrates that the CMZ follows the abundance
vs.\ [\ion{Ne}{3}] / [\ion{Ne}{2}] trend observed in normal star-forming regions
in nearby galaxies. This implies that the hardness of the stellar energy
distribution in the GC is similar to those found in nuclear and extra-nuclear
star-forming environments, assuming that the nebular oxygen abundance [O/H] is
close to what was measured from stars in the GC.

\subsection{AGN vs.\ Normal Star-Forming Activities in the GC}\label{sec:agn}

Empirical evidence suggests that some mid-IR line ratios are useful diagnostic
tools for discriminating between normal star-forming regions and AGNs
\citep{lutz:98,sturm:06,dale:09}.  In this section, we use several of these
emission-line diagnostics, such as [\ion{Si}{2}] $34.82\ \mu$m / [\ion{S}{3}]
$33.48\ \mu$m and [\ion{O}{4}] $25.89\ \mu$m / [\ion{Ne}{2}] $12.81\ \mu$m, to
study properties of gas clouds in the CMZ.  We inspect these line ratios in our
GC spectra, both separately and as a whole, and compare results with those
observed in other nearby galaxies.

\begin{figure*}
\epsscale{0.72}
\plotone{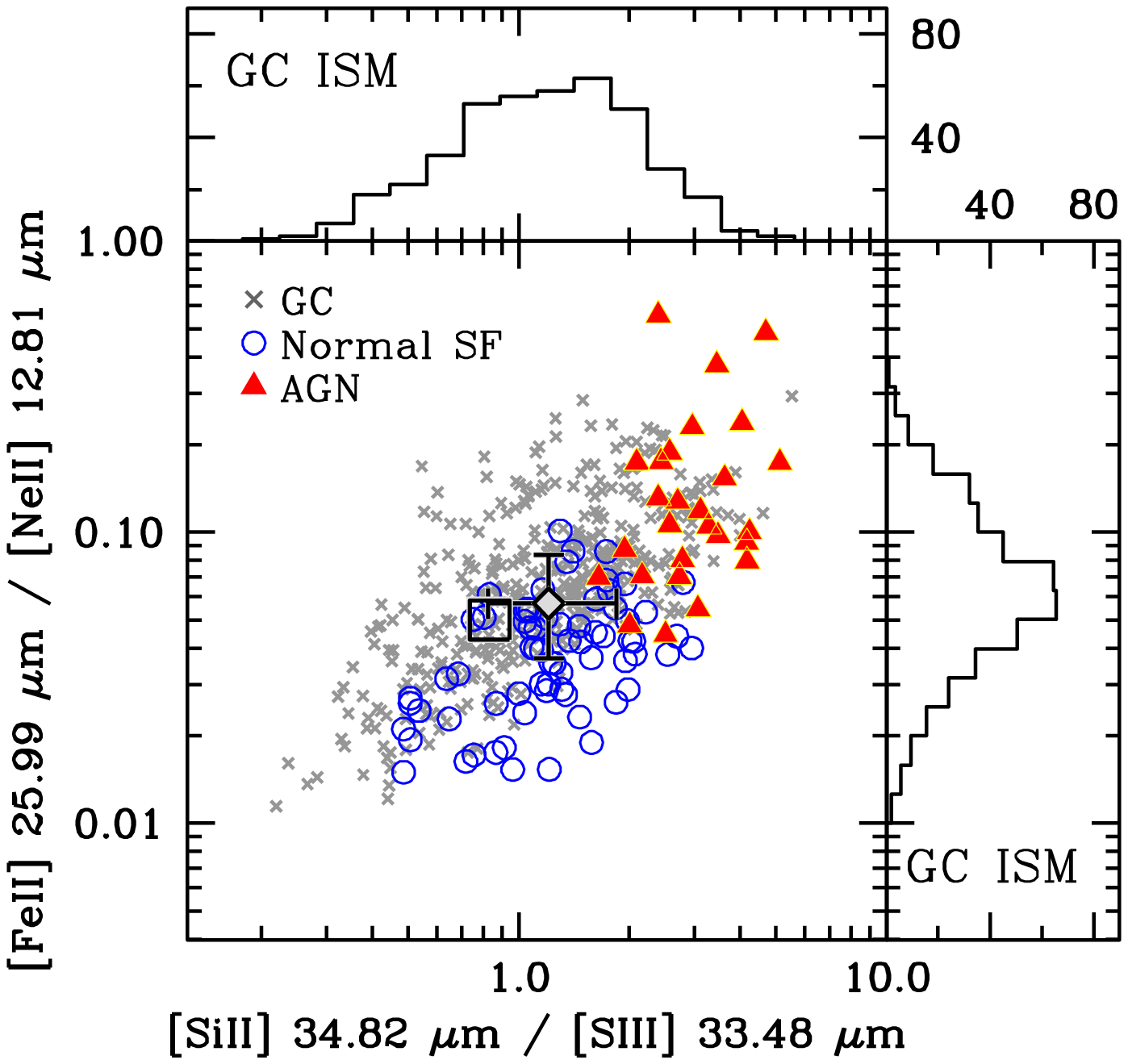}
\caption{Modification of Figure~$6$ in \citet{dale:09}, showing [\ion{Si}{2}]
$34.82\ \mu$m / [\ion{S}{3}] $33.48\ \mu$m vs.\ [\ion{Fe}{2}] $25.99\ \mu$m /
[\ion{Ne}{2}] $12.81\ \mu$m. Normal star-forming regions (both nuclear and
extra-nuclear) and AGNs are displayed as open blue circles and filled red
triangles, respectively.  Grey points are individual GC spectra, and their
number distributions are shown on each axis. Filled diamond point represents
median line ratios in the CMZ, with interquartile ranges of line ratios
indicated by error bars.  Open box shows the line flux ratio from the coadded GC
spectrum.
\label{fig:Si2S3vsFe2Ne2}}
\end{figure*}

\begin{figure*}
\epsscale{0.72}
\plotone{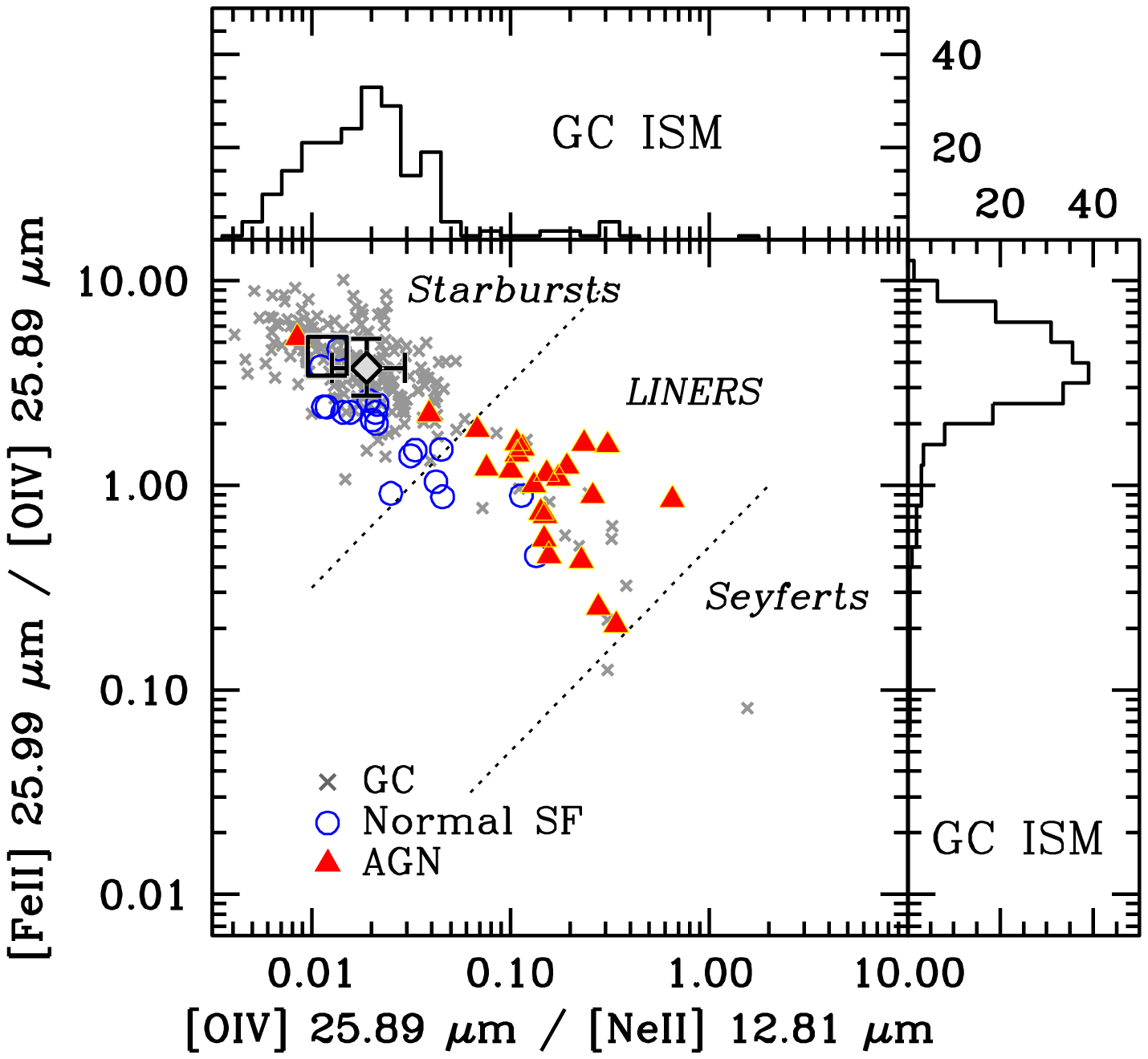}
\caption{Same as in Figure~\ref{fig:Si2S3vsFe2Ne2}, but for [\ion{O}{4}] $25.89\
\mu$m / [\ion{Ne}{2}] $12.81\ \mu$m vs.\ [\ion{Fe}{2}] $25.99\ \mu$m /
[\ion{O}{4}] $25.89\ \mu$m. Two dashed diagonal lines show approximate divisions
from \citet{sturm:06} into values for starbursts, LINERs (or pure shocks), and
Seyferts (see their Figure~2).
\label{fig:O4Ne2vsFe2O4}}
\end{figure*}

\begin{figure*}
\epsscale{0.72}
\plotone{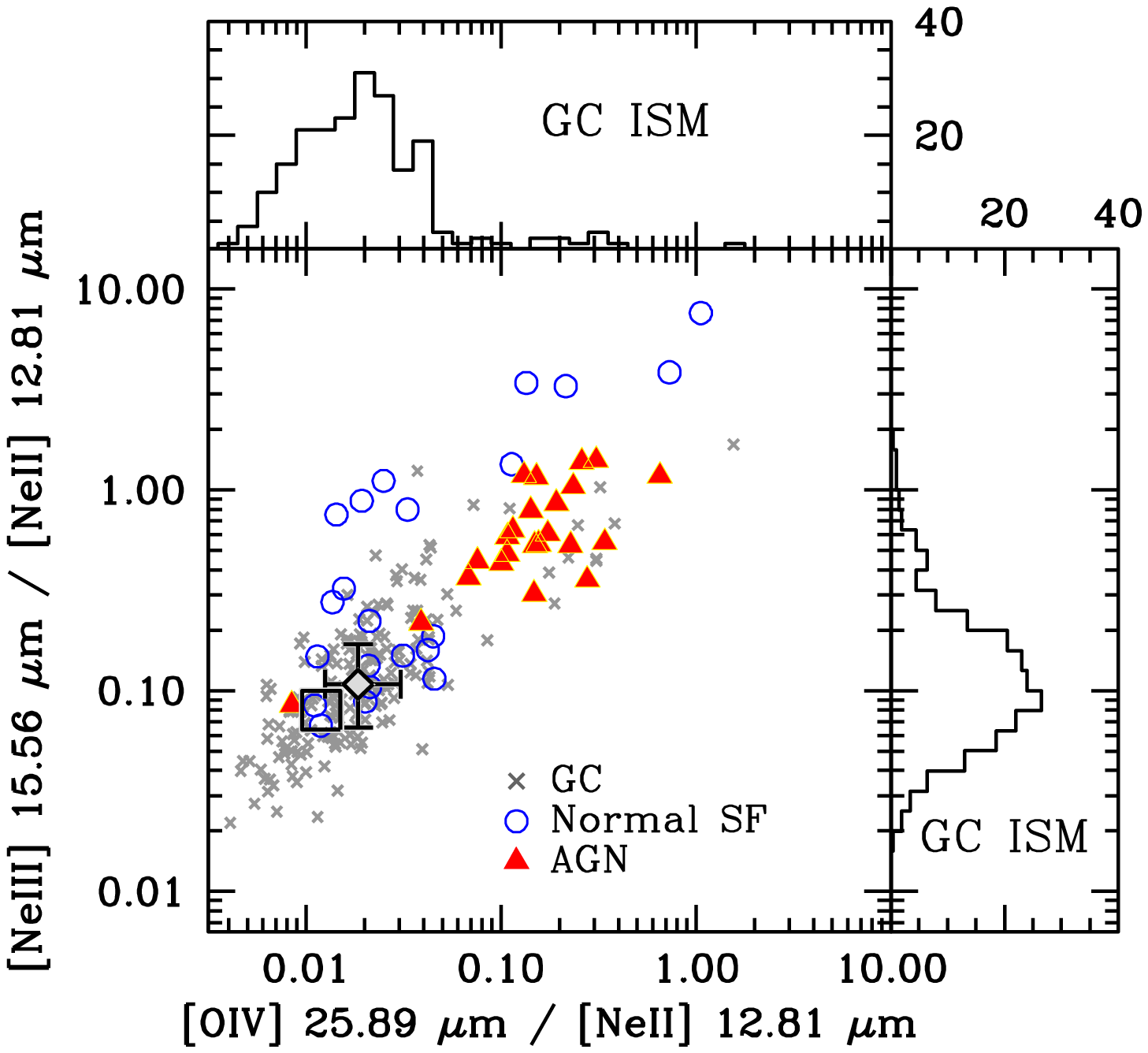}
\caption{Same as in Figure~\ref{fig:Si2S3vsFe2Ne2}, but for [\ion{O}{4}] $25.89\
\mu$m / [\ion{Ne}{2}] $12.81\ \mu$m vs.\ [\ion{Ne}{3}] $15.56\ \mu$m /
[\ion{Ne}{2}] $12.81\ \mu$m.
\label{fig:O4Ne2vsNe3Ne2}}
\end{figure*}

\begin{figure*}
\epsscale{0.72}
\plotone{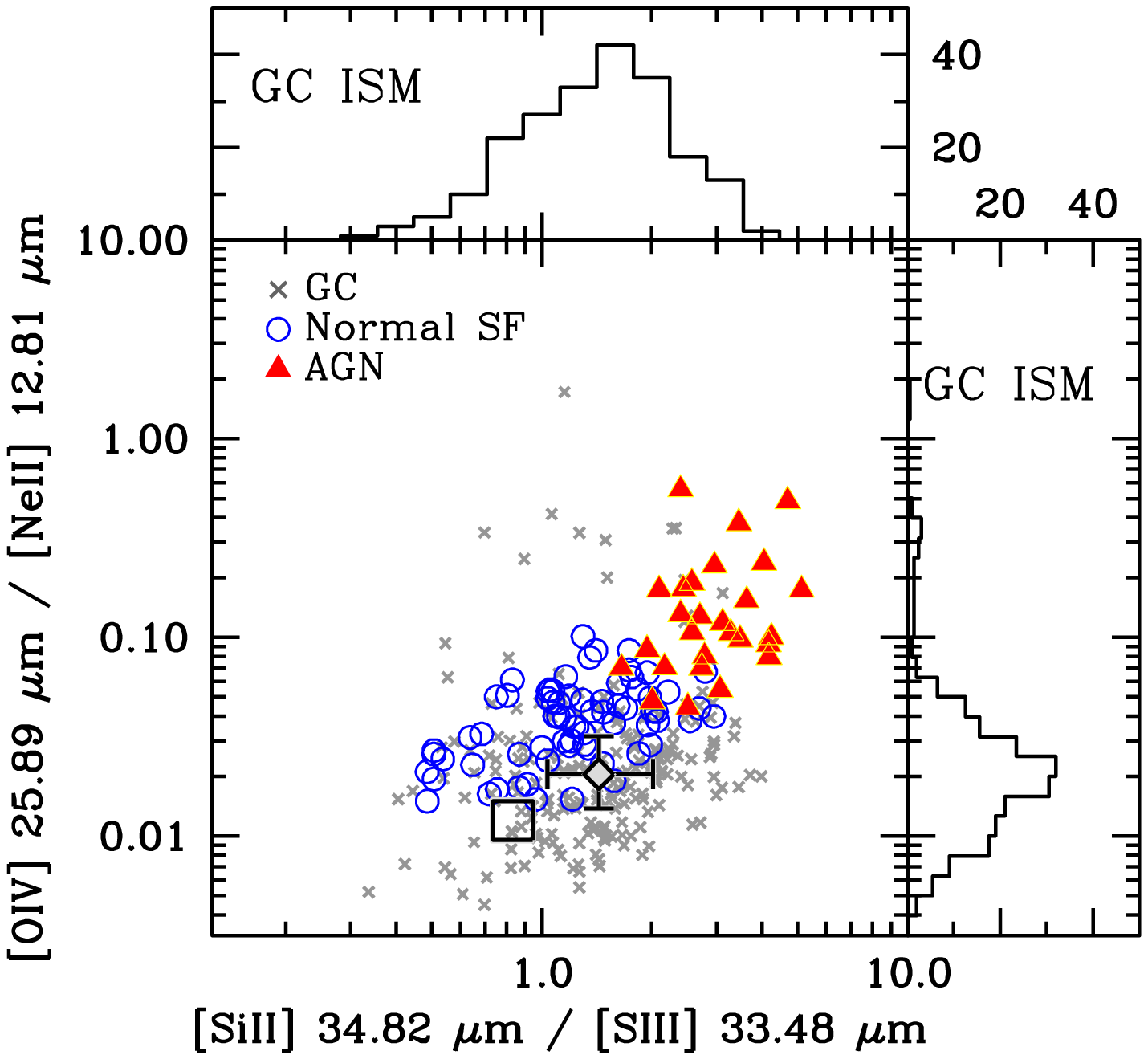}
\caption{Same as in Figure~\ref{fig:Si2S3vsFe2Ne2}, but for [\ion{Si}{2}]
$34.82\ \mu$m / [\ion{S}{3}] $33.48\ \mu$m vs.\ [\ion{O}{4}] $25.89\ \mu$m /
[\ion{Ne}{2}] $12.81\ \mu$m.
\label{fig:Si2S3vsO4Ne2}}
\end{figure*}

\subsubsection{[\ion{Si}{2}]/[\ion{S}{3}] and [\ion{Fe}{2}]/[\ion{Ne}{2}]}

Figure~\ref{fig:Si2S3vsFe2Ne2}, which is a modification of Figure~$6$ in
\citet{dale:09}, is one of such diagnostic tools, showing [\ion{Si}{2}] $34.82\
\mu$m / [\ion{S}{3}] $33.48\ \mu$m vs.\ [\ion{Fe}{2}] $25.99\ \mu$m /
[\ion{Ne}{2}] $12.81\ \mu$m from SINGS.  Both nuclear and extra-nuclear
star-forming regions are shown as blue circles, and AGNs are shown as red filled
triangles. Previous theoretical work showed that the [\ion{Si}{2}] $34.82\ \mu$m
line can be emitted either from PDRs or \ion{H}{2} regions, while [\ion{S}{3}]
$33.48\ \mu$m is mostly from \ion{H}{2} regions \citep{kaufman:06}. 

In Figure~\ref{fig:Si2S3vsFe2Ne2}, normal star-forming regions show a strong
correlation between [\ion{Si}{2}] $34.82\ \mu$m / [\ion{S}{3}] $33.48\ \mu$m and
[\ion{Fe}{2}] $25.99\ \mu$m / [\ion{Ne}{2}] $12.81\ \mu$m, the latter being
another useful line diagnostic of the ionization parameter. However, AGNs
generally show higher values of these line ratios than values from normal
star-forming nuclear and extra-nuclear regions in nearby galaxies, and a
combination of these two IR line ratios separates AGNs and normal star-forming
regions relatively well.  Physical reasons for this empirical division are
debated, but must include the low ionization potentials of Si$^+$ and Fe$^+$
ions ($8.15$~eV and $7.90$~eV, respectively) and the higher ionization
potentials of Ne$^+$ and S$^{++}$ ($21.56$~eV and $23.34$~eV, respectively).
\citet{dale:09} proposed a number of possible physical mechanisms for the
observed separation between AGNs and normal star-forming regions, which include
(1) enhanced dust destruction and sublimation of refractory elements (Si and Fe)
in the harsh AGN environment, (2) extended low-ionization volumes in AGNs
produced by X-ray photo-ionization processes, and/or (3) enhanced line emission
from [\ion{Si}{2}] $34.82\ \mu$m and [\ion{Fe}{2}] $25.99\ \mu$m due to high gas
density in PDRs and/or X-ray dominated regions in AGNs.

Gray cross points in Figure~\ref{fig:Si2S3vsFe2Ne2} display line ratios measured
from individual GC spectra, shown only for those detected at more than a
$3\sigma$ significance. Their number distributions are shown in a histogram on
each axis. The filled diamond point represents the median line ratio from the
individual spectra of the entire CMZ ([\ion{Si}{2}] / [\ion{S}{3}] $=1.21$ and
[\ion{Fe}{2}] / [\ion{Ne}{2}] $=0.06$), with error bars indicating interquartile
ranges. The values from the coadded CMZ spectrum are shown as an open box
symbol; they are different from the median GC values because we only plot
$3\sigma$ detections. Clearly, the vast majority of the GC measurements and the
median GC values are found in an area that is mostly populated by normal
star-forming regions in nearby galaxies. Line ratios on both axes are
insensitive to extinction corrections.

\subsubsection{[\ion{O}{4}]/[\ion{Ne}{2}] , [\ion{Fe}{2}]/[\ion{O}{4}], and
[\ion{Ne}{3}]/[\ion{Ne}{2}]}

Another set of mid-IR line diagnostics to separate AGNs from normal star-forming
regions are displayed in Figure~\ref{fig:O4Ne2vsFe2O4}, which plots [\ion{O}{4}]
$25.89\ \mu$m / [\ion{Ne}{2}] $12.81\ \mu$m vs.\ [\ion{Fe}{2}] $25.99\ \mu$m /
[\ion{O}{4}] $25.89\ \mu$m, and in Figure~\ref{fig:O4Ne2vsNe3Ne2}, which shows
[\ion{O}{4}] $25.89\ \mu$m / [\ion{Ne}{2}] $12.81\ \mu$m vs.\ [\ion{Ne}{3}]
$15.56\ \mu$m / [\ion{Ne}{2}] $12.81\ \mu$m.  Both diagrams, which involve line
emission from [\ion{O}{4}] $25.89\ \mu$m, separate AGNs (red filled triangles)
from normal star-forming regions (blue open circles) relatively well; see
\citet{dale:09} for the SINGS targets used in these diagrams.  In addition, two
diagonal dashed lines in Figure~\ref{fig:O4Ne2vsFe2O4} show approximate
divisions between starbursts, LINERs (Low Ionization Nuclear Emission Regions)
and pure shocks, and Seyfert galaxies from \citet[see their Figure~2]{sturm:06}.
Figure~\ref{fig:O4Ne2vsNe3Ne2} shows that there are a few normal star-forming
regions from SINGS (blue circles) with [\ion{O}{4}] $25.89\ \mu$m /
[\ion{Ne}{2}] $12.81\ \mu$m greater than $\sim0.10$.  These are low-metallicity
star-forming regions, as revealed from the comparison between
Figure~\ref{fig:lineratioOH} and Figure~\ref{fig:O4Ne2vsNe3Ne2}, where strong
[\ion{O}{4}] $25.89\ \mu$m emission is due to a harder spectrum for the stellar
ionizing photons in low-metallicity environments.

Gray cross points in Figures~\ref{fig:O4Ne2vsFe2O4}--\ref{fig:O4Ne2vsNe3Ne2}
represent our measurements from individual GC spectra.  As in
Figure~\ref{fig:Si2S3vsFe2Ne2}, we only included emission lines with more than
$3\sigma$ detections.  Number distributions of ionic line ratios are shown on
each axis.  The median values from the individual spectra of the CMZ are
indicated by a grey diamond point; error bars represent the interquartile ranges
of distributions of individual GC spectra.  The line ratios for the coadded GC
spectrum is displayed as an open box.
Figures~\ref{fig:O4Ne2vsFe2O4}--\ref{fig:O4Ne2vsNe3Ne2} show that the
emission-line properties of the CMZ are similar to those of extragalactic
star-forming regions (blue circles) and starburst galaxies (upper left corner in
Figure~\ref{fig:O4Ne2vsFe2O4}, fenced with a dashed diagonal line).  This agrees
with our conclusion based on [\ion{Si}{2}] / [\ion{S}{3}] vs.\ [\ion{Fe}{2}] /
[\ion{Ne}{2}] diagnostics in Figure~\ref{fig:Si2S3vsFe2Ne2}.

The line ratios for the coadded spectrum of the CMZ in
Figure~\ref{fig:O4Ne2vsNe3Ne2} ([\ion{O}{4}]/[\ion{Ne}{2}]$\sim0.01$ and
[\ion{Ne}{3}]/[\ion{Ne}{2}]$\sim0.08$) are found within the range observed among
starburst galaxies \citep[][see their Figure~2]{lutz:98}.  \citet{lutz:98}
suggested that shocks and/or hot stars are the likely origins of [\ion{O}{4}]
observed in these galaxies, rather than buried AGNs.  \citet{simpson:07} also
reported widespread detection of [\ion{O}{4}] in the GC; they attributed this to
shocks in the CMZ \citep[see also][]{contini:09}.  \citet{simpson:07} further
argued for the existence of shocked gas inside of the Radio Bubble based on the
high Fe abundance there, since grain destruction in shocks can return Fe to the
ISM.

If the above result implies the importance of shocked gas in the GC, widespread
detections of highly ionized ions such as [\ion{O}{4}] (Figure~\ref{fig:elines})
may support a large scale origin of turbulent motions in the GC. In other words,
the relatively uniform intensity distributions of these lines in our mapping
suggest that conditions for producing the lines are common throughout the CMZ.
One likely explanation is a dissipation of supersonic turbulence ($\ga10$~km/s
in the CMZ) induced by differential Galactic rotation or shearing motion of
clouds in the GC \citep[e.g.,][]{wilson:82,guesten:85}. However, shocks created
by massive stars are also likely; see \citet{rf:04} for discussion of other
potential heating mechanisms, such as cosmic rays, magnetic heatings, or X-ray
dominated regions.

While most of the GC points (gray crosses) in Figure~\ref{fig:Si2S3vsFe2Ne2} are
found in an area occupied by normal star-forming regions, including the coadded
spectrum of the CMZ, about $10\%$ of the detections fall into AGN territory.  As
shown in the mapping results in Figure~\ref{fig:lineratio}, both [\ion{Si}{2}]
$34.82\ \mu$m / [\ion{S}{3}] $33.48\ \mu$m and [\ion{Fe}{2}] $25.99\ \mu$m /
[\ion{Ne}{2}] $12.81\ \mu$m line ratios are peaked near Sgr~A and the
north-western rim of Sgr~B. The scatter of the [\ion{Fe}{2}] $25.99\ \mu$m /
[\ion{Ne}{2}] $12.81\ \mu$m points in the moving average plot is larger than
that of [\ion{Si}{2}] $34.82\ \mu$m / [\ion{S}{3}] $33.48\ \mu$m because
[\ion{Fe}{2}] $25.99\ \mu$m is relatively weak and blended with [\ion{O}{4}]
$25.89\ \mu$m. Nevertheless, the observed systematic trend in Galactic longitude
(right panels in Figure~\ref{fig:lineratio}) is larger than the random scatter
of data points, which means that those GC points in the upper right corner in
Figure~\ref{fig:Si2S3vsFe2Ne2} are not entirely produced by a statistical
fluctuation.

Figures~\ref{fig:O4Ne2vsFe2O4} and \ref{fig:O4Ne2vsNe3Ne2} show that about 10\%
of individual GC spectra fall in the AGN (red filled triangles) region of these
line ratio diagrams, almost independent of the choice of extinction correction.
Note again that a few SINGS targets from extragalactic star-forming regions are
also found in the area denoted as AGNs in Figure~\ref{fig:O4Ne2vsNe3Ne2},
because of their low metal abundance (see above). On the other hand, the GC
spectra with strong [\ion{O}{4}] $25.89\ \mu$m / [\ion{Ne}{2}] $12.81\ \mu$m are
not due to a low metallicity environment, because they follow the AGN sequence
in Figure~\ref{fig:O4Ne2vsNe3Ne2} rather than that of metal-poor, star-forming
regions, and because the GC has a supersolar oxygen abundance \citep{cunha:07,
davies:09}.

If high values of [\ion{O}{4}] $25.89\ \mu$m / [\ion{Ne}{2}] $12.81\ \mu$m and
low values of [\ion{Fe}{2}] $25.99\ \mu$m / [\ion{O}{4}] $25.89\ \mu$m in
Figures~\ref{fig:O4Ne2vsFe2O4} are due to ionization by a power-law continuum
source (i.e., an AGN), then the empirical division in
Figure~\ref{fig:Si2S3vsFe2Ne2} leads to a conclusion that these same spectra
should also have high values of the ratios of  [\ion{Si}{2}] $34.82\ \mu$m /
[\ion{S}{3}] $33.48\ \mu$m and [\ion{Fe}{2}] $25.99\ \mu$m / [\ion{Ne}{2}]
$12.81\ \mu$m.  To answer the question of whether AGN-like points in
Figures~\ref{fig:Si2S3vsFe2Ne2}--\ref{fig:O4Ne2vsNe3Ne2} are from the same
individual GC spectra, we plot [\ion{Si}{2}] $34.82\ \mu$m / [\ion{S}{3}]
$33.48\ \mu$m vs.\ [\ion{O}{4}] $25.89\ \mu$m / [\ion{Ne}{2}] $12.81\ \mu$m in
Figure~\ref{fig:Si2S3vsO4Ne2}. We also plot these line ratios measured in the
coadded CMZ spectrum, and compare to the line ratios measured in star-forming
galaxies and AGNs \citep{dale:09}.

We find that only $3\%$ of GC points fall in the AGN-like region of
Figure~\ref{fig:Si2S3vsO4Ne2}. Virtually all GC data points have ratios of
[\ion{Si}{2}] $34.82\ \mu$m / [\ion{S}{3}] $33.48\ \mu$m and [\ion{O}{4}]
$25.89\ \mu$m / [\ion{Ne}{2}] $12.81\ \mu$m agreeing with the line ratios
observed in star-forming galaxies. The few points that are not similar to
star-forming galaxies are those at high Galactic latitudes ($l\sim+0.4\arcdeg$,
$b\sim+0.2\arcdeg$ and $l\sim+0.6\arcdeg$, $b\sim+0.2\arcdeg$).

We considered whether the positions with AGN-like line ratios could be due
to having been irradiated by Sgr~A* if it was in a more active state in the
past.  This idea has been suggested to explain bright \ion{Fe}{1}~K$\alpha$
emission at $6.4$~keV in Sgr~B2 and other molecular clouds in the GC
\citep{sunyaev:93,koyama:96,murakami:01}.  The \ion{Fe}{1}~K$\alpha$ emission in
the GC varies spatially and temporily \citep{terrier:10}. \citet{capelli:12}
analyzed the \ion{Fe}{1}~K$\alpha$ emission to derive a light curve for the
X-ray luminosity of Sgr~A* over the last several hundred years.  Apparent
superluminal motion has also been observed in $6.4$~keV \ion{Fe}{1}~K$\alpha$
emission \citep{ponti:10}.  Others argue, however, that widespread
\ion{Fe}{1}~K$\alpha$ emission in the GC can be explained by cosmic rays
\citep{yz:07,yz:13,chernyshov:12,tatischeff:12}.  \citet{yz:13} show, in their
Figure~8a, an image of the equivalent width of \ion{Fe}{1}~K$\alpha$ emission,
covering $-0.8\arcdeg < l < +0.7\arcdeg$ and $\pm0.4\arcdeg$ in $b$.  We
compared this image to our Figure~\ref{fig:lineratio_b}, an image of the
[\ion{O}{4}]/[\ion{Ne}{2}] line ratio in the GC ISM.  We find no correlation
between GC ISM positions with AGN-like line ratios and regions of enhanced
\ion{Fe}{1}~K$\alpha$ emission.  Our data do not support -- but neither do they
rule out -- the idea that strong $6.4$~keV \ion{Fe}{1}~K$\alpha$ emission is
caused by higher X-ray luminosity of Sgr A* in the past.

We conclude that our observations of mid-IR line emission in the central
$210$~pc $\times 60$~pc of the Galaxy show no evidence of excitation by a
power-law continuum.  The few points that are not similar to star-forming
galaxies are outliers, such as those excited by very hot sources (e.g.,
Wolf-Rayet stars, planetary nebulae, or X-ray binaries), rather than belonging
to an AGN-like trend.  Our results agree with those of \citet{simpson:07} and
\citet{contini:09}, who conclude that the mid-IR line ratios in a smaller {\it
Spitzer} data set can be explained by photoionization by hot stars, combined
with shocks.

Figure~\ref{fig:Si2S3vsO4Ne2} shows a significant offset between the median
value of [\ion{O}{4}] $25.89\ \mu$m / [\ion{Ne}{2}] $12.81\ \mu$m measured from
individual GC spectra and the value from the coadded CMZ spectrum.  Emission
from [\ion{Ne}{2}] $12.81\ \mu$m, [\ion{Si}{2}] $34.82\ \mu$m, and [\ion{S}{3}]
$33.48\ \mu$m is detected in virtually all  individual GC spectra. The offset,
then, is due to faint [\ion{O}{4}] $25.89\ \mu$m emission which is missing from
the number distribution for individual spectra (we only plot 3$\sigma$
detections) but which contributes to the coadded spectrum.  Because the
[\ion{O}{4}] $25.89\ \mu$m / [\ion{Ne}{2}] $12.81\ \mu$m line ratio from the
coadded spectrum is at the lower end of the range of GC line ratios on
Figure~\ref{fig:Si2S3vsO4Ne2}, we can be confident that excluding regions of
faint [\ion{O}{4}] $25.89\ \mu$m  emission does not change our conclusion that
the GC is similar to star-forming galaxies and not similar to AGNs.

\section{Summary}\label{sec:summary}

We present a mid-IR spectroscopic survey of $428$ positions in the ISM of the
CMZ using the {\it Spitzer}/IRS, and construct mid-IR emission line maps for
several forbidden and molecular hydrogen lines over the CMZ.  We derive both
line strengths and radial velocities from individual lines, and compute line
flux ratios as a probe of physical conditions in the GC.  Our mapping is
superior to previous survey results in terms of the total area covered in the
GC. We also construct a CMZ spectrum by coadding individual spectra after
correcting each for extinction.

Mid-IR emission lines from the pure rotational transitions of molecular
hydrogen, S(0), S(1), and S(2), are observed in almost all lines of sight to the
CMZ.  Their intensity distribution in the GC is relatively uniform; their radial
velocity distributions are poorly correlated with those from ionic species, with
worse correlation for S(0) than S(2). We view this as evidence that most of the
H$_2$ S(0) emission, and some of the S(1) and S(2) emission, arises from PDRs
along the line of sight to the GC, rather than PDRs associated with ionized gas
in the GC.

The radiation field hardness traced by the [\ion{Ne}{3}] $15.56\ \mu$m /
[\ion{Ne}{2}] $12.81\ \mu$m line ratio indicates that the highest excitation gas
clouds are found in the Radio Bubble region and Quintuplet cluster, and the mean
value in the GC is consistent with a recent burst of star formation in the last
few million years.  The hardness of the ionization spectrum from hot stars is
tied to the metal abundance of the star-forming clouds, and our GC spectra,
combined with the published GC stellar oxygen abundance, show that the hardness
of the GC exciting radiation is similar to that found in normal star-forming
regions in nearby galaxies.

We present mid-IR line-ratio diagrams such as [\ion{Si}{2}] $34.82\ \mu$m /
[\ion{S}{3}] $33.48\ \mu$m vs.\ [\ion{Fe}{2}] $25.99\ \mu$m / [\ion{Ne}{2}]
$12.81\ \mu$m, [\ion{O}{4}] $25.89\ \mu$m / [\ion{Ne}{2}] $12.81\ \mu$m vs.\
[\ion{Fe}{2}] $25.99\ \mu$m / [\ion{O}{4}] $25.89\ \mu$m, [\ion{O}{4}] $25.89\
\mu$m / [\ion{Ne}{2}] $12.81\ \mu$m vs.\ [\ion{Ne}{3}] $15.56\ \mu$m /
[\ion{Ne}{2}] $12.81\ \mu$m, and [\ion{Si}{2}] $34.82\ \mu$m / [\ion{S}{3}]
$33.48\ \mu$m vs.\ [\ion{O}{4}] $25.89\ \mu$m / [\ion{Ne}{2}] $12.81\ \mu$m.  We
compare properties of individual GC spectra to those observed in nearby
extragalactic star-forming regions and AGNs.  These diagrams show that the
individual GC spectra and the mean GC spectrum are consistent with normal star
forming activity, where emission from highly ionized species such as
[\ion{O}{4}] is likely produced by shocks and/or turbulence prevalent in the CMZ
clouds.  Our GC line ratios do not agree with line ratios observed for LINER
galaxies or AGNs.

\acknowledgements

We thank Janet Simpson for a detailed set of comments, which greatly improved
our manuscript.  DA thanks Daniel Jaffe for useful discussions, and Daniel Dale
for generously providing SINGS data used in this paper. D.A.\ was supported by
Basic Science Research Program through the National Research Foundation of Korea
(NRF) funded by the Ministry of Education, Science and Technology
(2010--0025122). This work is based on observations made with the Spitzer Space
Telescope, which is operated by the Jet Propulsion Laboratory, California
Institute of Technology under a contract with NASA. Support for this work was
provided by NASA through an award issued by JPL/Caltech.  This research has made
use of the SIMBAD database, operated at CDS, Strasbourg, France.

\appendix

\section{[\ion{S}{3}] $18.71\ \mu$m / [\ion{S}{3}] $33.48\ \mu$m}\label{sec:appendix}

Line ratios from the same ionized species such as [\ion{S}{3}] $18.71\ \mu$m /
$33.48\ \mu$m are generally sensitive to the electron gas density ($n_e$) with a weak
dependence on the electron gas temperature \citep[e.g.,][]{rubin:89,martin:02,dale:09}.
But [\ion{S}{3}] $18.71\ \mu$m is found in the middle of the strong and broad 
$18\ \mu$m silicate feature, while [\ion{S}{3}] $33.48\ \mu$m is almost 
unaffected by foreground absorption (see the bottom panel in Figure~\ref{fig:spectra}).
Thus we, like \citet{simpson:07}, find that the [\ion{S}{3}] $18.71\ \mu$m / $33.48\ \mu$m 
ratio is most useful for checking our foreground extinction correction.

The grey dashed line for the [\ion{S}{3}] $18.71\ \mu$m and $33.48\ \mu$m ratio
in the right panel of Figure~\ref{fig:lineratio} indicates the observed line intensity ratios
without any extinction correction.
These values are located below the lower limit set by the theoretical value 
($0.4$; dotted horizontal line) as $n_e \rightarrow 0$ cm$^{-3}$ \citep{rubin:89}.\footnote{
New cross sections of \ion{S}{3} \citep{hudson:12} produce a lower limit of
$\sim0.5$ for this line ratio (J.\ Simpson 2012, private communication).}
On the other hand,
mean line ratios after extinction corrections using either the \citet{simpson:07}
method (red solid line) or the \citet{schultheis:09} map (green dashed line) are
higher than the theoretical limit, clearly showing a strong dependence of
the line ratio on the foreground extinction correction.

Over most parts of the GC, the [\ion{S}{3}] $18.71\ \mu$m / [\ion{S}{3}] $33.48\ \mu$m
ratios are close to unity after the extinction correction. 
This line ratio value corresponds to $n_e \sim 500$~cm$^{-3}$ \citep{martin:02,dale:09}.
Mean line ratios from either of the extinction estimation techniques used above
are in excellent agreement with each other, and are relatively uniform across
the CMZ. However, these average line ratios become lower than the minimum theoretical
value ($\approx0.4$) near Sgr~B2. Sgr~B2 has very high extinction (Figure~\ref{fig:tau}),
so we view this as due to unrecognized systematic errors in our extinction estimates from
the \citet{simpson:07} and \citet{schultheis:09} approaches that greatly underestimated
$\tau_{9.7}$ in this region.
\citet{goicoechea:04} found $n_e \sim 240\ {\rm cm^{-1}}$ in Sgr~B2 based 
on ISO observations.

\end{document}